\documentclass[aps,prd,onecolumn,nofootinbib,preprintnumbers,superscriptaddress]{revtex4}
\usepackage[mathscr]{eucal}
\usepackage{color}
\usepackage[dvips]{graphicx}
\usepackage{epsf}
\usepackage{bm}

\usepackage{dcolumn}
\usepackage{bm}
\usepackage{amsmath}
\usepackage{amssymb}
\usepackage{amsfonts}
\usepackage{amsthm}
\usepackage{amscd}
\usepackage{graphicx}

\def\slashchar#1{\setbox0=\hbox{$#1$}           
   \dimen0=\wd0                                     
   \setbox1=\hbox{/} \dimen1=\wd1                   
   \ifdim\dimen0>\dimen1                            
      \rlap{\hbox to \dimen0{\hfil/\hfil}}          
      #1                                            
   \else                                            
      \rlap{\hbox to \dimen1{\hfil$#1$\hfil}}       
      /                                             
   \fi}

\begin{document}

\title{Holographic fermions in external magnetic fields}

\author{E. Gubankova}
\altaffiliation{Also at ITEP, Moscow, Russia.}
\affiliation{Institute for Theoretical Physics, J. W. Goethe-University,\\
D-60438 Frankfurt am Main, Germany}
\author{J.~Brill}
\author{M. \v{C}ubrovi\'{c}}
\author{K. Schalm}
\author{P. Schijven}
\altaffiliation{Address from October 1st: Albert-Ludwigs-Universit\"{a}t Freiburg
D-79104 Freiburg, Germany.}
\author{J. Zaanen}
\affiliation{Instituut Lorentz, Leiden University, Niels Bohrweg 2,\\
2300 RA Leiden, Netherlands}

\begin{abstract}
We study the Fermi level structure of $2+1$-dimensional strongly
interacting electron systems in external magnetic field using the
AdS/CFT correspondence. The gravity dual of a finite density
fermion system is a Dirac field in the background of the dyonic
AdS-Reissner-Nordstr\"{o}m black hole. In the probe limit the
magnetic system can be reduced to the non-magnetic one, with
Landau-quantized momenta and rescaled thermodynamical variables.
We find that at strong enough magnetic fields, the Fermi surface
vanishes and the quasiparticle is lost either through a crossover
to conformal regime or through a phase transition to an unstable
Fermi surface. In the latter case, the vanishing Fermi velocity at
the critical magnetic field triggers the non-Fermi liquid regime
with unstable quasiparticles and a change in transport properties
of the system. We associate it with a metal-"strange metal" phase
transition.
Next we compute compute the DC Hall and longitudinal
conductivities using the gravity-dressed fermion propagators. For
dual fermions with a large charge, many different Fermi surfaces
contribute and the Hall conductivity is quantized as expected for
integer Quantum Hall Effect (QHE). At strong magnetic fields, as
additional Fermi surfaces open up, new plateaus typical for the
fractional QHE appear. The somewhat irregular pattern in the
length of fractional QHE plateaus resemble the outcomes of
experiments on thin graphite in a strong magnetic field. Finally,
motivated by the absence of the sign problem in holography, we
suggest a lattice approach to the
AdS calculations of finite density systems.\\
\\
\\
\\
Keywords: AdS/CFT, strongly correlated electrons, transport,
lattice
\end{abstract}


\maketitle

\section{Introduction}

The study of strongly interacting fermionic systems at finite
density and temperature is a challenging task in condensed matter
and high energy physics. Analytical methods are limited or not
available for strongly coupled systems, and numerical simulation
of fermions at finite density breaks down because of the sign
problem \cite{signs}. There has been an increased activity in
describing finite density fermionic matter by a gravity dual using
the holographic AdS/CFT correspondence \cite{review}. The
gravitational solution dual to the finite chemical potential
system is the electrically charged AdS-Reissner-Nordstr\"{o}m (RN)
black hole, which provides a background where only the metric and
Maxwell fields are nontrivial and all matter fields vanish. In the
classical gravity limit, the decoupling of the Einstein-Maxwell
sector holds and leads to universal results, which is an appealing
feature of applied holography. Indeed, the celebrated result for
the ratio of the shear viscosity over the entropy density
\cite{visc} is identical for many strongly interacting theories
and has been considered a robust prediction of the AdS/CFT
correspondence.

However, an extremal black hole alone is not enough to describe
finite density systems as it does not source the matter fields. In
holography, at leading order, the Fermi surfaces are not evident
in the gravitational geometry, but can only be detected by
external probes; either probe D-branes \cite{review} or probe bulk
fermions \cite{Lee:2008,Cubrovic:2009,Vegh:2009,Faulkner:2009}.
Here we shall consider the latter option, where the free Dirac
field in the bulk carries a finite charge density
\cite{Hartnoll:2010}. We ignore electromagnetic and gravitational
backreaction of the charged fermions on the bulk spacetime
geometry (probe approximation). At large temperatures, $T\gg \mu$,
this approach provides a reliable hydrodynamic description of
transport at a quantum criticality (in the vicinity of
superfluid-insulator transition) \cite{Hartnoll:2007}. At small
temperatures, $T\ll \mu$, in some cases sharp Fermi surfaces
emerge with either conventional Fermi-liquid scaling
\cite{Cubrovic:2009} or of a non-Fermi liquid type
\cite{Vegh:2009} with scaling properties that differ significantly
from those predicted by the Landau Fermi liquid theory. The
non-trivial scaling behavior of these non-Fermi liquids has been
studied semi-analytically in \cite{Faulkner:2009} and is of great
interest as  high-$T_c$ superconductors and metals near the
critical point are believed to represent non-Fermi liquids.


What we shall study is the effects of magnetic field on the
holographic fermions. A magnetic field is a probe of finite
density matter at low temperatures, where the Landau level physics
reveals the Fermi level structure. The gravity dual system is
described by a AdS dyonic black hole with electric and magnetic
charges $Q$ and $H$, respectively, corresponding to a
$2+1$-dimensional field theory at finite chemical potential in an
external magnetic field \cite{HallCondBH}. Probe fermions in the
background of the dyonic black hole have been considered in
\cite{Basu:2008,Albash:2010}; and probe bosons in the same
background have been studied in \cite{AJMagneticSC}. Quantum
magnetism is considered in \cite{Iqbal:2010}.

The Landau quantization of momenta due to the magnetic field found
there, shows again that the AdS/CFT correspondence has a powerful
capacity to unveil that certain quantum properties known from
quantum gases have a much more ubiquitous status than could be
anticipated theoretically. A first highlight is the demonstration
\cite{Faulkner:2010} that the Fermi surface of the Fermi gas
extends way beyond the realms of its perturbative extension in the
form of the Fermi-liquid. In AdS/CFT it appears to be
gravitationally encoded in the matching along the scaling
direction between the 'bare' Dirac waves falling in from the 'UV'
boundary, and the true IR excitations living near the black hole
horizon. This IR physics can insist on the disappearance of the
quasiparticle but, if so, this 'critical Fermi-liquid' is still
organized 'around' a Fermi surface. The Landau quantization, the
organization of quantum gaseous matter in quantized energy bands
(Landau levels) in a system of two space dimensions pierced by a
magnetic field oriented in the orthogonal spatial direction, is a
second such quantum gas property. We shall describe here following
\cite{Basu:2008}, that despite the strong interactions in the
system, the holographic computation reveals the same strict
Landau-level quantization. Arguably, it is the mean-field nature
imposed by large $N$ limit inherent in AdS/CFT that explains this.
The system is effectively non-interacting to first order in $1/N$.
The Landau quantization is not manifest from the geometry, but as
we show this statement is straightforwardly encoded in the
symmetry correspondences associated with the conformal
compactification of $AdS$ on its flat boundary (i.~e., in the UV
CFT).

An interesting novel feature in strongly coupled systems arises
from the fact that the background geometry is only sensitive to
the total energy density $Q^2+H^2$ contained in the electric and
magnetic fields sourced by the dyonic black hole. Dialing up the
magnetic field is effectively similar to a process where the
dyonic black hole loses its electric charge. At the same time, the
fermionic probe with charge $q$ is essentially only sensitive to
the Coulomb interaction $gqQ$. As shown in \cite{Basu:2008}, one
can therefore map a magnetic to a non-magnetic system with
rescaled parameters (chemical potential, fermion charge) and same
symmetries and equations of motion, as long as the
Reissner-Nordstr\"{o}m geometry is kept.

Translated to more experiment-compatible language, the above
magnetic-electric mapping means that the spectral functions at
nonzero magnetic field $h$ are identical to the spectral function
at $h=0$ for a reduced value of the coupling constant (fermion
charge) $q$, provided the probe fermion is in a Landau level
eigenstate. A striking consequence is that the spectrum shows
conformal invariance for arbitrarily high magnetic fields, as long
as the system is at negligible to zero density. Specifically, a
detailed analysis of the fermion spectral functions reveals that
at strong magnetic fields the Fermi level structure changes
qualitatively. There exists a critical magnetic field at which the
Fermi velocity vanishes. Ignoring the Landau level quantization,
we show that this corresponds to an effective tuning of the system
from a regular Fermi liquid phase with linear dispersion and
stable quasiparticles to a non-Fermi liquid liquid with fractional
power law dispersion and unstable excitations. This phenomenon can
be interpreted as a transition from metallic phase to a "strange
metal" at the critical magnetic field and corresponds to the
change of the infrared conformal dimension from $\nu>1/2$ to
$\nu<1/2$ while the Fermi momentum stays nonzero and the Fermi
surface survives. Increasing the magnetic field further, this
transition is followed by a "strange-metal"-conformal crossover
and eventually, for very strong fields, the system always has
near-conformal behavior where $k_F=0$ and the Fermi surface
disappears.

For some Fermi surfaces, this surprising metal-"strange metal"
transition is not physically relevant as the system prefers to
directly enter the conformal phase. Whether a fine tuned system
exists that does show a quantum critical phase transition from a
FL to a non-FL is determined by a Diophantine equation for the
Landau quantized Fermi momentum as a function of the magnetic
field. Perhaps these are connected to the magnetically driven
phase transition found in AdS$_5$/CFT$_4$ \cite{D'Hoker:2010ij}.
We leave this subject for further work.

Overall, the findings of Landau quantization and "discharge" of
the Fermi surface are in line with the expectations: both
phenomena have been found in a vast array of systems
\cite{Auerbach} and are almost tautologically tied to the notion
of a Fermi surface in a magnetic field. Thus we regard them also
as a sanity check of the whole bottom-up approach of fermionic
AdS/CFT \cite{Lee:2008,Vegh:2009,Cubrovic:2009,Faulkner:2010},
giving further credit to the holographic Fermi surfaces as having
to do with the real world.

Next we use the information of magnetic effects the Fermi surfaces
extracted from holography to calculate the quantum Hall and
longitudinal conductivities. Generally speaking, it is difficult
to calculate conductivity holographically beyond the
Einstein-Maxwell sector, and extract the contribution of
holographic fermions. In the semiclassical approximation, one-loop
corrections in the bulk setup involving charged fermions have been
calculated \cite{Faulkner:2010}. In another approach, the
backreaction of charged fermions on the gravity-Maxwell sector has
been taken into account and incorporated in calculations of the
electric conductivity \cite{Hartnoll:2010}. We calculate the
one-loop contribution on the CFT side, which is equivalent to the
holographic one-loop calculations as long as vertex corrections do
not modify physical dependencies of interest
\cite{Faulkner:2010,Gubankova:2010}. As we dial the magnetic
field, the Hall plateau transition happens when the Fermi surface
moves through a Landau level. One can think of a difference
between the Fermi energy and the energy of the Landau level as a
gap, which vanishes at the transition point and the
$2+1$-dimensional theory becomes scale invariant. In the
holographic D3-D7 brane model of the quantum Hall effect, plateau
transition occurs as D-branes move through one another
\cite{Davis:2008}. In the same model, a dissipation process has
been observed as D-branes fall through the horizon of the black
hole geometry, that is associated with the quantum Hall insulator
transition. In the holographic fermion liquid setting, dissipation
is present through interaction of fermions with the horizon of the
black hole. We have also used the analysis of the conductivities
to learn more about the metal-strange metal phase transition as
well as the crossover back to the conformal regime at high
magnetic fields.

We conclude with the remark that the findings summarized above are
in fact somewhat puzzling when contrasted to the conventional
picture of quantum Hall physics. It is usually stated that the
quantum Hall effect requires three key ingredients: Landau
quantization, quenched disorder \footnote{Quenched disorder means
that the dynamics of the impurities is "frozen", i.~e. they can be
regarded as having infinite mass. When coupled to the Fermi
liquid, they ensure that below some
scale the system behaves as if consisting
of non-interacting quasiparticles only.} and (spatial)
boundaries, i.~e., a finite-sized sample \cite{IntroQHE}. The
first brings about the quantization of conductivity, the second
prevents the states from spilling between the Landau levels
ensuring the existence of a gap and the last one in fact allows
the charge transport to happen (as it is the boundary states that
actually conduct). In our model, only the first condition is
satisfied. The second is put by hand by assuming that the gap is
automatically preserved, i.~e. that there is no mixing between the
Landau levels. There is, however, no physical explanation as to
how the boundary states are implicitly taken into account by
AdS/CFT.

The paper is organized as follows. We outline the holographic
setting of the dyonic black hole geometry and bulk fermions in the
section \ref{section:1}. In section \ref{section:2} we prove the
conservation of conformal symmetry in the presence of the magnetic
fields. Section \ref{section:3} is devoted to the holographic
fermion liquid, where we obtain the Landau level quantization,
followed by a detailed study of the Fermi surface properties at
zero temperature in section \ref{section:4}. We calculate the DC
conductivities in section \ref{section:5}, and compare the results
with available data in graphene. In section \ref{section:6}, we
show that the fermion sign problem is absent in the holographic
setting, therefore allowing lattice simulations of finite density
matter in principle.

\section{Holographic fermions in a dyonic black hole}\label{section:1}

We first describe the holographic setup with the dyonic black
hole, and the dynamics of Dirac fermions in this background. In
this paper, we exclusively work in the probe limit, i.~e., in the
limit of large fermion charge $q$.

\subsection{Dyonic black hole}

We consider the gravity dual of $3$-dimensional conformal field
theory (CFT) with global $U(1)$ symmetry. At finite charge density
and in the presence of magnetic field, the system can be described
by a dyonic black hole in $4$-dimensional anti-de Sitter
space-time, $AdS_4$, with the current $J_{\mu}$ in the CFT mapped
to a $U(1)$ gauge field $A_M$ in $AdS$. We use $\mu,\nu,
\rho,\ldots$ for the spacetime indices in the CFT and $M,N,\ldots$
for the global spacetime indices in $AdS$.

The action for a vector field $A_M$ coupled to $AdS_4$ gravity can be written as
\begin{equation}
S_g=\frac{1}{2\kappa^2}\int d^4x \sqrt{-g}\left( {\mathcal R} +\frac{6}{R^2} -\frac{R^2}{g_F^2}F_{MN}F^{MN}\right),
\label{action-g}
\end{equation}
where $g_F^2$ is an effective dimensionless gauge coupling and $R$
is the curvature radius of $AdS_4$. The equations of motion
following from eq. (\ref{action-g}) are solved by the geometry
corresponding to a dyonic black hole, having both electric and
magnetic charge:
\begin{equation}
ds^2=g_{MN}dx^Mdx^N =
\frac{r^2}{R^2}\left(-fdt^2+dx^2+dy^2\right)+\frac{R^2}{r^2}\frac{dr^2}{f}.
\label{ads4-metric1}
\end{equation}
The redshift factor $f$ and the vector field $A_M$ reflect the
fact that the system is at a finite charge density and in an
external magnetic field:
\begin{eqnarray}
 f &=& 1+\frac{Q^2+H^2}{r^4}-\frac{M}{r^3},\nonumber\\
A_t &=& \mu\left(1-\frac{r_0}{r}\right),\;\; A_y =
hx,\;\;A_x=A_r=0, \label{ads4-metric2}
\end{eqnarray}
where $Q$ and $H$ are the electric and magnetic charge of the
black hole, respectively. Here we chose the Landau gauge; the
black hole chemical potential $\mu$ and the magnetic field $h$ are
given by
\begin{equation}
 \mu=\frac{g_FQ}{R^2r_0},\;\; h = \frac{g_F H}{R^4},
\end{equation}
with $r_0$ is the horizon radius determined by the largest
positive root of the redshift factor $f(r_0)=0$:
\begin{eqnarray}
M=r_0^3 + \frac{Q^2+H^2}{r_0}.
\end{eqnarray}
The boundary of the $AdS$ is reached for $r\rightarrow\infty$. The
geometry described by eqs. (\ref{ads4-metric1}-\ref{ads4-metric2})
describes the boundary theory at finite density, i.~e., a system
in a charged medium at the chemical potential
$\mu=\mu_\mathrm{bh}$ and in transverse magnetic field
$h=h_\mathrm{bh}$, with charge, energy, and entropy densities
given, respectively, by
\begin{eqnarray}
\rho = 2\frac{Q}{\kappa^2R^2g_F},\;\;
\epsilon =\frac{M}{\kappa^2R^4},\;\;
s=\frac{2\pi}{\kappa^2}\frac{r_0^2}{R^2}.
\end{eqnarray}
The temperature of the system is identified with the Hawking
temperature of the black hole, $T_H\sim |f^{\prime}(r_0)|/4\pi$,
\begin{equation}
T=\frac{3r_0}{4\pi R^2}\left(1-\frac{Q^2+H^2}{3r_0^4}\right).
\end{equation}
Since $Q$ and $H$ have dimensions of $[L]^2$, it is convenient to
parametrize them as
\begin{equation}
 Q^2 = 3r_{*}^4, \;\;
Q^2+H^2 = 3r_{**}^4.
\label{q-and-h}
\end{equation}
In terms of $r_0$, $r_{*}$ and $r_{**}$ the above expressions
become
\begin{eqnarray}
f &=& 1+\frac{3r_{**}^4}{r^4}-\frac{r_0^3+3r_{**}^4/r_0}{r^3},
\label{factor}
\end{eqnarray}
with
\begin{equation}
\mu=\sqrt{3}g_F\frac{r_{*}^2}{R^2r_0},\;\;
h=\sqrt{3}g_F\frac{\sqrt{r_{**}^4-r_{*}^4}}{R^4}.
\end{equation}
The expressions for the charge, energy and entropy densities, as
well as for the temperature are simplified as
\begin{eqnarray}
\rho&=&\frac{2\sqrt{3}}{\kappa^2g_F}\frac{r_{*}^2}{R^2},\;\;
\epsilon=\frac{1}{\kappa^2}\frac{r_0^3+3r_{**}^4/r_0}{R^4},\;\;
s=\frac{2\pi}{\kappa^2}\frac{r_0^2}{R^2},\nonumber\\
T&=& \frac{3}{4\pi}\frac{r_0}{R^2}\left(1-\frac{r_{**}^4}{r_0^4}\right).
\end{eqnarray}
In the zero temperature limit, i.~e., for an extremal black hole,
we have
\begin{equation}
 T=0 \;\; \rightarrow  \;\; r_0=r_{**},
\label{zeroT}
\end{equation}
which in the original variables reads $Q^2+H^2=3r_0^4$. In the
zero temperature limit (\ref{zeroT}), the redshift factor $f$ as
given by eq. (\ref{factor}) develops a double zero at the horizon:
\begin{equation}
 f= 6\frac{(r-r_{**})^2}{r_{**}^2}+\mathcal{O}\left(\left(r-r_{**}\right)^3\right) \,.
\end{equation}
As a result, near the horizon the $AdS_4$ metric reduces to $AdS_2 \times \mathbb{R}^2$
with the curvature radius of $AdS_2$ given by
\begin{equation}
 R_2=\frac{1}{\sqrt{6}}R.
\end{equation}
This is a very important property of the metric, which
considerably simplifies the calculations, in particular in the
magnetic field.


In order to scale away the $AdS_4$ radius $R$ and
the horizon radius $r_0$,
we introduce dimensionless variables
\begin{eqnarray}
&& r\rightarrow r_0 r,\;\;
r_{*}\rightarrow r_0 r_{*},\;\; r_{**}\rightarrow r_0 r_{**},
\nonumber\\
&& M\rightarrow r_0^3 M,\;\; Q\rightarrow r_0^2 Q,\;\;, H\rightarrow r_0^2H,
\label{dimensionless1}
\end{eqnarray}
and
\begin{eqnarray}
&& (t,\vec{x})\rightarrow \frac{R^2}{r_0}(t,\vec{x}),\;\;
A_{M}\rightarrow \frac{r_0}{R^2}A_M,\;\; \omega\rightarrow \frac{r_0}{R^2}\omega,\;\;
\nonumber\\
&& \mu\rightarrow \frac{r_0}{R^2}\mu,\;\;
h\rightarrow \frac{r_0^2}{R^4}h,\;\; T\rightarrow \frac{r_0}{R^2}T,
\nonumber\\
&& ds^2\rightarrow R^2 ds^2.
\label{dimensionless2}
\end{eqnarray}
Note that the scaling factors in the above equation that describes
the quantities of the boundary field theory involve the curvature
radius of $AdS_4$, not $AdS_2$.

In the new variables we have
\begin{eqnarray}
T&=&\frac{3}{4\pi}\left(1-r_{**}^4\right)=\frac{3}{4\pi}\left(1-\frac{Q^2+H^2}{3}\right),\;\;
f=1+\frac{3r_{**}^4}{r^4}-\frac{1+3r_{**}^4}{r^3},
\nonumber\\
A_t&=&\mu\left(1-\frac{1}{r}\right),\;\;
\mu=\sqrt{3}g_Fr_{*}^2=g_F Q,\;\; h = g_F H,
\label{dimensionless3}
\end{eqnarray}
and the metric is given by
\begin{equation}
ds^2=r^2\left(-fdt^2+dx^2+dy^2\right)+\frac{1}{r^2}\frac{dr^2}{f},
\label{dim-metric}
\end{equation}
with the horizon at $r=1$, and the conformal boundary at $r\rightarrow \infty$.

At $T=0$, $r_{**}$ becomes unity, and the redshift factor develops
the double zero near the horizon,
\begin{equation}
f= \frac{(r-1)^2(r^2+2r+3)}{r^4}.
\end{equation}
As mentioned before, due to this fact the metric near the horizon
reduces to $AdS_2\times \mathbb{R}^2$ where the analytical calculations are
possible for small frequencies \cite{Faulkner:2009}. However, in
the chiral limit $m=0$, analytical calculations are also possible
in the bulk $AdS_4$ \cite{Hartman:2010}, which we utilize in this
paper.

\subsection{Holographic fermions}

To include the bulk fermions, we consider a spinor field $\psi$ in
the $AdS_4$ of charge $q$ and mass $m$, which is dual to an
operator ${\mathcal O}$ in the boundary $CFT_3$ of charge $q$ and
dimension
\begin{equation}
\Delta = \frac{3}{2} + mR,
\label{conformal-dimension}
\end{equation}
with $mR\geq -\frac{1}{2}$ and in dimensionless units corresponds
to $\Delta=\frac{3}{2}+m$. In the black hole geometry, eq.
(\ref{ads4-metric1}), the quadratic action for $\psi$ reads as
\begin{equation}
S_{\psi} = i\int d^4x \sqrt{-g}\left(\bar{\psi}\Gamma^{M}{\mathcal D}_{M}\psi-m\bar{\psi}\psi\right),
\label{action-psi}
\end{equation}
where $\bar{\psi}=\psi^{\dagger}\Gamma^{\underline t}$, and
\begin{equation}
{\mathcal D}_M=\partial_M +\frac{1}{4}\omega_{abM}\Gamma^{ab}-iqA_M,
\end{equation}
where $\omega_{abM}$ is the spin connection, and
$\Gamma^{ab}=\frac{1}{2}[\Gamma^a,\Gamma^b]$. Here, $M$ and $a,b$
denote the bulk space-time and tangent space indices respectively,
while $\mu,\nu$ are indices along the boundary directions, i.~e.
$M=(r,\mu)$. Gamma matrix basis (Minkowski signature) is given by
eq. (\ref{matrices}) as in \cite{Faulkner:2009}.

We will be interested in spectra and response functions of the
boundary fermions in the presence of magnetic field. This requires
solving the Dirac equation in the bulk
\cite{Vegh:2009,Cubrovic:2009}:
\begin{equation}
 \left(\Gamma^M{\mathcal D}_M-m\right)\psi=0.
\label{direq}
\end{equation}
From the solution of the Dirac equation at small $\omega$, an analytic expression for the
retarded fermion Green's function of the boundary CFT at zero
magnetic field has been obtained in \cite{Faulkner:2009}. Near the
Fermi surface it reads as \cite{Faulkner:2009}:
\begin{equation}
G_R(\Omega,k)=\frac{(-h_1v_F)}{\omega-v_Fk_{\perp}-\Sigma(\omega,T)},
\label{green-function2}
\end{equation}
where $k_{\perp}=k-k_F$ is the perpendicular distance from the
Fermi surface in momentum space, $h_1$ and $v_F$ are real
constants calculated below, and the self-energy
$\Sigma=\Sigma_1+i\Sigma_2$ is given by \cite{Faulkner:2009}
\begin{eqnarray}
\Sigma(\omega,T)/v_F=T^{2\nu}g(\frac{\omega}{T})=
(2\pi T)^{2\nu}h_2{\rm e}^{i\theta-i\pi\nu}\frac{\Gamma(\frac{1}{2}+\nu-\frac{i\omega}{2\pi T}+\frac{i\mu_q}{6})}
{\Gamma(\frac{1}{2}-\nu-\frac{i\omega}{2\pi T}+\frac{i\mu_q}{6})},
\end{eqnarray}
where $\nu$ is the zero temperature conformal dimension at the
Fermi momentum, $\nu\equiv \nu_{k_F}$, given by eq. (\ref{nu}),
$\mu_q\equiv \mu q$, $h_2$ is a positive constant and the phase
$\theta$ is such that the poles of the Green's function are
located in the lower half of the complex frequency plane.
These poles correspond to
quasinormal modes of the Dirac Equation \eqref{direq} and they can be found numerically solving
$F(\omega_{*})=0$ \cite{Denef:2009}, with
\begin{eqnarray}
F(\omega)=\frac{k_{\perp}}{\Gamma(\frac{1}{2}+\nu-\frac{i\omega}{2\pi T}+\frac{i\mu_q}{6})}-
\frac{h_2{\rm e}^{i\theta-i\pi\nu}(2\pi T)^{2\nu}}{\Gamma(\frac{1}{2}-\nu-\frac{i\omega}{2\pi T}+\frac{i\mu_q}{6})},
\end{eqnarray}
The solution gives the full motion of the quasinormal poles
$\omega_{*}^{(n)}(k_{\perp})$ in the complex $\omega$ plane as a
function of $k_{\perp}$. It has been found in \cite{Denef:2009,Faulkner:2009},
that, if the charge of the fermion is large enough compared to its
mass, the pole closest to the real $\omega$ axis bounces off the
axis at $k_{\perp}=0$ (and $\omega=0$). Such behavior is
identified with the existence of the Fermi momentum $k_F$
indicative of an underlying strongly coupled Fermi surface.

At $T=0$, the self-energy becomes $T^{2\nu}g(\omega/T)\rightarrow
c_k\omega^{2\nu}$, and the Green's function obtained from the
solution to the Dirac equation reads \cite{Faulkner:2009}
\begin{equation}
G_R(\Omega,k)=\frac{(-h_1v_F)}{\omega-v_Fk_{\perp}-h_2v_F{\rm e}^{i\theta-i\pi\nu}\omega^{2\nu}},
\label{green-function}
\end{equation}
where $k_{\perp}=\sqrt{k^2}-k_F$.
 The last term 
is
determined by the IR $AdS_2$ physics near the horizon. Other terms
are determined by the UV physics of the $AdS_4$ bulk.

The solutions to (\ref{direq}) have been studied in detail in
\cite{Vegh:2009,Cubrovic:2009,Faulkner:2009}. Here we simply
summarize the novel aspects due to the background magnetic field
(formal details can be found in the Appendix).
\begin{itemize}
\item The background magnetic field $h$ introduces a
discretization of the momentum (see Appendix A for details):
\begin{equation}
k\to k_{\rm{eff}}=\sqrt{2|qh|l},\;\;{\rm with}\;\; l\in N,
\label{rule}
\end{equation}
with Landau level index $l$ \cite{Denef:2009,Albash:2010}.
These discrete values of $k$ are the
analogue of the well-known Landau levels that occur in magnetic
systems. \item There exists a (non-invertible) mapping on the
level of Green's functions, from the magnetic system to the
non-magnetic one by sending
\begin{equation}
\label{eq:GFMapping}
\left(H,Q,q\right)\mapsto
\left(0,\sqrt{Q^2+H^2},q\sqrt{1 - \frac{H^2}{Q^2+H^2}}\right).
\end{equation}
The Green's functions in a magnetic system are thus equivalent to
those in the absence of magnetic fields. To better appreciate
that, we reformulate eq. (\ref{eq:GFMapping}) in terms of the
boundary quantities:
\begin{equation}
\label{GFMapping2}
\left(h,\mu_q,T\right)\mapsto
\left(0,\mu_q,T\left(1-\frac{h^2}{12\mu^2}\right)\right),
\end{equation}
where we used dimensionless variables defined in eqs.
(\ref{dimensionless1},\ref{dimensionless3}). The magnetic field
thus effectively decreases the coupling constant $q$ and increases
the chemical potential $\mu=g_FQ$ such that the combination
$\mu_q\equiv \mu q$ is preserved \cite{Basu:2008}. This is an
important point as the equations of motion actually only depend on
this combination and not on $\mu$ and $q$ separately
\cite{Basu:2008}. In other words, eq. (\ref{GFMapping2}) implies
that the additional scale brought about by the magnetic field can
be understood as changing $\mu$ and $T$ independently in the
effective non-magnetic system instead of only tuning the ratio
$\mu/T$. This point is important when considering the
thermodynamics. \item The discrete momentum
$k_{\mathrm{eff}}=\sqrt{2|qh|l}$ must be held fixed in the
transformation (\ref{eq:GFMapping}). The bulk-boundary relation is
particularly simple in this case, as the Landau levels can readily
be seen in the bulk solution, only to remain identical in the
boundary theory. \item Similar to the non-magnetic system
\cite{Basu:2008}, the IR physics is controlled by the near horizon
$AdS_2\times \mathbb{R}^2$ geometry, which indicates the existence of an IR
CFT, characterized by operators $\mathcal{O}_l$, $l\in N$ with
operator dimensions $\delta=1/2+\nu_l$:
\begin{equation}
\nu_l=\frac{1}{6}\sqrt{6\left(m^2+\frac{2|qh|l}{r_{**}^2}\right)-\frac{\mu_q^2}{r_{**}^4}},
\end{equation}
in dimensionless notation, and $\mu_q\equiv \mu q$. At $T=0$, when
$r_{**}=1$, it becomes
\begin{equation}
\nu_l=\frac{1}{6}\sqrt{6\left(m^2+2|qh|l\right)-\mu_q^2}.
\label{redefnul}
\end{equation}
The Green's function for these operators $\mathcal{O}_l$ is found
to be $\mathcal{G}_l^R(\omega) \sim \omega^{2\nu_l}$ and the
exponents $\nu_l$ determines the dispersion properties of the
quasiparticle excitations. For $\nu>1/2$ the system has a stable
quasiparticle and a linear dispersion, whereas for $\nu\leq 1/2$
one has a non-Fermi liquid with power-law dispersion and an
unstable quasiparticle.
\end{itemize}

\section{Magnetic fields and conformal invariance}\label{section:2}

Despite the fact that a magnetic field introduces a scale, in the
absence of a chemical potential, all spectral functions are
essentially still determined by conformal symmetry. To show this,
we need to establish certain properties of the near-horizon
geometry of a Reissner-Nordstr\"{o}m black hole. This leads to the
$AdS_2$ perspective that was developed in \cite{Faulkner:2009}.
The result relies on the conformal algebra and its relation to the
magnetic group, from the viewpoint of the infrared CFT that was
studied in \cite{Faulkner:2009}. Later on we will see that the
insensitivity to the magnetic field also carries over to $AdS_4$
and the UV CFT in some respects. To simplify the derivations, we
consider the case $T=0$.

\subsection{The near-horizon limit and Dirac equation in $AdS_2$}

It was established in \cite{Faulkner:2009} that an electrically
charged extremal $AdS$-Reissner-Nordstr\"{o}m black hole has an
$AdS_2$ throat in the inner bulk region. This conclusion carries
over to the magnetic case with some minor differences. We will now
give a quick derivation of the $AdS_2$ formalism for a dyonic
black hole, referring the reader to \cite{Faulkner:2009} for more
details (that remain largely unchanged in the magnetic field).

Near the horizon $r=r_{**}$ of the black hole described by the
metric (\ref{ads4-metric1}), the redshift factor $f(r)$ develops a
double zero:
\begin{equation}
f(r)=6\frac{(r-r_{**})^2}{r_{**}^2}+\mathcal{O}((r-r_{**})^3).
\end{equation}
Now consider the scaling limit
\begin{equation}
\label{eq:scalinglimit}
r-r_{**}=\lambda\frac{R_2^2}{\zeta},\;\;
t=\lambda^{-1}\tau,\;\; \lambda\to
0\;\;\mathrm{with}\;\;\tau,\zeta\,\mathrm{finite}.
\end{equation}
In this limit, the metric (\ref{ads4-metric1}) and the gauge field
reduce to
\begin{eqnarray}
ds^2 &=& \frac{R_2^2}{\zeta^2}\left(-d\tau^2+d\zeta^2\right)
+\frac{r_{**}^2}{R^2}\left(dx^2+dy^2\right)
\nonumber\\
A_{\tau} &=& \frac{\mu R_2^2r_0}{r_{**}^2}\frac{1}{\zeta},\;\; A_x=Hx
\label{ads2-metric}
\end{eqnarray}
where $R_2=\frac{R}{\sqrt{6}}$. The geometry described by this
metric is indeed $AdS_2\times \mathbb{R}^2$. Physically, the scaling limit
given in eq. (\ref{eq:scalinglimit}) with finite $\tau$ corresponds to the
long time limit of the original time coordinate $t$, which
translates to the low frequency limit of the boundary theory:
\begin{equation}
\frac{\omega}{\mu}\rightarrow 0,
\end{equation}
where $\omega$ is the frequency conjugate to $t$. (One can think
of $\lambda$ as being the frequency $\omega$). Near the $AdS_4$
horizon, we expect the $AdS_2$ region of an extremal dyonic black
hole to have a $\mathrm{CFT}_1$ dual. We refer to
\cite{Faulkner:2009} for an account of this $AdS_2/\mathrm{CFT}_1$
duality. The horizon of $AdS_2$ region is at $\zeta\rightarrow
\infty$ (coefficient in front of $d\tau$ vanishes at the horizon
in eq. (\ref{ads2-metric})) and the infrared CFT (IR CFT) lives at
the $AdS_2$ boundary at $\zeta=0$. The scaling picture given by
eqs. (\ref{eq:scalinglimit}-\ref{ads2-metric}) suggests that in
the low frequency limit, the $2$-dimensional boundary theory is
described by this IR CFT (which is a $\mathrm{CFT}_1$). The
Green's function for the operator ${\mathcal O}$ in the boundary
theory is obtained through a small frequency expansion and a
matching procedure between the two different regions (inner and
outer) along the radial direction, and can be expressed through
the Green's function of the IR CFT \cite{Faulkner:2009}.

The explicit form for the Dirac equation (\ref{dir}) in the
magnetic field is of little interest for the analytical results
that follow; for completeness, we give it in the Appendix. Of
primary interest is its limit in the IR region with metric given
by eq. (\ref{ads2-metric}):
\begin{eqnarray}
\left(-\frac{1}{\sqrt{g_{\zeta\zeta}}}\sigma^3\partial_{\zeta}- m
+\frac{1}{\sqrt{-g_{\tau\tau}}}\sigma^1\left(\omega+\frac{\mu_qR_2^2r_0}{r_{**}^2\zeta}\right)
-\frac{1}{\sqrt{g_{ii}}i\sigma^2\lambda_l}\right)F^{\left(l\right)}=0,\nonumber\\
\label{dir2}
\end{eqnarray}
where the effective momentum of the $l$-th Landau level is
$\lambda_l=\sqrt{2|qh|l}$, $\mu_q\equiv \mu q$ and we omit the
index of the spinor field. To obtain eq. (\ref{dir2}), it is
convenient to pick the gamma matrix basis as
$\Gamma^{\hat{\zeta}}=-\sigma_3$, $\Gamma^{\hat{\tau}}=i\sigma_1$
and $\Gamma^{\hat{i}} =-\sigma_2$. We can write explicitly:
\begin{equation}
\left(\begin{array}{cc}
\frac{\zeta}{R_2}\partial_{\zeta}+m &  -\frac{\zeta}{R_2}(\omega+
\frac{\mu_q R_2^2r_0}{r_{**}^2\zeta})+\frac{R}{r_{**}}\lambda_l
\\
\frac{\zeta}{R_2}(\omega+\frac{\mu_q R_2^2r_0}{r_{**}^2\zeta})
+\frac{R}{r_{**}}\lambda_l &
\frac{\zeta}{R_2}\partial_{\zeta} -m
\end{array}
\right)
\left(\begin{array}{c}
y \\
z
\end{array}
\right) =0.
\end{equation}
Note that the $AdS_2$ radius $R_2$ enters for the $(\tau,\zeta)$
directions. At the $AdS_2$ boundary, $\zeta\rightarrow 0$, the
Dirac equation to the leading order is given by
\begin{equation}
\zeta\partial_{\zeta}F^{\left(l\right)}=-UF^{\left(l\right)},\quad
U=R_2\left(\begin{array}{cc} m &  -\frac{\mu_qR_2r_0}{r_{**}^2}
+\frac{R}{r_{**}}\lambda_l
\\
\frac{\mu_qR_2r_0}{r_{**}^2}+\frac{R}{r_{**}}\lambda_l
 & -m
\end{array}
\right)
\label{ads2-dirac}
\end{equation}
The solution to this equation is given by the scaling function
$F^{(l)}= Ae_{+}\zeta^{-\nu_l}+Be_{-}\zeta^{\nu_l}$ where
$e_{\pm}$ are the real eigenvectors of $U$ and the exponent is
\begin{eqnarray}
\nu_l=\frac{1}{6}\sqrt{6\left(m^2+\frac{R^2}{r_{**}^2}2|qh|l\right)R^2-\frac{\mu_q^2R^4r_0^2}{r_{**}^4}}.
\label{nuexp}
\end{eqnarray}
The conformal dimension of the operator ${\mathcal O}$ in the
${\rm IR\; CFT}$ is $\delta_l=\frac{1}{2}+\nu_l$. Comparing eq.
(\ref{nuexp}) to the expression for the scaling exponent in
\cite{Faulkner:2009}, we conclude that the scaling properties and
the $AdS_2$ construction are unmodified by the magnetic field,
except that the scaling exponents are now fixed by the Landau
quantization. This "quantization rule" was already exploited in
\cite{Denef:2009} to study de Haas-van Alphen oscillations.

\section{Spectral functions}\label{section:3}

In this section we will explore some of the properties of the
spectral function, in both plane wave and Landau level basis. We
first consider some characteristic cases in the plane wave basis
and make connection with the ARPES measurements.

\subsection{Relating to the ARPES measurements}

In reality, ARPES measurements cannot be performed in magnetic
fields so the holographic approach, allowing a direct insight into
the propagator structure and the spectral function, is especially
helpful. This follows from the observation that the spectral
functions as measured in ARPES are always expressed in the plane
wave basis of the photon, thus in a magnetic field, when the
momentum is not a good quantum number anymore, it becomes
impossible to perform the photoemission spectroscopy.

In order to compute the spectral function, we have to choose a
particular fermionic plane wave as a probe. Since the separation
of variables is valid throughout the bulk, the basis
transformation can be performed at every constant $r$-slice. This
means that only the $x$ and $y$ coordinates have to be taken into
account (the plane wave probe lives only at the CFT side of the
duality). We take a plane wave propagating in the $+x$ direction
with spin up along the $r$-axis. In its rest frame such a particle
can be described by
\begin{equation}
\label{planwaveq}
\Psi_{\rm{probe}}=e^{i\omega t-ip_{x}x}
\left(\begin{array}{c}
\xi\\
\xi\end{array}\right),\qquad
\xi=\left(\begin{array}{c}
1\\
0\end{array}\right).
\end{equation}
Near the boundary (at $r_b\rightarrow\infty$) we can rescale our
solutions of the Dirac equation making use of eqs.
(\ref{solution1}-\ref{solution2}, \ref{GreenFunction}):
\begin{eqnarray}
F_{l}=\left(\begin{array}{c}
   \zeta^{(1)}_l(\tilde{x})\\
   \xi_{+}^{(l)}(r_b)\zeta^{(1)}_l(\tilde{x})\\
   \zeta^{(2)}_l(\tilde{x})\\
   -\xi_{+}^{(l)}(r_b)\zeta^{(2)}_l(\tilde{x})
  \end{array}\right),\quad
\tilde{F}_{l}=\left(\begin{array}{c}
   \zeta^{(1)}_l(\tilde{x})\\
   \xi_{-}^{(l)}(r_b)\zeta^{(1)}_l(\tilde{x})\\
   -\zeta^{(2)}_l(\tilde{x})\\
   \xi_{-}^{(l)}(r_b)\zeta^{(2)}_l(\tilde{x})
  \end{array}\right),
\label{boundary-solutions}
 \end{eqnarray}
with rescaled $\tilde{x}$ defined after eq. (\ref{eqx}). This
representation is useful since we calculate the components
$\xi_{\pm}(r_b)$ related to the retarded Green's function in our
numerics (we keep the notation of \cite{Faulkner:2009}).

Let $\mathcal{O}_l$ and $\tilde{\mathcal{O}}_l$ be the CFT
operators dual to respectively $F_{l}$ and
$\tilde{F}_{l}$, and $c_k^{\dagger}$, $c_k$ be the creation and
annihilation operators for the plane wave state
$\Psi_{\rm{probe}}$. Since the states $F$ and
$\tilde{F}$ form a complete set in the bulk, we can write
\begin{equation}
c_p^{\dagger}(\omega)=\sum_l \left(U_{l}^{*},\tilde{U}_l^{*}\right)
\left(\begin{array}{c}
\mathcal{O}_l^{\dagger}(\omega)\\
\tilde{\mathcal{O}}_l^{\dagger}(\omega)
      \end{array}
\right)=\sum_l
\left(U_{l}^{*}\mathcal{O}_l^{\dagger}(\omega)
+\tilde{U}_{l}^{*}\tilde{\mathcal{O}}_l^{\dagger}(\omega)\right)
\label{represent}
\end{equation}
where the overlap coefficients $U_l(\omega)$ are given by the
inner product between $\Psi_{\rm{probe}}$ and $F$:
\begin{eqnarray}
U_l(p_x)=\int dx F_l^{\dagger}i\Gamma^0\Psi_{\rm{probe}}
=-\int dx e^{-i p_x x}\xi_{+}(r_b)\left(\zeta^{(1)\dagger}_l(\tilde{x})-\zeta^{(2)\dagger}_l(\tilde{x})\right),
\end{eqnarray}
with $\bar{F}=F^{\dagger}i\Gamma^{0}$, and similar expression for
$\tilde{U}_l$ involving $\xi_{-}(r_b)$. The constants $U_l$ can be
calculated analytically using the numerical value of
$\xi_{\pm}(r_b)$, and by noting that the Hermite functions are
eigenfunctions of the Fourier transform. We are interested in the
retarded Green's function, defined as
\begin{eqnarray}
G^R_{\mathcal{O}_l}(\omega,p) &=& -i\int d^xdte^{i\omega t-i p\cdot x}\theta(t)G^R_{\mathcal{O}_l}(t,x)\nonumber \\
G^R_{\mathcal{O}_l}(t,x) &=& \langle 0\vert\left[\mathcal{O}_l(t,x),\bar{\mathcal{O}}_l(0,0)\right]\vert 0\rangle\nonumber\\
G^R &=& \left(\begin{array}{cc}
           G_{\mathcal{O}} & 0\\
           0 & \tilde{G}_{\mathcal{O}}
          \end{array}
\right),
\end{eqnarray}
where $\tilde{G}_{\mathcal{O}}$ is the retarded Green's function
for the operator $\tilde{\mathcal{O}}$.

Exploiting the orthogonality of the spinors created by
$\mathcal{O}$ and $\mathcal{O}^{\dagger}$ and using eq.
(\ref{represent}), the Green's function in the plane wave basis
can be written as
\begin{equation}
\label{PlaneWaveGreens}
 G^R_{c_p}(\omega,p_x)=\sum_l{\rm tr}\left(\begin{array}{c}
                                            U\\ \tilde{U}
                                           \end{array}\right)
\left(U^{*},\tilde{U}^{*}\right)G^R=
\left(\vert U_l(p_x)\vert^2 G^R_{\mathcal{O}_l}(\omega,l)+
\vert \tilde{U}_l(p_x)\vert^2 \tilde{G}^R_{\mathcal{O}_l}(\omega,l)\right)
\end{equation}
In practice, we cannot perform the sum in eq.
(\ref{PlaneWaveGreens}) all the way to infinity, so we have to
introduce a cutoff Landau level $l_\mathrm{cut}$. In most cases we
are able to make $l_\mathrm{cut}$ large enough that the behavior
of the spectral function is clear.

Using the above formalism, we have produced spectral functions for
two different conformal dimensions and fixed chemical potential
and magnetic field (Fig. 1). Using the plane wave basis allows us
to directly detect the Landau levels. The unit used for plotting
the spectra (here and later on in the paper) is the effective
temperature $T_\mathrm{eff}$ \cite{Cubrovic:2009}:
\begin{equation}
\label{teffeq}T_\mathrm{eff}=\frac{T}{2}\left(1+\sqrt{1+\frac{3\mu^2}{\left(4\pi
T\right)^2}}\right).
\end{equation}
This unit interpolates between $\mu$ at $T/\mu=0$ and $T$ and is
of or $T/\mu\to\infty$, and is convenient for the reason that the
relevant quantities (e.~g., Fermi momentum) are of order unity for
any value of $\mu$ and $h$.

\begin{figure}[!ht]
\includegraphics[width=0.4\textwidth]{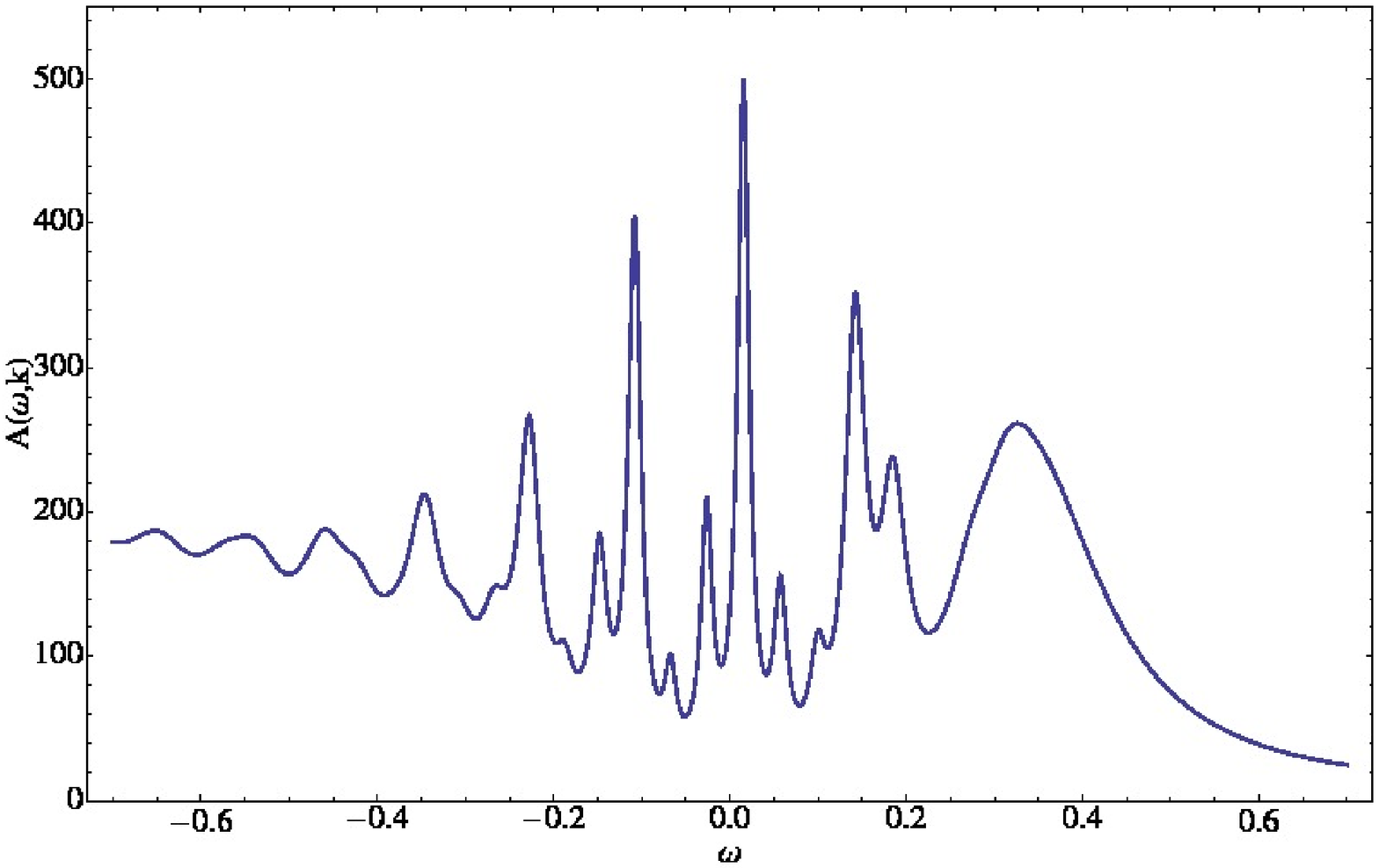}
\includegraphics[width=0.4\textwidth]{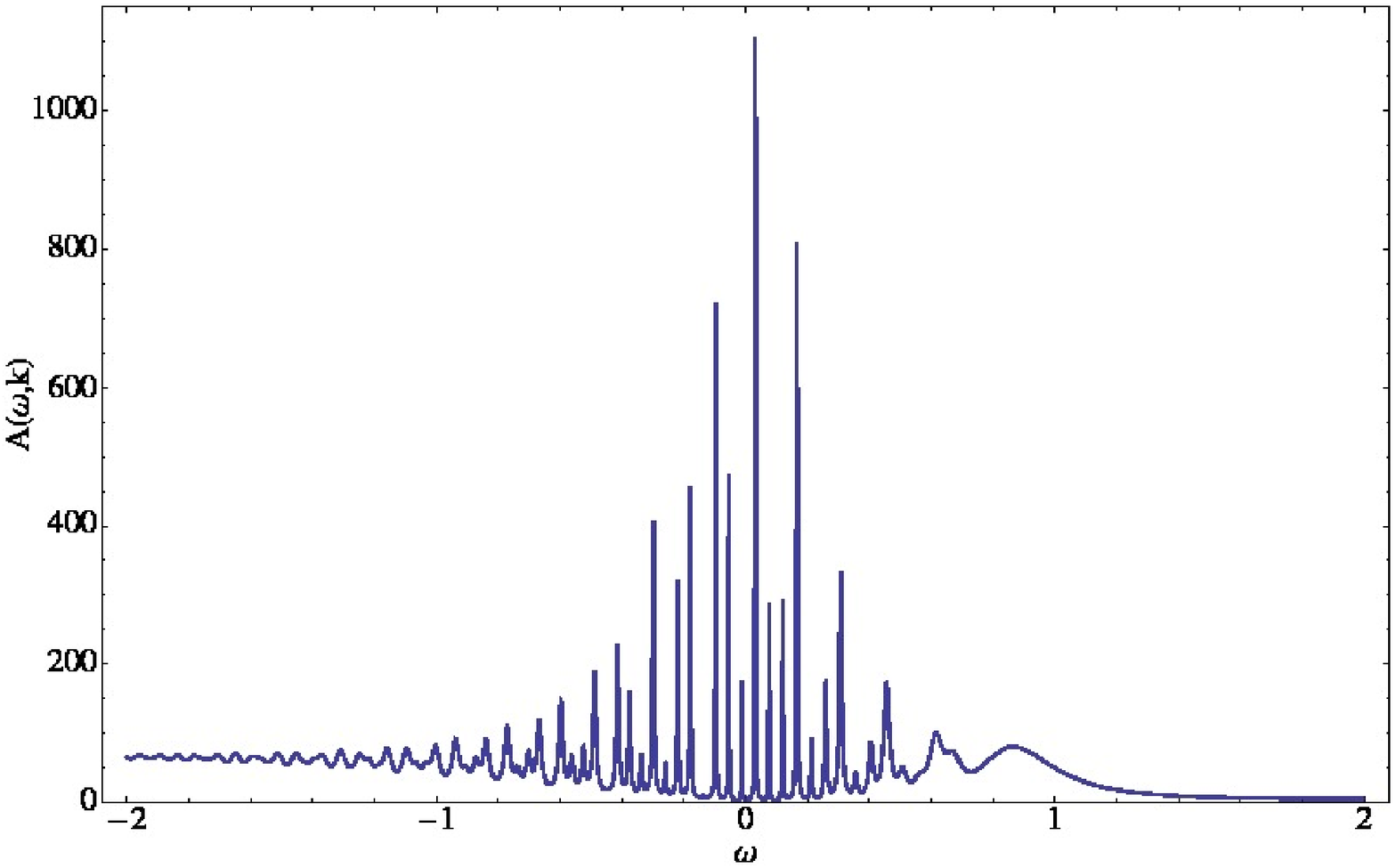}
\caption{\label{spwfig} Two examples of spectral functions in the
plane wave basis for $\mu/T = 50$ and $h/T$ = 1. The conformal
dimension is $\Delta=5/4$ (left) and $\Delta=3/2$ (right).
Frequency is in the units of effective temperature
$T_\mathrm{eff}$. The plane wave momentum is chosen to be $k=1$.
Despite the convolution of many Landau levels, the presence of the
discrete levels is obvious.}
\end{figure}

\subsection{Magnetic crossover and disappearance of the quasiparticles}

Theoretically, it is more convenient to consider the spectral
functions in the Landau level basis. For definiteness let us pick
a fixed conformal dimension $\Delta=\frac{5}{4}$ which corresponds
to $m=-\frac{1}{4}$. In the limit of weak magnetic fields, $h/T\to
0$, we should reproduce the results that were found in
\cite{Cubrovic:2009}.

\begin{figure*}
\begin{center}
(A)\includegraphics[width=0.4\textwidth]{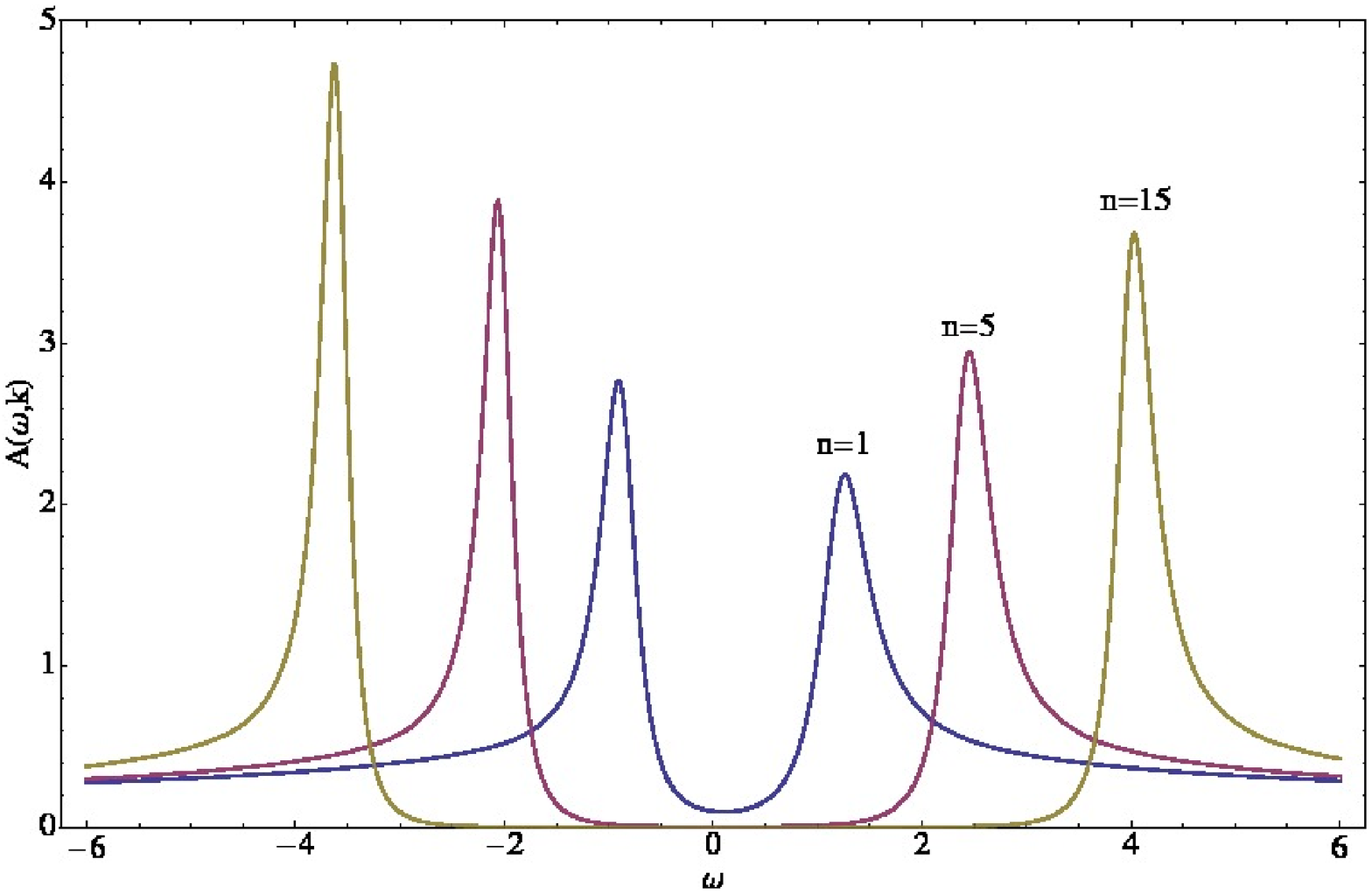}\label{fig:LLspectm025mt1ht1}
(B)\includegraphics[width=0.4\textwidth]{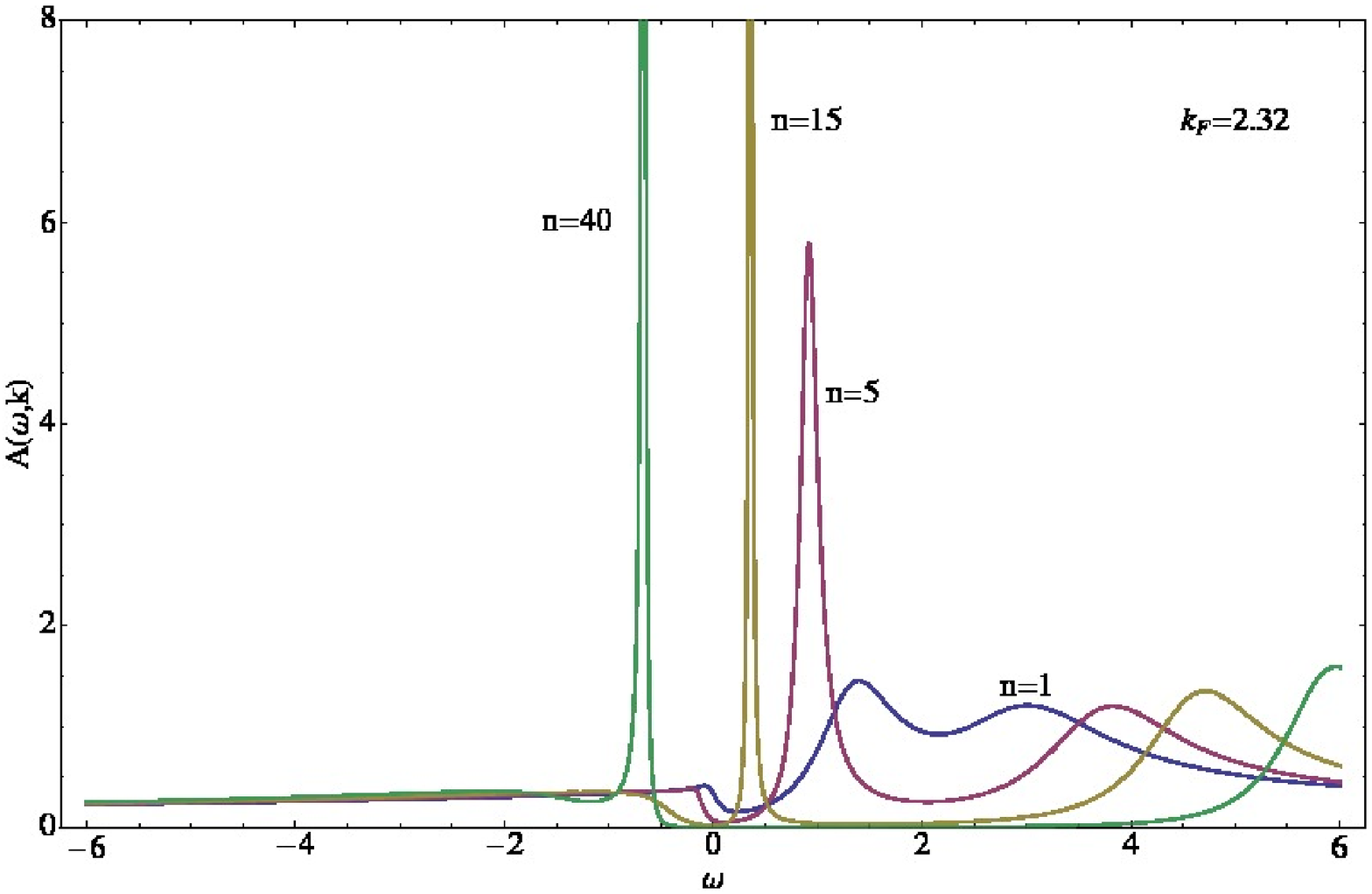}\label{fig:LLspectm025mt50ht1}
(C)\includegraphics[width=0.4\textwidth]{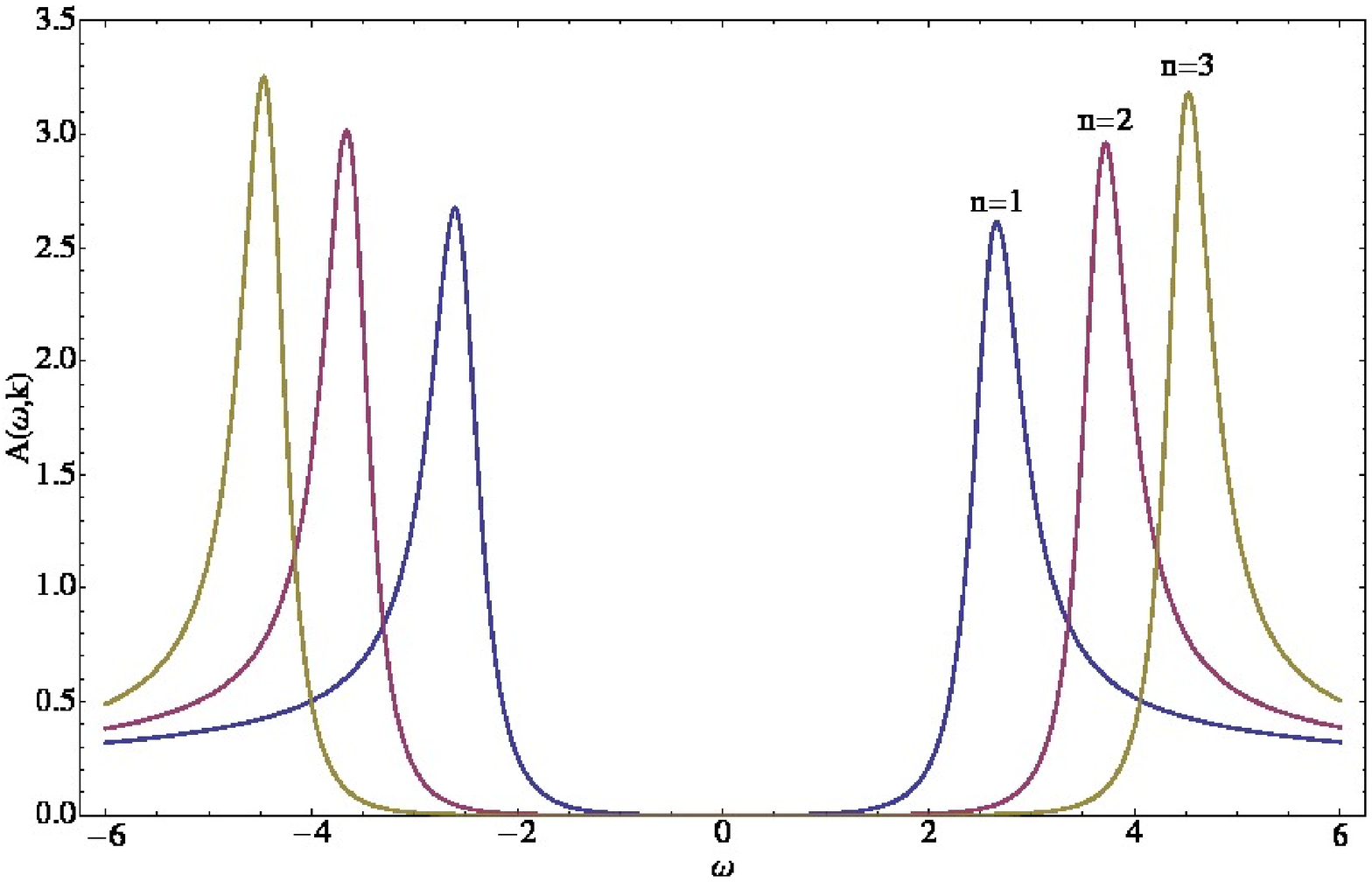}\label{fig:LLspectm025mt1ht50}
(D)\includegraphics[width=0.4\textwidth]{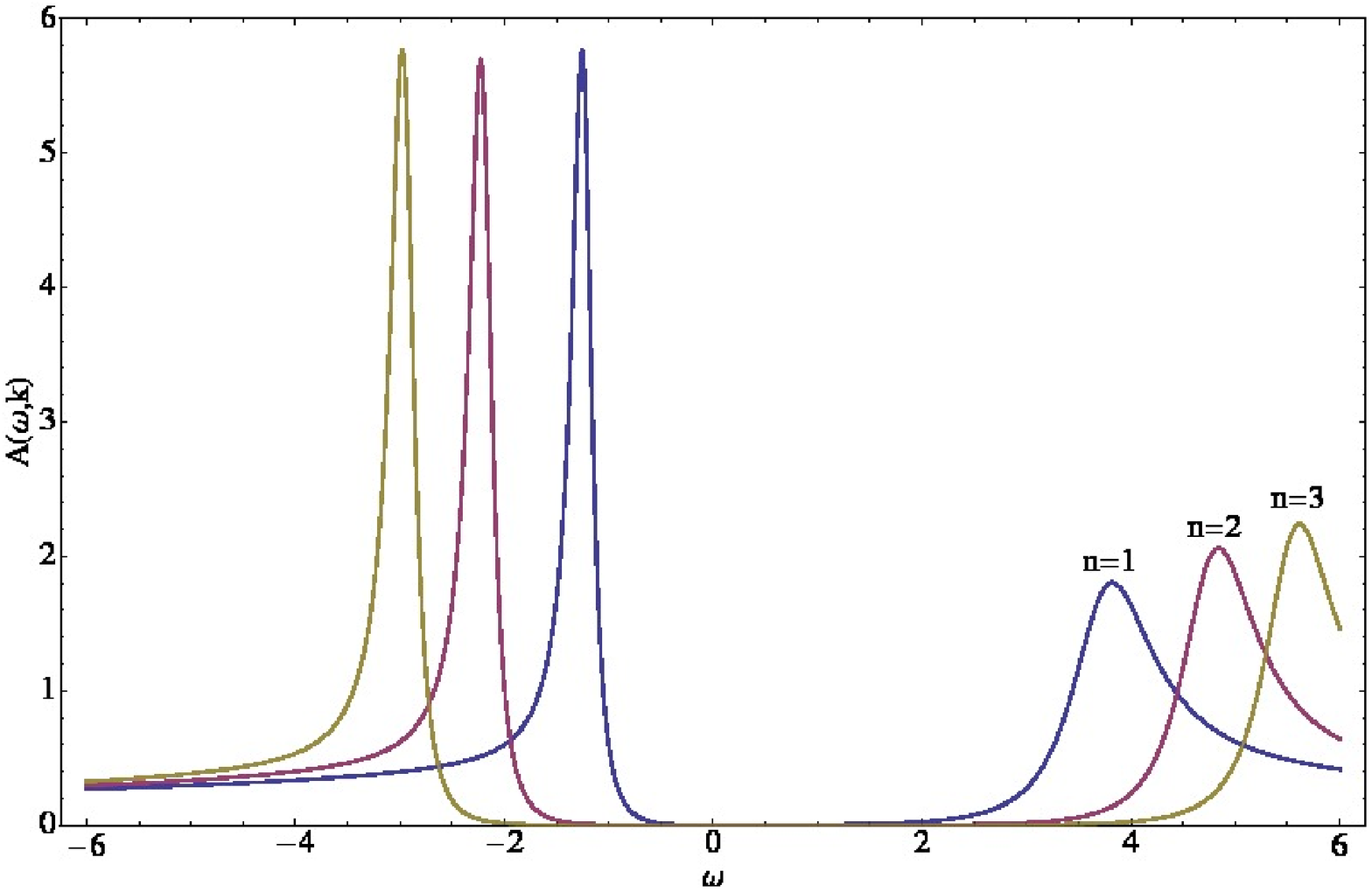}\label{fig:LLspectm025mt50ht50}
 \vspace{3mm}
 \rule{0.9\textwidth}{0.1mm}
 \vspace{2mm}
(E){\includegraphics[width=0.4\textwidth]{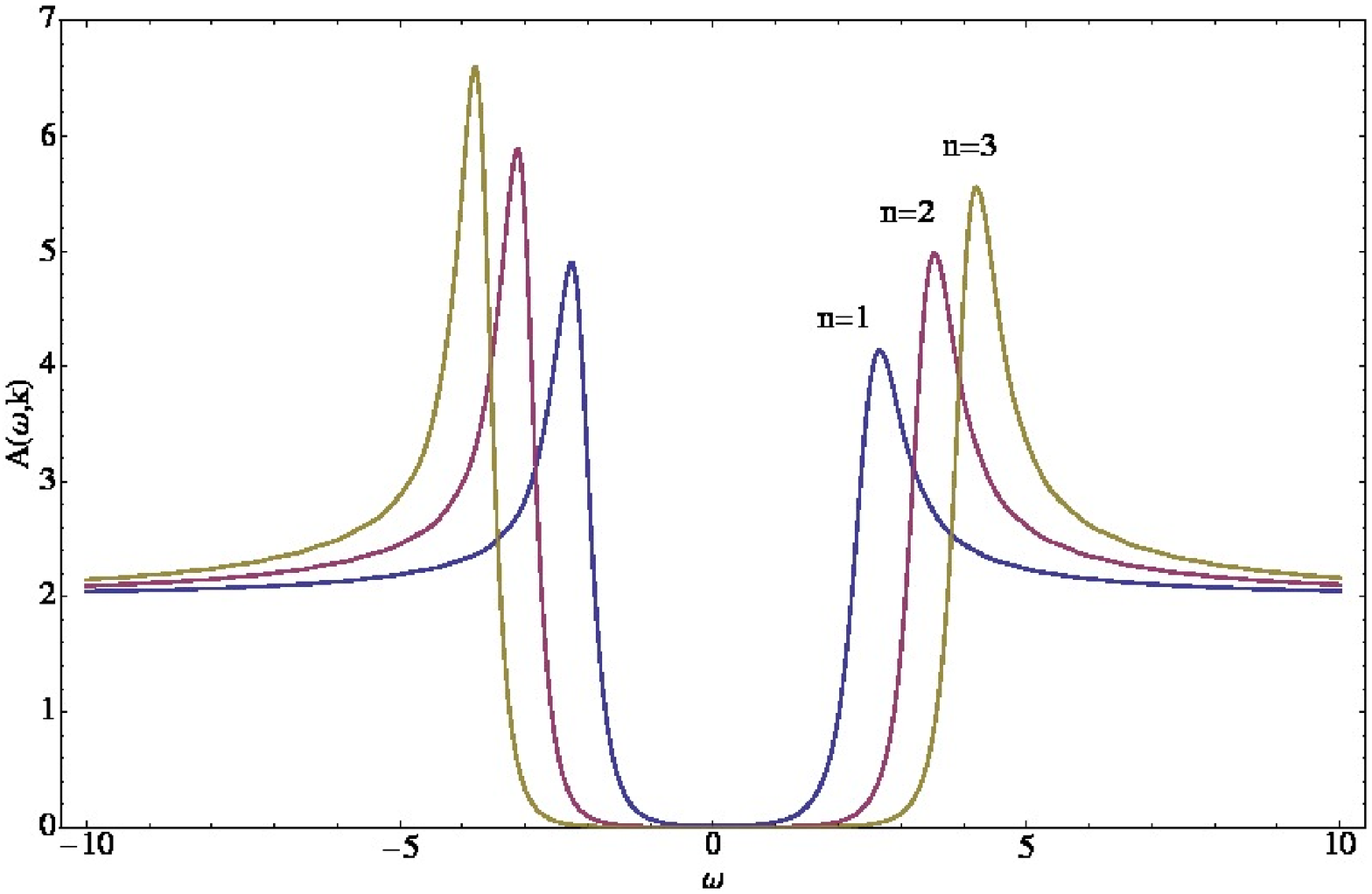}}\label{fig:LLspectm0mt1ht1}
(F){\includegraphics[width=0.4\textwidth]{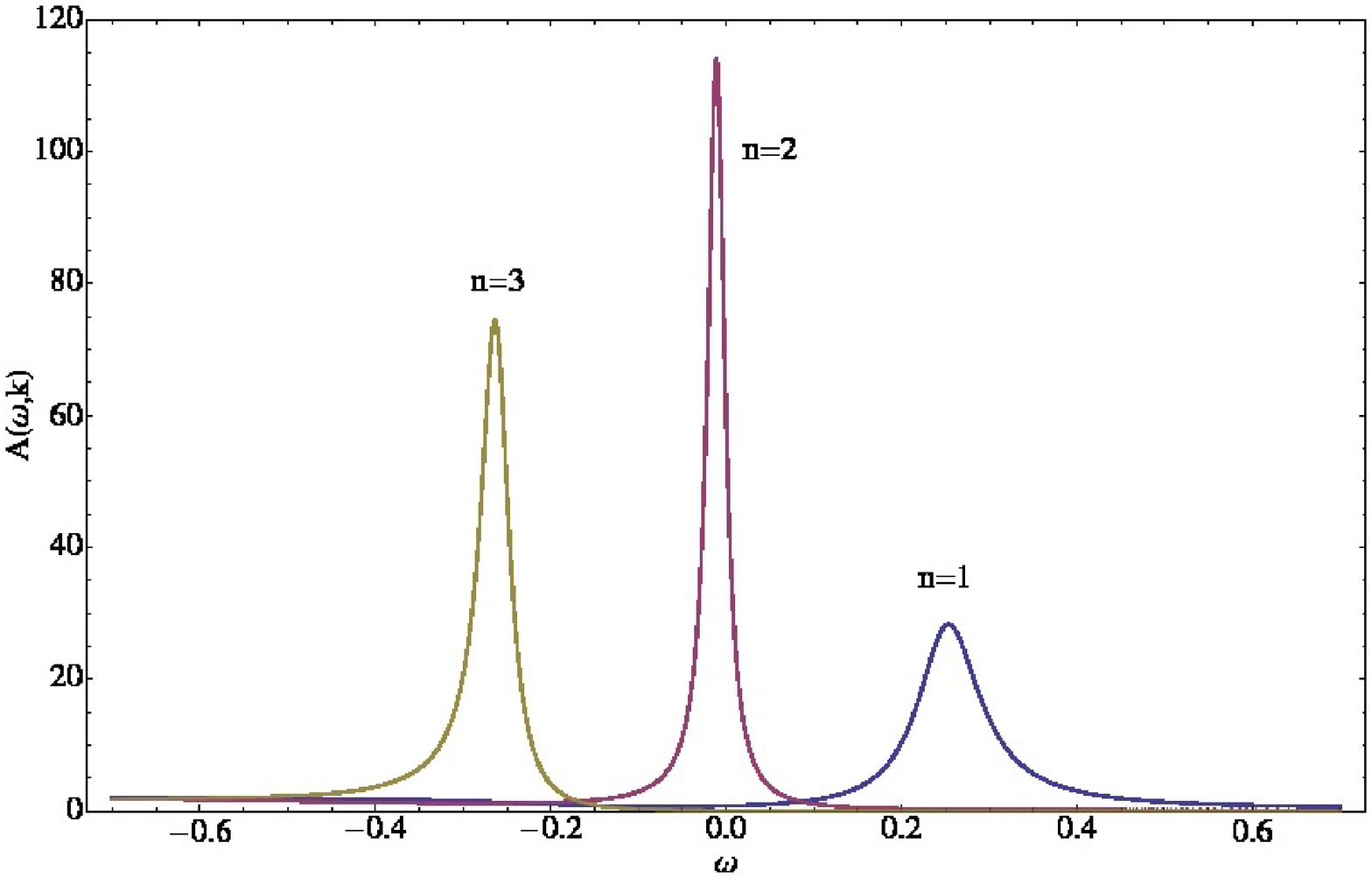}}\label{fig:LLspectm0mt50ht1}
(G){\includegraphics[width=0.4\textwidth]{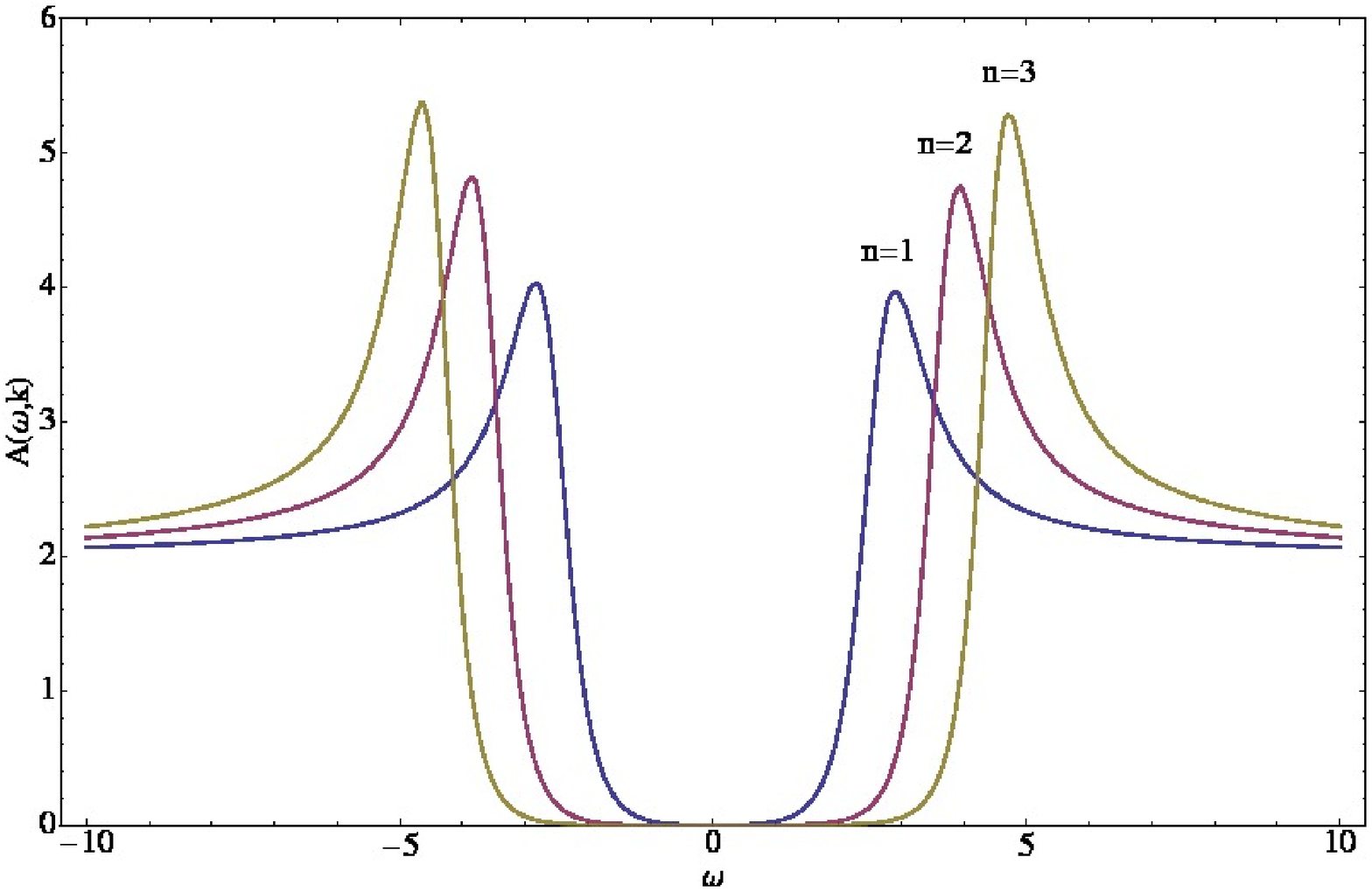}}\label{fig:LLspectm0mt1ht50}
(H){\includegraphics[width=0.4\textwidth]{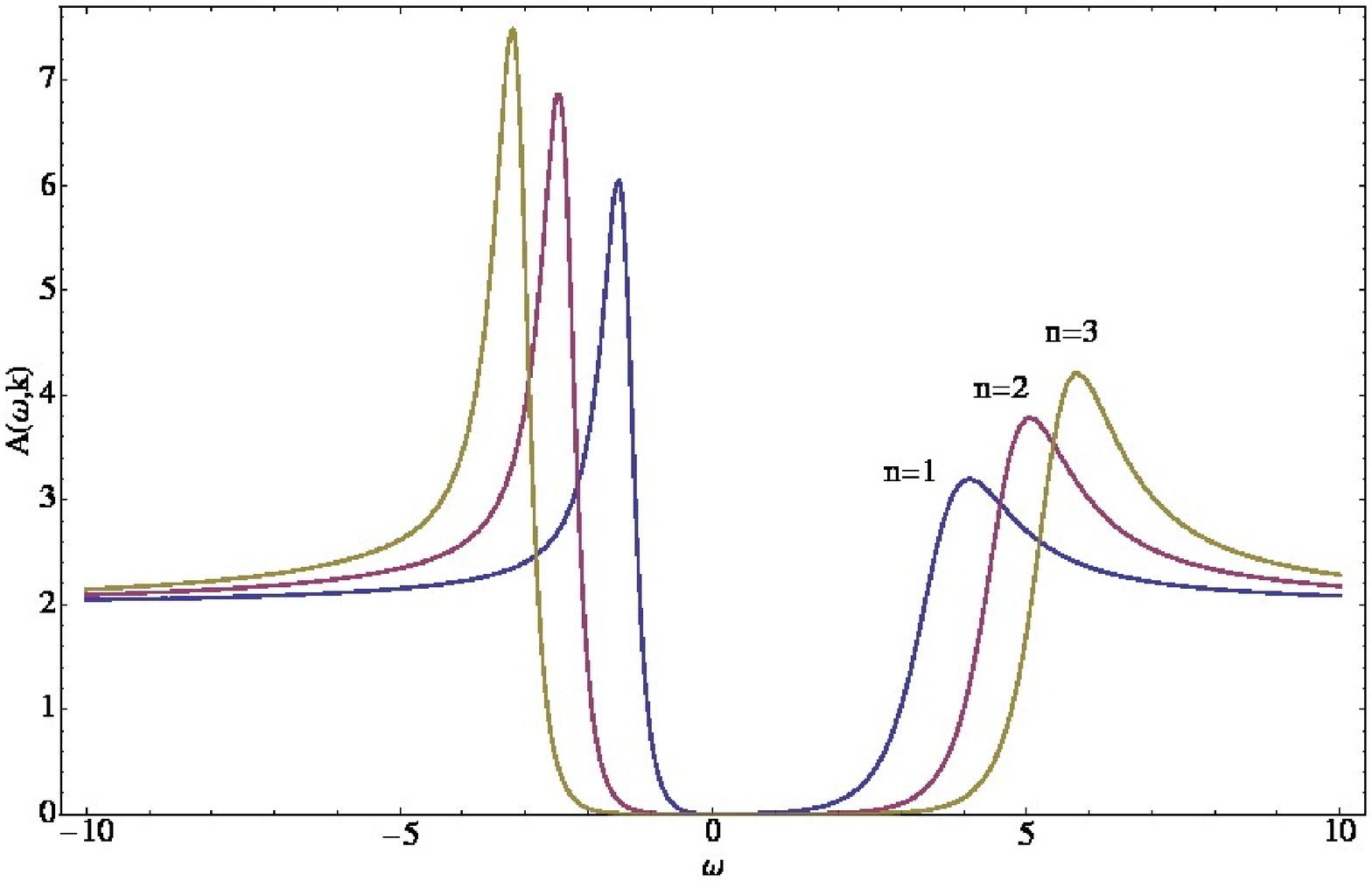}}\label{fig:LLspectm0mt50ht50}
\end{center}
\caption{ Some typical examples of spectral functions
$A(\omega,k_{\rm{eff}})$ vs. $\omega$ in the Landau basis,
$k_{\rm{eff}}=\sqrt{2|qh|n}$. The top four correspond to a
conformal dimension $\Delta=\frac{5}{4}$ $m=-\frac{1}{4}$ and the
bottom four to $\Delta=\frac{3}{2}$ ($m=0$). In each plot we show
different Landau levels, labelled by index $n$, as a function of
$\mu/T$ and $h/T$. The ratios take values
$(\mu/T,h/T)=(1,1),(50,1),(1,50),(50,50)$ from left to right.
Conformal case can be identified when $\mu/T$ is small regardless
of $h/T$ (plots in the left panel). Nearly conformal behavior is
seen when both $\mu/T$ and $h/T$ are large. This confirms our
analytic result that the behavior of the system is primarily
governed by $\mu$. Departure from the conformality and sharp
quasiparticle peaks are seen when $\mu/T$ is large and $h/T$ is
small in (2.B) and (2.F). Multiple quasiparticle peaks arise
whenever $k_{\rm{eff}}=k_F$. This suggests the existence of a
critical magnetic field, beyond which the quasiparticle
description becomes invalid and the system exhibits a
conformal-like behavior. As before, the frequency $\omega$ is in
units of $T_\mathrm{eff}$.}
\end{figure*}

In Fig. (2.A) 
we indeed see that the spectral function, corresponding to a low
value of $\mu/T$, behaves as expected for a nearly conformal
system. The spectral function is approximately symmetric about
$\omega = 0$, it vanishes for $\left|\omega\right| < k$, up to a
small residual tail due to finite temperature, and for
$\left|\omega\right|\gg k$ it scales as $\omega^{2m}$.

\begin{figure}
\begin{center}
\label{fig:DispersionDensity}
(A)\includegraphics[width=0.4\textwidth]{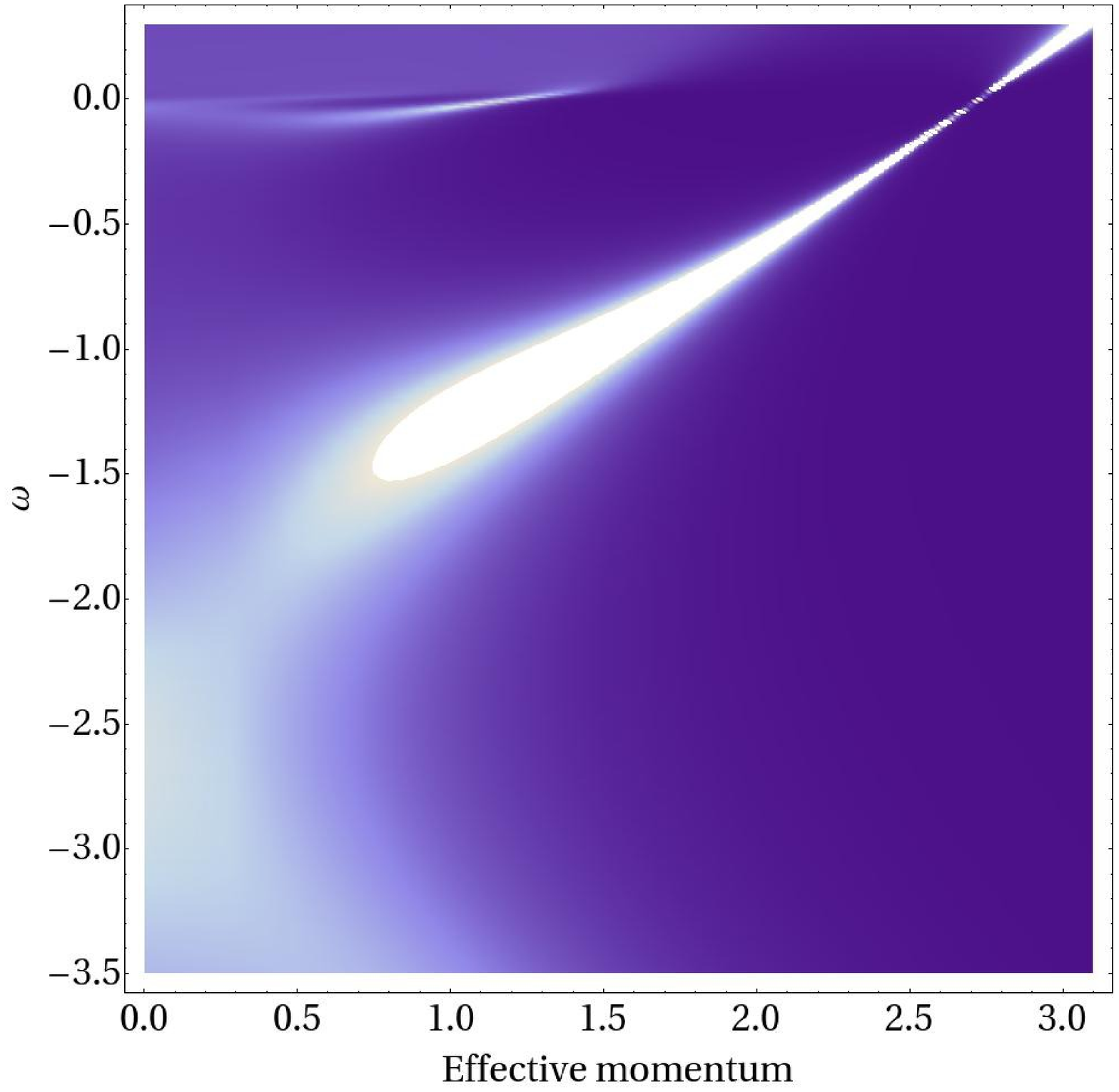}
(B)\includegraphics[width=0.4\textwidth]{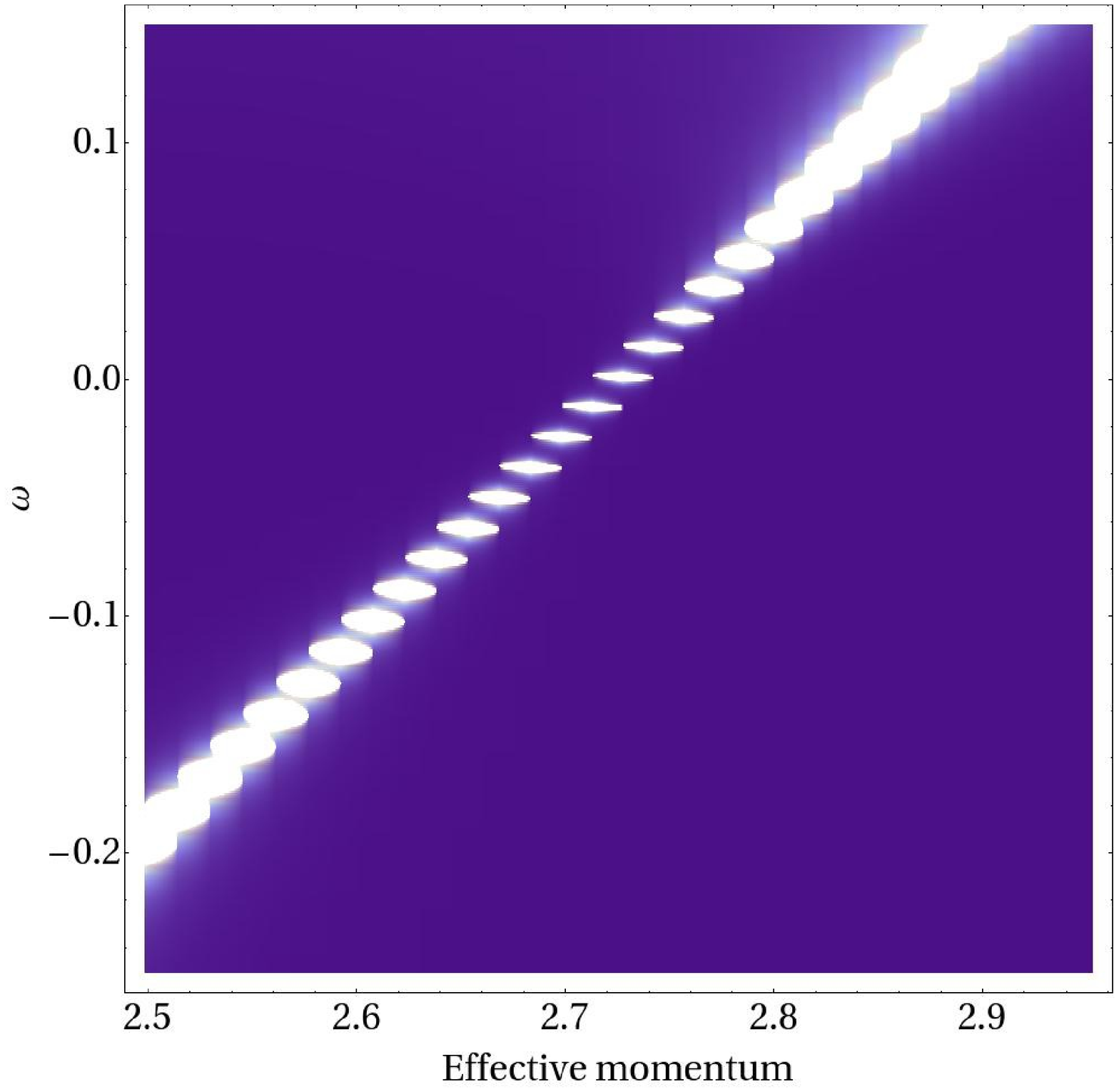}
 \end{center}
\caption{ Dispersion relation $\omega$ vs. $k_{\rm{eff}}$ for
$\mu/T=50$, $h/T=1$ and $\Delta=\frac{5}{4}$ ($m=-\frac{1}{4}$).
The spectral function $A(\omega,k_{\rm{eff}})$ is displayed as a
density plot. (A) On a large energy and momentum scale, we clearly
sees that the peaks disperse almost linearly ($\omega\approx
v_Fk$), indicating that we are in the stable quasiparticle regime.
(B) A zoom-in near the location of the Fermi surface shows clear
Landau quantization.}
\end{figure}

In Fig. (2.B), 
which corresponds to a high value of $\mu/T$, we see the emergence
of a sharp quasiparticle peak. This peak becomes the sharpest when
the Landau level $l$ corresponding to an effective momentum
$k_\mathrm{eff}=\sqrt{2|qh|l}$ coincides with the Fermi momentum
$k_F$. The peaks also broaden out when $k_\mathrm{eff}$ moves away
from $k_F$. A more complete view of the Landau quantization in the
quasiparticle regime is given in Fig. \ref{fig:DispersionDensity},
where we plot the dispersion relation ($\omega$-$k$ map). Both the
sharp peaks and the Landau levels can be visually identified.

Collectively, the spectra in Fig. 2 show that conformality is only
broken by the chemical potential $\mu$ and not by the magnetic
field. Naively, the magnetic field introduces a new scale in the
system. However, this scale is absent from the spectral functions,
visually validating the the discussion in the previous section
that the scale $h$ can be removed by a rescaling of the
temperature and chemical potential.

One thus concludes that there is some value $h_c^{\prime}$ of the
magnetic field, depending on $\mu/T$, such that the spectral
function loses its quasiparticle peaks and displays near-conformal
behavior for $h>h_c^{\prime}$. The nature of the transition and
the underlying mechanism depends on the parameters
$(\mu_q,T,\Delta)$. One mechanism, obvious from the rescaling in
eq. (\ref{eq:GFMapping}), is the reduction of the effective
coupling $q$ as $h$ increases. This will make the influence of the
scalar potential $A_0$ negligible and push the system back toward
conformality. Generically, the spectral function shows no sharp
change but is more indicative of a crossover.


A more interesting phenomenon is the disappearance of coherent
quasiparticles at high effective chemical potentials. For the
special case $m=0$, we can go beyond numerics and study this
transition analytically, combining the exact $T=0$ solution found
in \cite{Hartman:2010} and the mapping (\ref{GFMapping2}). In the
next section, we will show that the transition is controlled by
the change in the dispersion of the quasiparticle and corresponds
to a sharp phase transition. Increasing the magnetic field leads
to a decrease in phenomenological control parameter $\nu_{k_F}$.
This can give rise to a transition to a non-Fermi liquid when
$\nu_{k_F}\leq 1/2$, and finally to the conformal regime at
$h=h_c^\prime$ when $\nu_{k_F}=0$ and the Fermi surface vanishes.

\subsection{Density of states}

As argued at the beginning of this section, the spectral function
can look quite different depending on the particular basis chosen.
Though the spectral function is an attractive quantity to consider
due to connection with ARPES experiments, we will also direct our
attention to basis-independent and manifestly gauge invariant
quantities. One of them is the density of states (DOS), defined by
\begin{equation}
\label{eq:DOS}
  D(\omega) = \sum_{l} A(\omega,l),
\end{equation}
where the usual integral over the momentum is replaced by a sum
since only discrete values of the momentum are allowed.

\begin{figure}
 \begin{center}
(A){\includegraphics[width=0.4\textwidth]{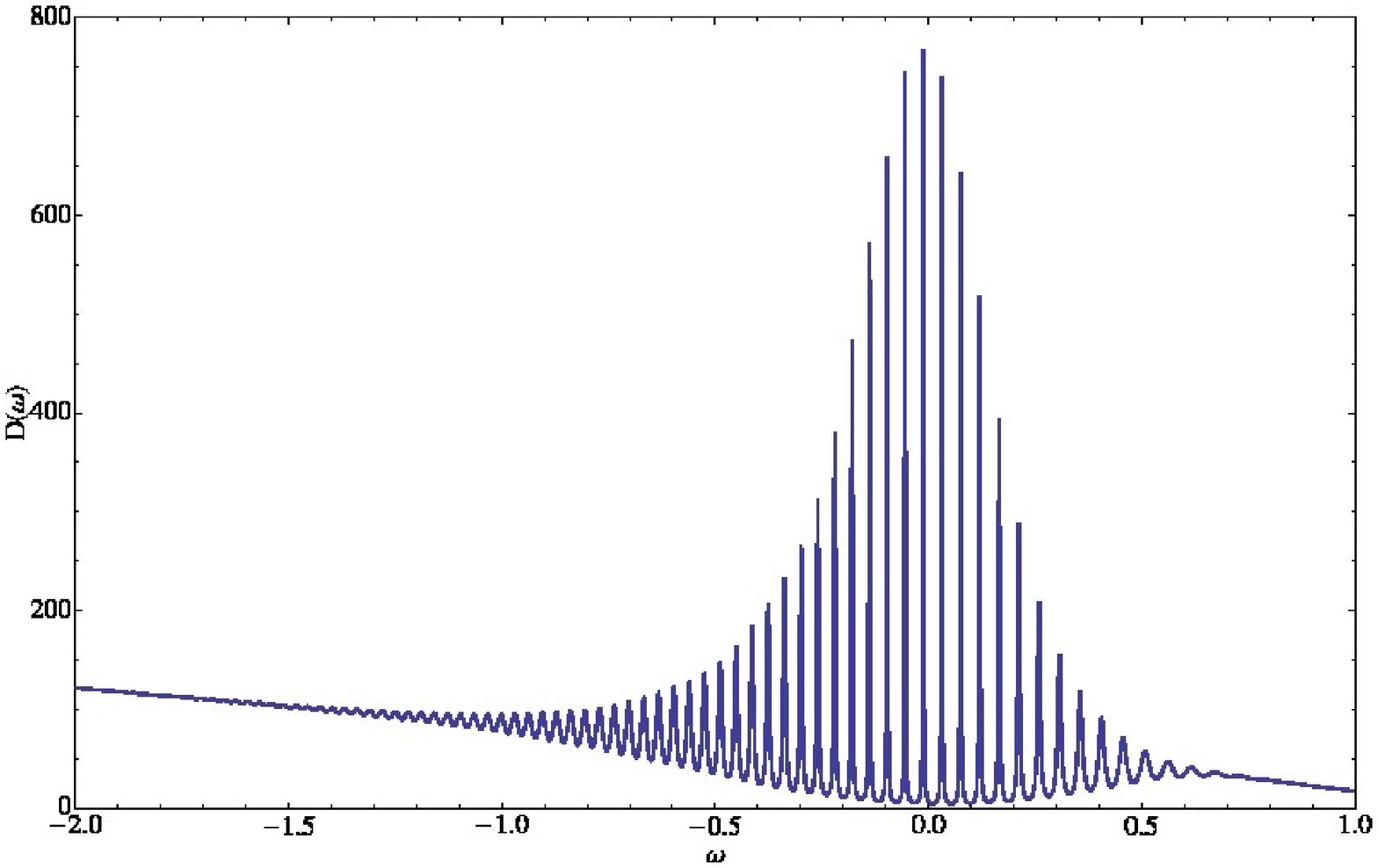}}\label{fig:DOSmass025}
(B){\includegraphics[width=0.4\textwidth]{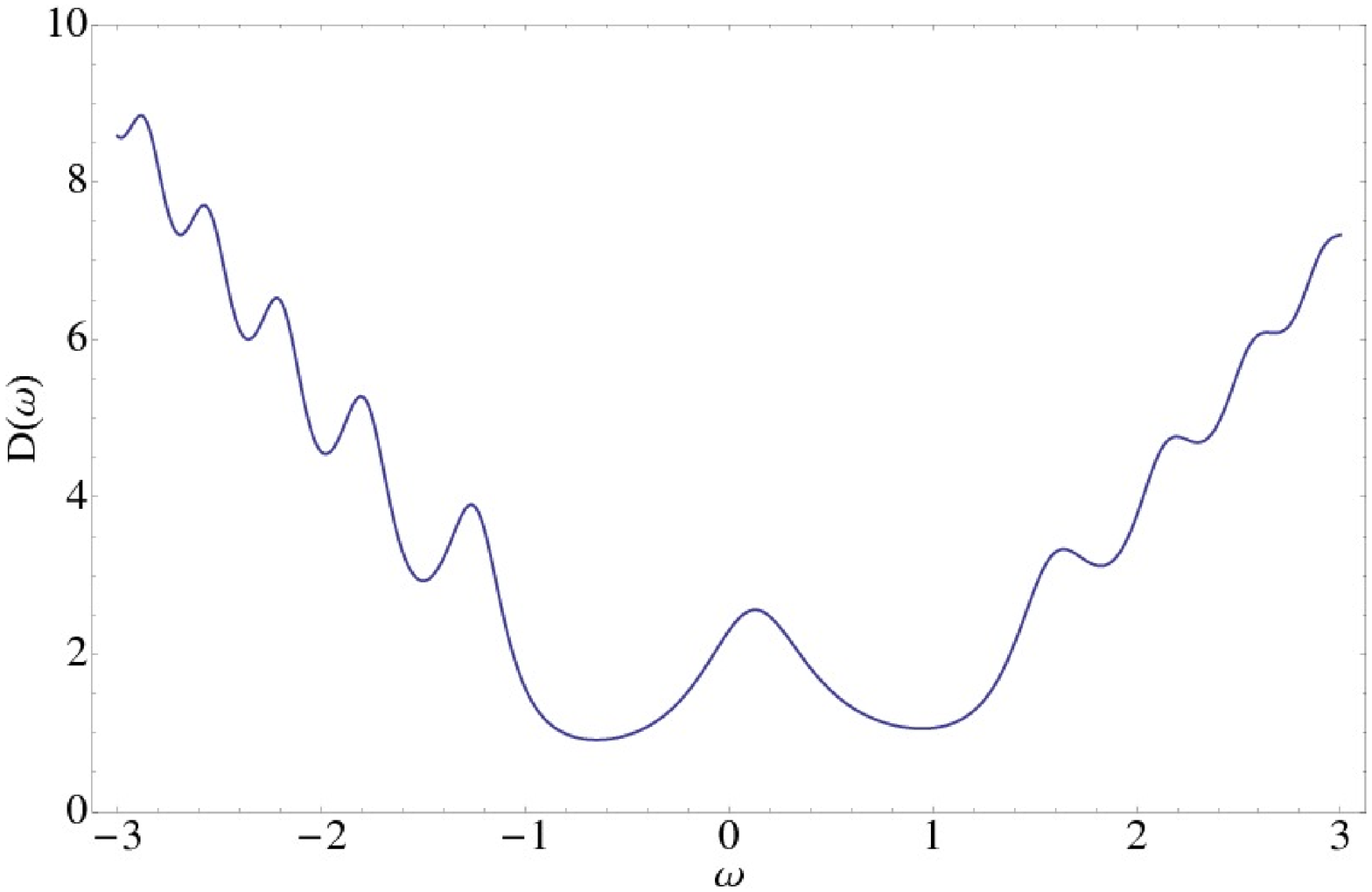}}\label{fig:DOSmass0}
 \end{center}
\caption{ Density of states $D(\omega)$ for $m=-\frac{1}{4}$ and
(A) $\mu/T = 50$, $h/T=1$, and (B) $\mu/T=1$, $h/T=1$. Sharp
quasiparticle peaks from the splitting of the Fermi surface are
clearly visible in (A). The case (B) shows square-root level
spacing characteristic of a (nearly) Lorentz invariant spectrum
such as that of graphene. }
\end{figure}

In Fig. 4, we plot the density of states for two systems. We
clearly see the Landau splitting of the Fermi surface. A peculiar
feature of these plots is that the DOS seems to grow for negative
values of $\omega$. This, however, is an artefact of our
calculation. Each individual spectrum in the sum eq.
(\ref{eq:DOS}) has a finite tail that scales as $\omega^{2m}$ for
large $\omega$, so each term has a finite contribution for large
values of $\omega$. When the full sum is performed, this fact
implies that $\lim\limits_{\omega\to\infty}D(\omega)\rightarrow
\infty$. The relevant information on the density of states can be
obtained by regularizing the sum, which in practice is done by
summing over a finite number of terms only, and then considering
the peaks that lie on top of the resulting finite-sized envelope.
The physical point in Fig. 4A is the linear spacing of Landau
levels, corresponding to a non-relativistic system at finite
density. This is to be contrasted with Fig. 4B where the level
spacing behaves as $\propto\sqrt{h}$, appropriate for a Lorentz
invariant system and realized in graphene \cite{experiment}.

\section{Fermi level structure at zero temperature}\label{section:4}

In this section, we solve the Dirac equation in the magnetic field
for the special case $m=0$ ($\Delta=\frac{3}{2}$). Although there
are no additional symmetries in this case, it is possible to get
an analytic solution. Using this solution, we obtain Fermi level
parameters such as $k_F$ and $v_F$ and consider the process of
filling the Landau levels as the magnetic field is varied.

\subsection{Dirac equation with $m=0$}

In the case $m=0$, it is convenient to solve the Dirac equation
including the spin connection (eq. \ref{dirac-connection}) rather
than scaling it out:
\begin{eqnarray}
&& \hspace{-2cm} \left(-\frac{\sqrt{g_{ii}}}{\sqrt{g_{rr}}}\sigma^1\partial_r
-\frac{\sqrt{g_{ii}}}{\sqrt{-g_{tt}}}\sigma^3(\omega+qA_t)
+\frac{\sqrt{g_{ii}}}{\sqrt{-g_{tt}}}\sigma^1
\frac{1}{2}\omega_{\hat{t}\hat{r}t}\right.\nonumber\\
&-&\left.\sigma^1\frac{1}{2}\omega_{\hat{x}\hat{r}x}
-\sigma^1\frac{1}{2}\omega_{\hat{y}\hat{r}y}-\lambda_l\right)\otimes 1
\left(\begin{array}{c}
\psi_1\\
\psi_2
\end{array}\right)=0,
\label{dirac-equation}
\end{eqnarray}
where $\lambda_l = \sqrt{2|qh|l}$ are the energies of the Landau
levels $l=0,1,\dots$, $g_{ii}\equiv g_{xx}=g_{yy}$, $A_t(r)$ is
given by eq. (\ref{ads4-metric2}), and the gamma matrices are
defined in eq. (\ref{matrices}). In the basis of eq.
(\ref{matrices}), the two components $\psi_1$ and $\psi_2$
decouple. Therefore, in what follows we solve for the first
component only (we omit index $1$). Substituting the spin
connection, we have \cite{Gubankova:2010}:
\begin{equation}
\left(-\frac{r^2\sqrt{f}}{R^2}\sigma^1\partial_r
-\frac{1}{\sqrt{f}}\sigma^3(\omega+qA_t)
-\sigma^1\frac{r\sqrt{f}}{2R^2}(3+\frac{rf'}{2f})-\lambda_l\right)\psi=0,
\label{dirac-equationy12}
\end{equation}
with $\psi=(y_1,y_2)$. It is convenient to change to the basis
\begin{equation}
\left(\begin{array}{c}
\tilde{y_1}\\
\tilde{y_2}
\end{array}
\right) = \left(
\begin{array}{cc}
1 & -i \\
-i &  1
\end{array} \right)
\left(\begin{array}{c}
 y_1 \\
 y_2
\end{array}
\right),
\label{basis-rotation}
\end{equation}
which diagonalizes the system into a second order differential
equation for each component. We introduce the dimensionless
variables as in eqs. (\ref{dimensionless1}-\ref{dimensionless3}),
and make a change of the dimensionless radial variable:
\begin{equation}
r=\frac{1}{1-z},
\end{equation}
with the horizon now being at $z=0$, and the conformal boundary at
$z=1$. Performing these transformations in eq.
(\ref{dirac-equationy12}), the second order differential equations
for $\tilde{y}_1$ reads
\begin{eqnarray}
&&\hspace{-3cm} \left(f\partial^2_z+(\frac{3f}{1-z}+f')\partial_z+\frac{15f}{4(1-z)^2}
+\frac{3f'}{2(1-z)}+\frac{f''}{4}\right.\nonumber\\
&+&\left.\frac{1}{f}((\omega+q \mu z)\pm \frac{if'}{4})^2
-iq\mu-\lambda_l^2 \right)\tilde{y}_{1}=0,
\label{dirac-spinconnection}
\end{eqnarray}
The second component $\tilde{y}_2$ obeys the same equation with
$\mu\mapsto -\mu$.

At $T=0$,
\begin{eqnarray}
f &=& 3z^2(z-z_0)(z-\bar{z}_0),\;\; z_0=\frac{1}{3}(4+i\sqrt{2}).
\label{parameters}
\end{eqnarray}
The solution of this fermion system at zero magnetic field and
zero temperature $T=0$ has been found in \cite{Hartman:2010}. To
solve eq. (\ref{dirac-spinconnection}), we use the mapping to a
zero magnetic field system eq. (\ref{eq:GFMapping}). The
combination $\mu_q\equiv \mu q$ at non-zero $h$ maps to
$\mu_{q,\mathrm{eff}}\equiv \mu_{\mathrm{eff}}q_{\mathrm{eff}}$ at
zero $h$ as follows:
\begin{eqnarray}
\mu_q\mapsto q\sqrt{1-\frac{H^2}{Q^2+H^2}}\cdot g_F\sqrt{Q^2+H^2}=\sqrt{3}qg_F\sqrt{1-\frac{H^2}{3}}= \mu_{q,\mathrm{eff}}
\label{substitution}
\end{eqnarray}
where at $T=0$ we used $Q^2+H^2=3$. We solve eq.
(\ref{dirac-spinconnection}) for zero modes, i.~e. $\omega=0$, and
at the Fermi surface $\lambda=k$, and implement eq.
(\ref{substitution}).

Near the horizon ($z=0$, $f=6z^2$), we have
\begin{equation}
6z^2\tilde{y}''_{1;2}+12z\tilde{y}'_{1;2}+\left(\frac{3}{2}+\frac{\left(\mu_{q,{\mathrm
eff}}\right)^2}{6}-k_F^2\right)\tilde{y}_{1;2}=0,
\end{equation}
which gives the following behavior:
\begin{equation}
\tilde{y}_{1;2}\sim z^{-\frac{1}{2}\pm \nu_k},
\end{equation}
with the scaling exponent $\nu$ following from eq.
(\ref{redefnul}):
\begin{equation}
\nu=\frac{1}{6}\sqrt{6 k^2-(\mu_{q,{\mathrm eff}})^2}, \label{nu}
\end{equation}
at the momentum $k$. Using Maple, we find the zero mode solution
of eq. (\ref{dirac-spinconnection}) with a regular behavior
$z^{-\frac{1}{2}+\nu}$ at the horizon
\cite{Hartman:2010,Gubankova:2010}:
\begin{eqnarray}
\tilde{y}_{1}^{(0)} &=& N_{1}
(z-1)^{\frac{3}{2}}z^{-\frac{1}{2}+\nu}(z-\bar{z}_0)^{-\frac{1}{2}-\nu}
\left(\frac{z-z_0}{z-\bar{z}_0}\right)^{\frac{1}{4}(-1- \sqrt{2}\mu_{q,{\mathrm eff}}/z_0)},\nonumber\\
&\times&
{}_2F_1\left(\frac{1}{2}+\nu-\frac{\sqrt{2}}{3}\mu_{q,{\rm
eff}},\nu+ i\frac{\mu_{q,{\mathrm eff}}}{6},
1+2\nu,\frac{2i\sqrt{2}z}{3 z_0(z-\bar{z}_0)}\right),
\label{solution-y}
\end{eqnarray}
and
\begin{eqnarray}
\tilde{y}_{2}^{(0)} &=& N_{2}
(z-1)^{\frac{3}{2}}z^{-\frac{1}{2}+\nu}(z-\bar{z}_0)^{-\frac{1}{2}-\nu}
\left(\frac{z-z_0}{z-\bar{z}_0}\right)^{\frac{1}{4}(-1+ \sqrt{2}\mu_{q,{\mathrm eff}}/z_0)},\nonumber\\
&\times&
{}_2F_1\left(\frac{1}{2}+\nu+\frac{\sqrt{2}}{3}\mu_{q,{\rm
eff}},\nu- i\frac{\mu_{q,{\mathrm eff}}}{6},
1+2\nu,\frac{2i\sqrt{2}z}{3 z_0(z-\bar{z}_0)}\right),
\label{solution-y2}
\end{eqnarray}
where ${}_2F_1$ is the hypergeometric function and $N_1, N_2$ are
normalization factors. Since normalization factors are constants,
we find their relative weight by substituting solutions given in
eq. (\ref{solution-y}) back into the first order differential
equations at $z\sim 0$,
\begin{equation}
\frac{N_1}{N_2}=-\frac{6i\nu+\mu_{q,{\mathrm eff}}}{\sqrt{6}k}\left(\frac{z_0}
{\bar{z}_0}\right)^{\mu_{q,{\mathrm eff}}/\sqrt{2}z_0}.
\label{normalization}
\end{equation}
The same relations are obtained when calculations are done for any
$z$. The second solution $\tilde{\eta}_{1;2}^{(0)}$, with behavior
$z^{-\frac{1}{2}-\nu}$ at the horizon, is obtained by replacing
$\nu\rightarrow -\nu$ in eq.(\ref{solution-y}).

\begin{figure}[ht!]
\begin{center}
\includegraphics[width=10cm]{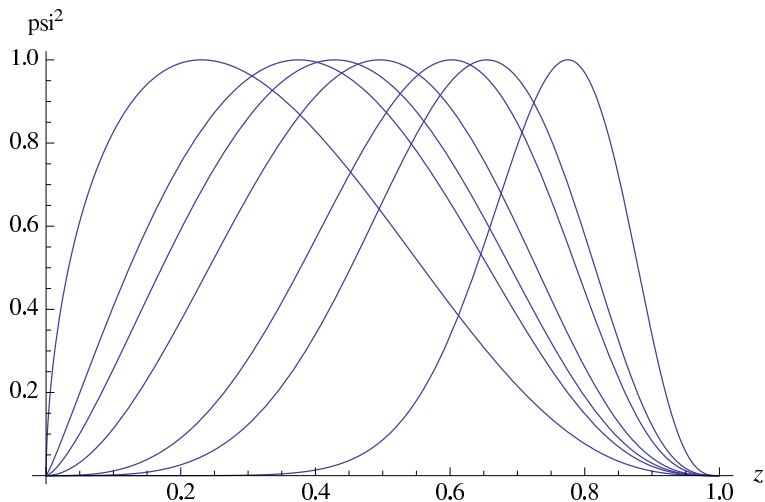}
\caption{Density of the zero mode $\psi^{0\dagger}\psi^{0}$ vs.
the radial coordinate $z$ (the horizon is at $z=0$ and the
boundary is at $z=1$) for different values of the magnetic field
$h$ for the first (with the largest root for $k_F$) Fermi surface.
We set $g_F=1$ ($h\rightarrow H$) and $q=\frac{15}{\sqrt{3}}$
($\mu_{q,{\rm eff}}\rightarrow 15\sqrt{1-\frac{H^2}{3}}$). From
right to left the values of the magnetic field are
$H=\{0,1.40,1.50,1.60,1.63,1.65,1.68\}$. The amplitudes of the
curves are normalized to unity. At weak magnetic fields, the wave
function is supported away from the horizon while at strong fields
it is supported near the horizon.} \label{plot-wave-function}
\end{center}
\end{figure}

To get insight into the zero-mode solution (\ref{solution-y}), we
plot the radial profile for the density function
$\psi^{(0)\dagger}\psi^{(0)}$ for different magnetic fields in
Fig. (\ref{plot-wave-function}). The momentum chosen is the Fermi
momentum of the first Fermi surface (see the next section). The
curves are normalized to have the same maxima. Magnetic field is
increased from right to left. At small magnetic field, the zero
modes are supported away from the horizon, while at large magnetic
field, the zero modes are supported near the horizon. This means
that at large magnetic field the influence of the black hole to
the Fermi level structure becomes more important.

\subsection{Magnetic effects on the Fermi momentum and Fermi velocity at $T=0$}

In the presence of a magnetic field there is only a true pole in
the Green's function whenever the Landau level crosses the Fermi
energy \cite{Denef:2009}
\begin{equation}
2l|qh|=k_F^2. \label{peak}
\end{equation}
As shown in
Fig. 2, whenever the equation (\ref{peak}) is satisfied the
spectral function $A(\omega)$ has a (sharp) peak. This is not
surprising since quasiparticles can be easily excited from the
Fermi surface. From eq. (\ref{peak}), the spectral function
$A(\omega)$ and the density of states on the Fermi surface
$D(\omega)$ are periodic in $\frac{1}{h}$ with the period
\begin{equation}
\Delta(\frac{1}{h})=\frac{2\pi q}{A_F},
\end{equation}
where $A_F=\pi k_F^2$ is the area of the Fermi surface
\cite{Denef:2009}. This is a manifestation of the de Haas-van
Alphen quantum oscillations. At $T=0$, the electronic properties
of metals depend on the density of states on the Fermi surface.
Therefore, an oscillatory behavior as a function of magnetic field
should appear in any quantity that depends on the density of
states on the Fermi energy. Magnetic susceptibility
\cite{Denef:2009} and magnetization together with the
superconducting gap \cite{Shovkovy:2007} have been shown to
exhibit quantum oscillations. Every Landau level contributes an
oscillating term and the period of the $l$-th level oscillation is
determined by the value of the magnetic field $h$ that satisfies
eq. (\ref{peak}) for the given value of $k_F$. Quantum
oscillations (and the quantum Hall effect which we consider later
in the paper) are examples of phenomena in which Landau level
physics reveals the presence of the Fermi surface. The
superconducting gap found in the quark matter in magnetic fields
\cite{Shovkovy:2007} is another evidence for the existence of the
(highly degenerate) Fermi surface and the corresponding Fermi
momentum.

Generally, a Fermi surface controls the occupation of energy
levels in the system: the energy levels below the Fermi surface
are filled and those above are empty (or non-existent). Here,
however, the association to the Fermi momentum can be obscured by
the fact that the fermions form highly degenerate Landau levels.
Thus, in two dimensions, in the presence of the magnetic field the
corresponding effective Fermi surface is given by a single point
in the phase space, that is determined by $n_F$, the Landau index
of the highest occupied level, i.e., the highest Landau level
below the chemical potential \footnote{We would like to thank Igor
Shovkovy for clarifying the issue with the Fermi momentum in the
presence of the magnetic field.}. Increasing the magnetic field,
Landau levels 'move up' in the phase space leaving only the lower
levels occupied, so that the effective Fermi momentum scales
roughly (excluding interactions) as a square root of the magnetic
field, $k_F\sim \sqrt{n_F}\sim k_F^{max}\sqrt{1-h/h_{max}}$. High
magnetic fields drive the effective density of the charge carriers
down, approaching the limit when the Fermi momentum coincides with
the lowest Landau level.

Many phenomena observed in the paper can thus be qualitatively
explained by Landau quantization. As discussed before, the notion
of the Fermi momentum is lost at very high magnetic fields. In
what follows, the quantitative Fermi level structure at zero
temperature, described by $k_F$ and $v_F$ values, is obtained as a
function of the magnetic field using the solution of the Dirac
equation given by eqs. (\ref{solution-y},\ref{solution-y2}). As in
\cite{Basu:2008}, we neglect first the discrete nature of the Fermi
momentum and velocity in order to obtain general understanding.
Upon taking the quantization into account,
the smooth curves become combinations of step functions following
the same trend as the smooth curves (without quantization). While
usually the grand canonical ensemble is used, where the fixed
chemical potential controls the occupation of the Landau levels
\cite{Shovkovy:2006}, in our setup, the Fermi momentum is allowed
to change as the magnetic field is varied, while we keep track of
the IR conformal dimension $\nu$.


\begin{figure}[ht!]
\begin{center}
\includegraphics[width=10cm]{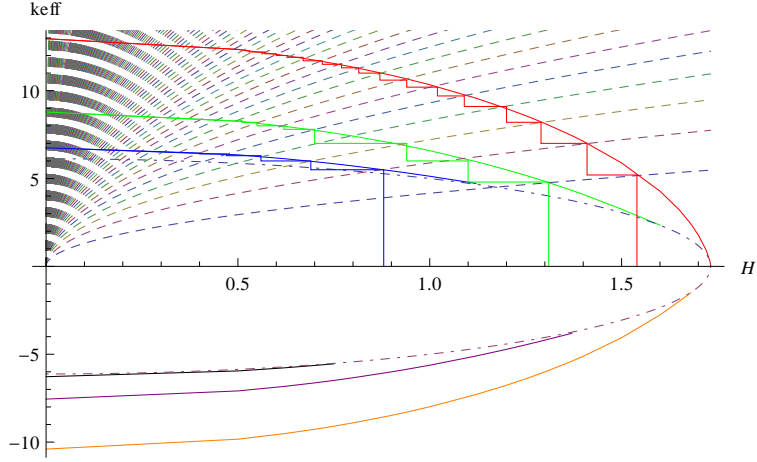}
\caption{Effective momentum $k_{eff}$ vs. the magnetic field $h\rightarrow
H$ (we set $g_F=1$, $q=\frac{15}{\sqrt{3}}$). As we increase magnetic field
the Fermi surface shrinks. Smooth solid curves represent situation as if momentum
is a continuous parameter (for convenience), stepwise solid functions are the real 
Fermi momenta which are discretized due to the Landau level quantization:
$k_F\to \sqrt{2|qh|l}$ with $l=1,2,\dots$ where $\sqrt{2|qh|l}$ are Landau levels given by dotted lines
(only positive discrete $k_F$ are shown). 
At a given $h$ there
are multiple Fermi surfaces. From right to left are the first,
second etc. Fermi surfaces. The dashed-dotted line is
$\nu_{k_F}=0$ where $k_F$ is terminated.
Positive and negative $k_{eff}$ correspond to Fermi
surfaces in two components of the Green's function.}
\label{plot-fermi-momentum}
\end{center}
\end{figure}

\begin{figure}[ht!]
\begin{center}
\includegraphics[width=10cm]{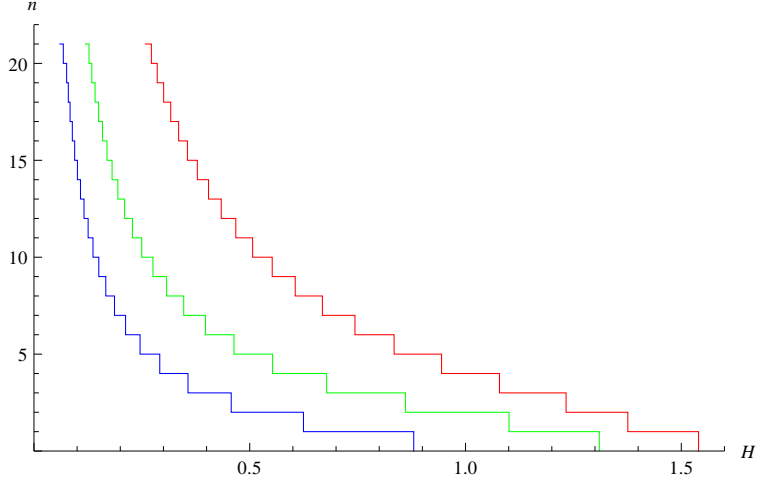}
\caption{Landau level numbers $n$ coresponding to the quantized Fermi momenta vs.
the magnetic field $h\rightarrow H$ for the three Fermi surfaces
with positive $k_F$. We set $g_F=1$, $q=\frac{15}{\sqrt{3}}$. From
right to left are the first, second and third Fermi surfaces.}
\label{plot-landau-level}
\end{center}
\end{figure}

The Fermi momentum is defined by the matching between IR and UV
physics \cite{Faulkner:2009}, therefore it is enough to know the
solution at $\omega=0$, where the matching is performed. To obtain
the Fermi momentum, we require that the zero mode solution is
regular at the horizon ($\psi^{(0)}\sim z^{-\frac{1}{2}+\nu}$) and
normalizable at the boundary. At the boundary $z\sim 1$, the wave
function behaves as
\begin{equation}
a(1-z)^{\frac{3}{2}-m}
\left(\begin{array}{c}
1\\
0
\end{array}\right)
+b(1-z)^{\frac{3}{2}+m}
\left(\begin{array}{c}
0\\
1
\end{array}\right).
\end{equation}
To require it to be normalizable is to set the first term $a=0$;
the wave function at $z\sim 1$ is then
\begin{equation}
\psi^{(0)}\sim (1-z)^{\frac{3}{2}+m} \left(\begin{array}{c}
0\\
1
\end{array}\right).
\label{normalizable}
\end{equation}
Eq. (\ref{normalizable}) leads to the condition
$\lim_{z\rightarrow
1}(z-1)^{-3/2}(\tilde{y}^{(0)}_2+i\tilde{y}^{(0)}_1)=0$, which,
together with eq. (\ref{solution-y}), gives the following equation
for the Fermi momentum as function of the magnetic field
\cite{Hartman:2010,Gubankova:2010}
\begin{equation}
\frac{{}_2F_1(1+\nu+\frac{i\mu_{q,{\mathrm
eff}}}{6},\frac{1}{2}+\nu-\frac{\sqrt{2}\mu_{q,{\mathrm
eff}}}{3},1+2\nu, \frac{2}{3}(1-i\sqrt{2}))}
{{}_2F_1(\nu+\frac{i\mu_{q,{\mathrm
eff}}}{6},\frac{1}{2}+\nu-\frac{\sqrt{2}\mu_{q,{\mathrm
eff}}}{3},1+2\nu, \frac{2}{3}(1-i\sqrt{2}))}=
\frac{6\nu-i\mu_{q,{\mathrm eff}}}{k_F(-2i+\sqrt{2})}, \label{kF2}
\end{equation}
with $\nu\equiv \nu_{k_F}$ given by eq. (\ref{nu}).
Using Mathematica to evaluate the hypergeometric functions, we
numerically solve the equation for the Fermi surface, 
which gives effective momentum as if it were continuous, i.e. when quantization is
neglected. 
The solutions of eq. (\ref{kF2}) are given in Fig.
(\ref{plot-fermi-momentum}). There are
multiple Fermi surfaces for a given magnetic field $h$. Here and
in all other plots we choose $g_F=1$, therefore $h\rightarrow H$,
and $q=\frac{15}{\sqrt{3}}$. In Fig.(\ref{plot-fermi-momentum}),
positive and negative $k_F$ correspond to the Fermi surfaces in
the Green's functions $G_1$ and $G_2$. The relation between two
components is $G_2(\omega,k)=G_1(\omega,-k)$ \cite{Vegh:2009},
therefore Fig.(\ref{plot-fermi-momentum}) is not symmetric with
respect to the x-axis. Effective momenta terminate at the dashed
line $\nu_{k_F}=0$. Taking into account Landau quantization of
$k_F\to \sqrt{2|qh|l}$ with $l=1,2\dots$, the plot consists of stepwise functions tracing the
existing curves (we depict only positive $k_F$). 
Indeed Landau quantiaztion can be also
seen from the dispersion relation
at Fig. (3), where only discrete values of effective momentum are allowed and the Fermi surface
has been chopped up as a result of it Fig. (3B).   
 
Our findings agree with the results for the
(largest) Fermi momentum in a three-dimensional magnetic system
considered in \cite{Shovkovy:2011}, compare the stepwise dependence $k_F(h)$
with Fig.(5) in \cite{Shovkovy:2011}.


In Fig.(\ref{plot-landau-level}), the Landau level index $l$ is
obtained from $k_F(h)=\sqrt{2|qh|l}$ where $k_F(h)$ is a numerical
solution of eq. (\ref{kF2}). Only those Landau levels which are
below the Fermi surface are filled. In Fig.(\ref{plot-fermi-momentum}),
as we decrease magnetic field first nothing happens until the next Landau level
crosses the Fermi surface which corresponds to a jump up to the next step. 
Therefore, at strong magnetic
fields, fewer states contribute to transport properties and the
lowest Landau level becomes more important (see the next section).
At weak magnetic fields, the sum over many Landau levels has to be
taken, ending with the continuous limit as $h\rightarrow 0$, when
quantization can be ignored.

\begin{figure}[ht!]
\begin{center}
\includegraphics[width=7cm]{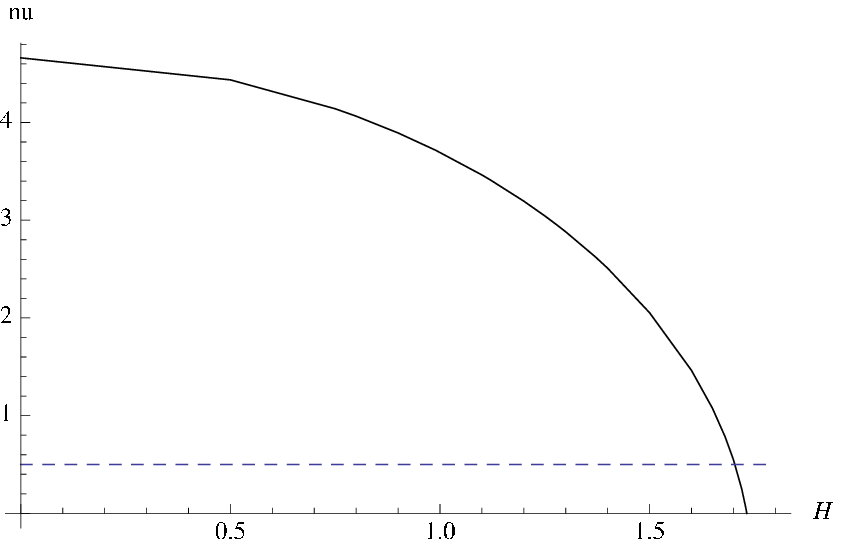}
\includegraphics[width=5cm]{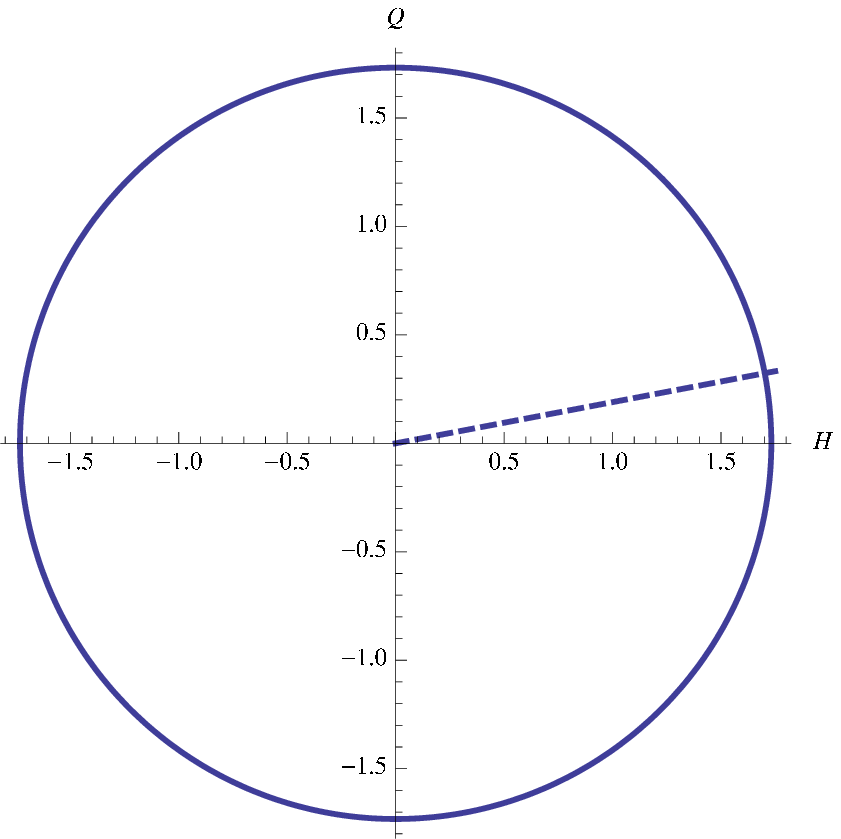}
\caption{Left panel. The IR conformal dimension $\nu\equiv
\nu_{k_F}$ calculated at the Fermi momentum vs. the magnetic field
$h\rightarrow H$ (we set $g_F$=1, $q=\frac{15}{\sqrt{3}}$).
Calculations are done for the first Fermi surface. Dashed line is
for $\nu=\frac{1}{2}$ (at $H_c=1.70$), which is the border between
the Fermi liquids $\nu>\frac{1}{2}$ and non-Fermi liquids
$\nu<\frac{1}{2}$. Right panel. Phase diagram in terms of the
chemical potential and the magnetic field $\mu^2+h^2=3$ (in
dimensionless variables $h=g_F H$, $\mu=g_F Q$; we set $g_F=1$).
Fermi liquids are above the dashed line ($H< H_c$) and non-Fermi
liquids are below the dashed line ($H> H_c$).}
\label{plot-conformal-dimension}
\end{center}
\end{figure}

In Fig. (\ref{plot-conformal-dimension}), we show the IR conformal
dimension as a function of the magnetic field. We have used the
numerical solution for $k_F$. Fermi liquid regime takes place at
magnetic fields $h<h_c$, while non-Fermi liquids exist in a narrow
band at $h_c<h<h_c^{\prime}$, and at $h_c^{\prime}$ the system
becomes near-conformal.

In this figure we observe the pathway of the possible phase
transition exhibited by the Fermi surface (ignoring Landau
quantization): it can vanish at the line $\nu_{k_F}=0$, undergoing
a crossover to the conformal regime, or cross the line
$\nu_{k_F}=1/2$ and go through a non-Fermi liquid regime, and
subsequently cross to the conformal phase. Note that the primary
Fermi surface with the highest $k_F$ and $\nu_{k_F}$ seems to
directly cross over to conformality, while the other Fermi
surfaces first exhibit a "strange metal" phase transition.
Therefore, all the Fermi momenta with $\nu_{k_F}>0$ contribute to
the transport coefficients of the theory. In particular, at high
magnetic fields when for the first (largest) Fermi surface
$k_F^{(1)}$ is nonzero but small, the lowest Landau level $n=0$
becomes increasingly important contributing to the transport with
half degeneracy factor as compared to the higher Landau levels.



\begin{figure}[ht!]
\begin{center}
\includegraphics[width=6cm]{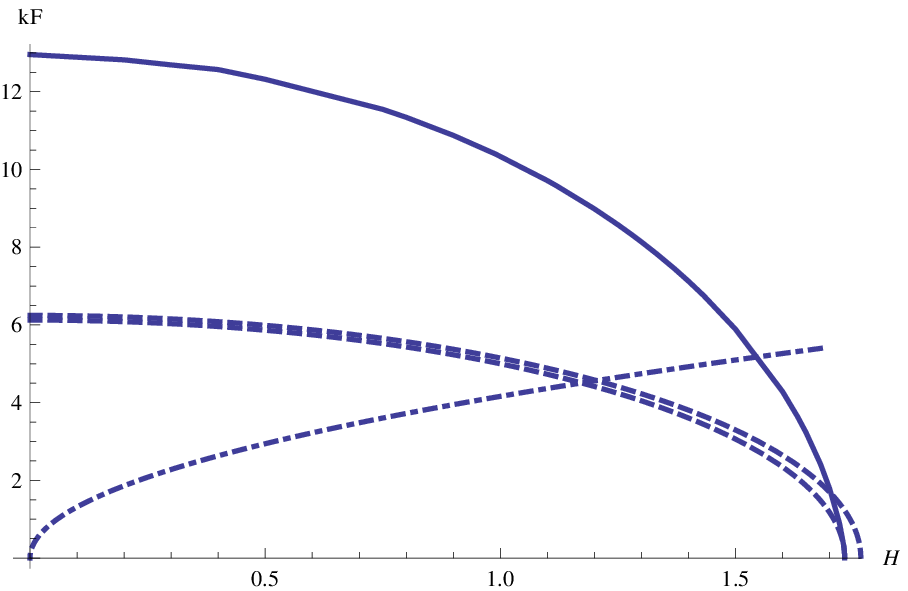}
\includegraphics[width=6cm]{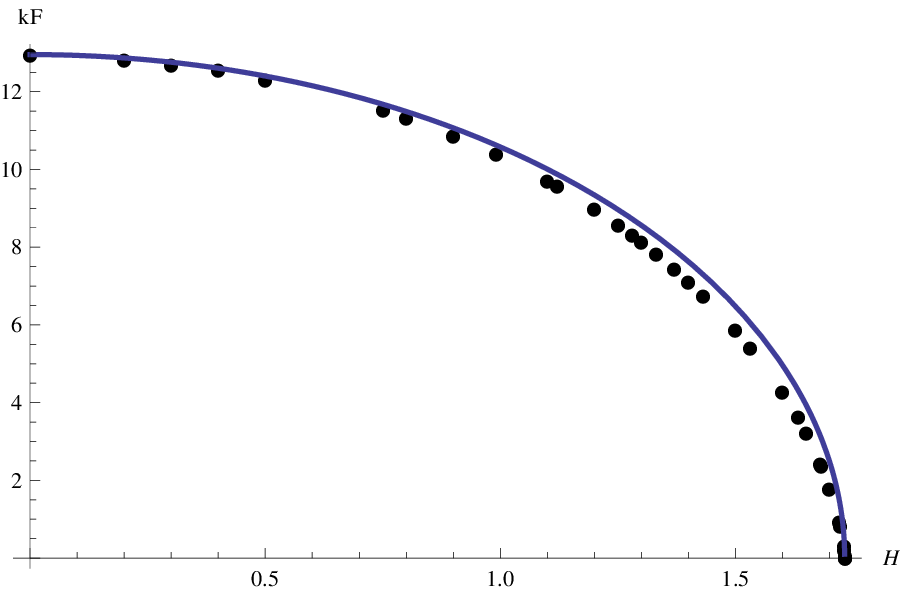}
\caption{Fermi momentum $k_F$ vs. the magnetic field $h\rightarrow
H$ (we set $g_F=1$, $q=\frac{15}{\sqrt{3}}$) for the first Fermi
surface. Left panel. The inner (closer to x-axis) dashed
line is $\nu_{k_F}=0$ and the outer dashed line is
$\nu_{k_F}=\frac{1}{2}$, the region between these lines
corresponds to non-Fermi liquids $0<\nu_{k_F}<\frac{1}{2}$. The
dashed-dotted line is for the first Landau level $k_1=\sqrt{2qH}$. The
first Fermi surface hits the border-line between a Fermi and
non-Fermi liquids $\nu=\frac{1}{2}$ at $H_c\approx 1.70$, and it
vanishes at $H_{max}=\sqrt{3}=1.73$. Right panel. Circles are the data points for 
the Fermi momentum
calculated analytically, solid line is a
fit function 
$k_F^{max}\sqrt{1-\frac{H^2}{3}}$ with $k_F^{max}=12.96$.}
\label{plot-fermi-momentum-one}
\end{center}
\end{figure}

In Fig. \ref{plot-fermi-momentum-one}, we plot the Fermi momentum
$k_F$ as a function of the magnetic field for the first Fermi
surface (the largest root of eq. (\ref{kF2})). Quantization is neglected here. 
At the left panel,
the relatively small region between the dashed lines corresponds
to non-Fermi liquids $0< \nu< \frac{1}{2}$. At large magnetic
field, the physics of the Fermi surface is captured by the near
horizon region (see also Fig. (\ref{plot-wave-function})) which is
${\rm AdS}_2\times\mathbb{R}^2$. At the maximum magnetic field,
$H_{max}=\sqrt{3}\approx 1.73$, when the black hole becomes pure
magnetically charged, the Fermi momentum vanishes when it crosses
the line $\nu_{k_F}=0$. This only happens for the first Fermi
surface. For the higher Fermi surfaces the Fermi momenta terminate
at the line $\nu_{k_F}=0$, Fig. (\ref{plot-fermi-momentum}). Note
the Fermi momentum for the first Fermi surface can be almost fully
described by a function $k_F=k_F^{max}\sqrt{1-\frac{H^2}{3}}$. It
is tempting to view the behavior $k_F\sim \sqrt{H_{max}-H}$ as a
phase transition in the system although it strictly follows from
the linear scaling for $H=0$ by using the mapping
(\ref{eq:GFMapping}). (Note that also $\mu=g_F
Q=g_F\sqrt{3-H^2}$.) Taking into account the discretization of
$k_F$, the plot will consist of an array of step functions tracing
the existing curve. Our findings agree with the results for the
Fermi momentum in a three dimensional magnetic system considered
in \cite{Shovkovy:2011}, compare with Fig.(5) there.

The Fermi velocity given in eq. (\ref{green-function}) is defined
by the UV physics; therefore solutions at non-zero $\omega$ are
required. The Fermi velocity is extracted from matching two
solutions in the inner and outer regions at the horizon. The Fermi
velocity as function of the magnetic field for $\nu> \frac{1}{2}$
is \cite{Hartman:2010,Gubankova:2010}
\begin{eqnarray}
v_F &=& \frac{1}{h_1}\left(\int_{0}^{1} dz \sqrt{g/g_{tt}}\psi^{(0)\dagger}\psi^{(0)}\right)^{-1}
\lim_{z\rightarrow 1}\frac{|\tilde{y}_{1}^{(0)}+i\tilde{y}_{2}^{(0)}|^2}{(1-z)^3},\nonumber\\
h_1 &=& \lim_{z\rightarrow 1}\frac{\tilde{y}_{1}^{(0)}+i\tilde{y}_{2}^{(0)}}
{\partial_{k}(\tilde{y_{2}^{(0)}}+i\tilde{y}_{1}^{(0)})},
\label{constant}
\end{eqnarray}
where the zero mode wavefunction is taken at $k_F$ eq.(\ref{solution-y}).

\begin{figure}[ht!]
\begin{center}
\includegraphics[width=10cm]{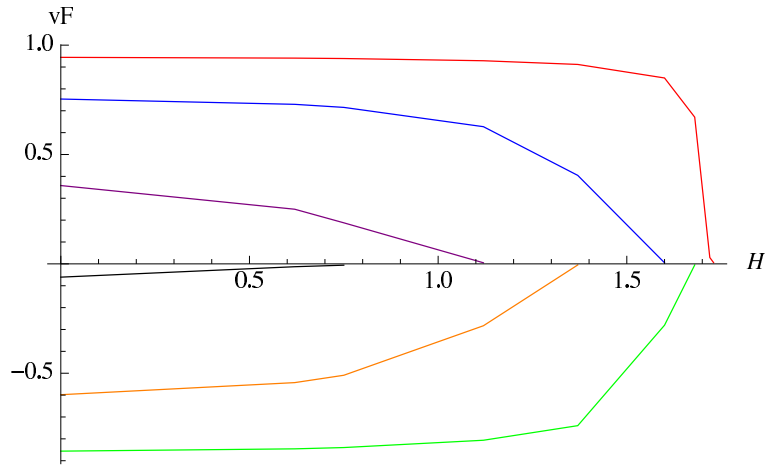}
\caption{Fermi velocity $v_F$ vs. the magnetic field $h
\rightarrow H$ (we set $g_F=1$, $q=\frac{15}{\sqrt{3}}$) for the
regime of Fermi liquids $\nu\geq\frac{1}{2}$. Fermi velocity
vanishes at $\nu_{k_F}=\frac{1}{2}$ (x-axis). The multiple lines
are for various Fermi surfaces in ascending order, with the first
Fermi surface on the right. The Fermi velocity $v_F$ has the same
sign as the Fermi momentum $k_F$. As above, positive and negative
$v_F$ correspond to Fermi surfaces in the two components of the
Green's function.} \label{plot-fermi-velocity}
\end{center}
\end{figure}

\begin{figure}[ht!]
\begin{center}
\includegraphics[width=7cm]{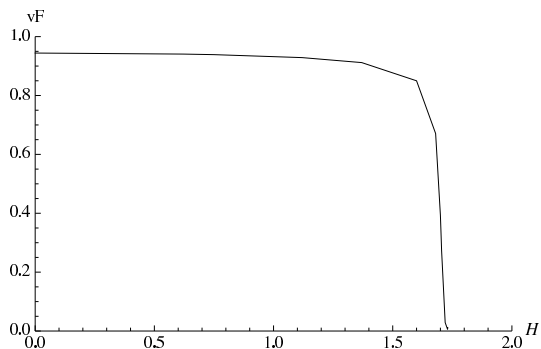}
\caption{Fermi velocity $v_F$ vs. the magnetic field $h
\rightarrow H$ (we set $g_F=1$, $q=\frac{15}{\sqrt{3}}$) for the
first Fermi surface. Fermi velocity vanishes at
$\nu_{k_F}=\frac{1}{2}$ at $H_c\approx 1.70$. The region $H< H_c$
corresponds to the Fermi liquids and quasiparticle description.}
\label{plot-fermi-velocity-one}
\end{center}
\end{figure}

We plot the Fermi velocity for several Fermi surfaces in Fig.
\ref{plot-fermi-velocity} and for the first Fermi surface in Fig.
\ref{plot-fermi-velocity-one}. Quantization is neglected here. 
The Fermi velocity is shown for
$\nu> \frac{1}{2}$. It is interesting that the Fermi velocity
vanishes when the IR conformal dimension is
$\nu_{k_F}=\frac{1}{2}$. Formally, it follows from the fact that
$v_F\sim (2\nu-1)$ \cite{Faulkner:2009}. The first Fermi surface
is at the far right. Positive and negative $v_F$ correspond to the
Fermi surfaces in the Green's functions $G_1$ and $G_2$,
respectively. The Fermi velocity $v_F$ has the same sign as the
Fermi momentum $k_F$. At small magnetic field values, the Fermi
velocity is very weakly dependent on $H$ and it is close to the
speed of light; at large magnetic field values, the Fermi velocity
rapidly decreases and vanishes (at $H_c=1.70$ for the first Fermi
surface Fig.(\ref{plot-fermi-velocity-one})). Geometrically, this
means that with increasing magnetic field the zero mode
wavefunction is supported near the black hole horizon
fig.(\ref{plot-wave-function}), where the gravitational redshift
reduces the local speed of light as compared to the boundary
value. It was also observed in \cite{Hartman:2010,Faulkner:2009}
at small fermion charge values.

\section{Hall and longitudinal conductivities}\label{section:5}

In this section, we calculate the contributions to Hall
$\sigma_{xy}$ and the longitudinal $\sigma_{xx}$ conductivities
directly in the boundary theory. This should be contrasted with
the standard holographic approach, where calculations are
performed in the (bulk) gravity theory and then translated to the
boundary field theory using the AdS/CFT dictionary. Specifically,
the conductivity tensor has been obtained in \cite{HallCondBH} by
calculating the on-shell renormalized action for the gauge field
on the gravity side and using the gauge/gravity duality
$A_M\rightarrow j_{\mu}$ to extract
 the $R$ charge current-current correlator
at the boundary. Here, the Kubo formula involving the
current-current correlator is used directly by utilizing the
fermion Green's functions extracted from holography in
\cite{Faulkner:2009}. Therefore, the conductivity is obtained for
the charge carriers described by the fermionic operators of the
boundary field theory.

The use of the conventional Kubo formulo to extract the
contribution to the transport due to fermions is validated in that
it also follows from a direct AdS/CFT computation of the one-loop
correction to the on-shell renormalized AdS action
\cite{Faulkner:2010}.
We study in particular stable quasiparticles with $\nu>
\frac{1}{2}$ and at zero temperature. This regime effectively
reduces to the clean limit where the imaginary part of the
self-energy vanishes ${\rm Im}\Sigma\rightarrow 0$. We use the
gravity-``dressed'' fermion propagator from eq.
(\ref{green-function}) and to make the calculations complete, the
``dressed'' vertex is necessary, to satisfy the Ward identities.
As was argued in \cite{Faulkner:2010}, the boundary vertex which
is obtained from the bulk calculations can be approximated by a
constant in the low temperature limit. Also, according to
\cite{Basagoiti:2002}, the vertex only contains singularities of
the product of the Green's functions. Therefore, dressing the
vertex will not change the dependence of the DC conductivity on
the magnetic field \cite{Basagoiti:2002}. In addition, the zero
magnetic field limit of the formulae for conductivity obtained
from holography \cite{Faulkner:2010} and from direct boundary
calculations \cite{Gubankova:2010} are identical.

\subsection{Integer quantum Hall effect}

Let us start from the ``dressed'' retarded and advanced fermion
propagators \cite{Faulkner:2009}: $G_R$ is given by eq.
(\ref{green-function}) and $G_A=G_R^{*}$. To perform the Matsubara
summation we use the spectral representation
\begin{eqnarray}
G(i\omega_n,\vec{k})=\int\frac{d\omega}{2\pi}\frac{A(\omega,\vec{k})}{\omega-i\omega_n},
\label{spectral}
\end{eqnarray}
with the spectral function defined as
$A(\omega,\vec{k})=-\frac{1}{\pi}{\rm
Im}G_R(\omega,\vec{k})=\frac{1}{2\pi
i}(G_R(\omega,\vec{k})-G_A(\omega,\vec{k}))$. Generalizing to a
non-zero magnetic field and spinor case \cite{Shovkovy:2006}, the
spectral function \cite{RealTimerespspinor} is
\begin{eqnarray}
A(\omega,\vec{k})&=& \frac{1}{\pi}{\rm e}^{-\frac{k^2}{|qh|}}
\sum_{l=0}^{\infty}
(-1)^l(-h_1v_F)\left(\frac{\Sigma_2(\omega,k_F)f(\vec{k})\gamma^0}{(\omega+\varepsilon_F+\Sigma_1(\omega,k_F)-E_l)^2+
\Sigma_2(\omega,k_F)^2}+ (E_l\rightarrow -E_l)\right),
\label{spectral-function}
\end{eqnarray}
where $\varepsilon_F=v_Fk_F$ is the Fermi energy,
$E_l=v_F\sqrt{2|qh|l}$ is the energy of the Landau level,
$f(\vec{k})=P_{-}L_{l}(\frac{2k^2}{|qh|})-P_{+}L_{l-1}(\frac{2k^2}{|qh|})$
with spin projection operators $P_{\pm}=(1\pm
i\gamma^1\gamma^2)/2$, we take $c=1$, the generalized Laguerre
polynomials are $L_n^{\alpha}(z)$ and by definition
$L_{n}(z)=L_{n}^{0}(z)$, (we omit the vector part
$\vec{k}\vec{\gamma}$, it does not contribute to the DC
conductivity), all $\gamma$'s are the standard Dirac matrices,
$h_1$, $v_F$ and $k_F$ are real constants (we keep the same
notations for the constants as in \cite{Faulkner:2009}). The
self-energy $\Sigma \sim \omega^{2\nu_{k_F}}$ contains the real
and imaginary parts, $\Sigma=\Sigma_1+i\Sigma_2$. The imaginary
part comes from scattering processes of a fermion in the bulk,
e.~g. from pair creation, and from the scattering into the black
hole. It is exactly due to inelastic/dissipative processes that we
are able to obtain finite values for the transport coefficients,
otherwise they are formally infinite.

Using the Kubo formula, the DC electrical conductivity tensor is
\begin{eqnarray}
\sigma_{ij}(\Omega)= \lim_{\Omega\rightarrow 0}\frac{{\rm
Im}\Pi_{ij}^{R}}{\Omega+i0^{+}},
\end{eqnarray}
where $\Pi_{ij}(i\Omega_m\rightarrow \Omega+i0^{+})$ is the
retarded current-current correlation function; schematically the
current density operator is
$j^i(\tau,\vec{x})=qv_F\sum_{\sigma}\bar{\psi}_{\sigma}(\tau,\vec{x})\gamma^i\psi_{\sigma}(\tau,\vec{x})$.
Neglecting the vertex correction, it is given by
\begin{eqnarray}
\Pi_{ij}(i\Omega_m)= q^2v_F^2T\sum_{n=-\infty}^{\infty}\int\frac{d^2k}{(2\pi)^2}
{\rm tr}\left(\gamma^{i}G(i\omega_n,\vec{k})\gamma^{j}G(i\omega_n+i\Omega_m,\vec{k})\right).
\end{eqnarray}
The sum over the Matsubara frequency is
\begin{eqnarray}
T\sum_{n}\frac{1}{i\omega_n-\omega_1}\frac{1}{i\omega_n+i\Omega_m-\omega_2}=
\frac{n(\omega_1)-n(\omega_2)}{i\Omega_m+\omega_1-\omega_2}.
\end{eqnarray}
Taking $i\Omega_m\rightarrow \Omega+i0^{+}$, the polarization
operator is now
\begin{eqnarray}
\Pi_{ij}(\Omega) = \frac{d\omega_1}{2\pi}\frac{d\omega_2}{2\pi}
\frac{n_\mathrm{FD}(\omega_1)-n_\mathrm{FD}(\omega_2)}{\Omega+\omega_1-\omega_2}
\int\frac{d^2k}{(2\pi)^2}{\rm tr}\left(\gamma^{i}
A(\omega_1,\vec{k})\gamma^{j}A(\omega_2,\vec{k})\right),
\end{eqnarray}
where the spectral function $A(\omega,\vec{k})$ is given by eq.
(\ref{spectral-function}) and $n_\mathrm{FD}(\omega)$ is the
Fermi-Dirac distribution function.  Evaluating the traces, we have
\begin{eqnarray}
\sigma_{ij} &=& -\frac{4q^2v_F^2(h_1v_F)^2|qh|}{\pi\Omega}{\rm Re}\sum_{l,k=0}^{\infty}
(-1)^{l+k+1}\left\{\delta_{ij}(\delta_{l,k-1}+\delta_{l-1,k})
+i\epsilon_{ij}{\rm sgn}(qh)(\delta_{l,k-1}-\delta_{l-1,k})\right\} \nonumber\\
&\times& \int \frac{d\omega_1}{2\pi} (\tanh\frac{\omega_1}{2T}-\tanh\frac{\omega_2}{2T})\nonumber\\
&\times& \left( \frac{\Sigma_2(\omega_1)}{(\tilde{\omega}_1-E_l)^2+\Sigma_2^2(\omega_1)}
+(E_l\rightarrow -E_l)\right)
\left( \frac{\Sigma_2(\omega_2)}{(\tilde{\omega}_2-E_k)^2+\Sigma_2^2(\omega_2)}
+ (E_k\rightarrow -E_k)\right),
\label{conductivity}
\end{eqnarray}
with $\omega_2=\omega_1+\Omega$. We have also introduced
$\tilde{\omega}_{1;2}\equiv\omega_{1;2}+\varepsilon_F+\Sigma_{1}(\omega_{1;2})$
with $\epsilon_{ij}$ being the antisymmetric tensor
($\epsilon_{12}=1$), and $\Sigma_{1;2}(\omega)\equiv
\Sigma_{1;2}(\omega,k_F)$. In the momentum integral, we use the
orthogonality condition for the Laguerre polynomials
$\int_0^{\infty} dx {\rm e}^{x}L_l(x)L_k(x)=\delta_{lk}$.

From eq. (\ref{conductivity}), the term symmetric/antisymmetric
with respect to exchange $\omega_1\leftrightarrow\omega_2$
contributes to the diagonal/off-dialgonal component of the
conductivity (note the antisymmetric term
$n_\mathrm{FD}(\omega_1)-n_\mathrm{FD}(\omega_2)$). The
longitudinal and Hall DC conductivities ($\Omega\rightarrow 0$)
are thus
\begin{eqnarray}
\sigma_{xx} &=& -\frac{2q^2(h_1v_F)^2|qh|}{\pi T}\int_{-\infty}^{\infty}\frac{d\omega}{2\pi}
\frac{\Sigma_2^2(\omega)}{\cosh^2\frac{\omega}{2T}}\nonumber\\
&\times& \sum_{l=0}^{\infty}\left(\frac{1}{(\tilde{\omega}-E_l)^2+\Sigma_2^2(\omega)}
+(E_l\rightarrow -E_l)\right)
\left(\frac{1}{(\tilde{\omega}-E_{l+1})^2+\Sigma_2^2(\omega)}+(E_{l+1}\rightarrow -E_{l+1})\right),
\label{conductivity22}\\
\sigma_{xy} &=& -\frac{q^2 (h_1v_F)^2 {\rm sgn}(qh)}{\pi}\nu_h,\nonumber\\
\nu_h &=& 2 \int_{-\infty}^{\infty}
\frac{d\omega}{2\pi}\tanh\frac{\omega}{2T}\;\Sigma_2(\omega)
\sum_{l=0}^{\infty}\alpha_l
\left(\frac{1}{(\tilde{\omega}-E_l)^2+\Sigma_2^2(\omega)}+(E_l\rightarrow -E_l)\right),
\label{conductivity2}
\end{eqnarray}
where $\tilde{\omega}=\omega+\varepsilon_F+\Sigma_1(\omega))$. The
filling factor $\nu_h$ is proportional to the density of carriers:
$|\nu_h|=\frac{\pi}{|qh|h_1v_F}n$ (we derive this relation
below eq. \eqref{result-n}).
The degeneracy factor of the Landau levels is
$\alpha_l$: $\alpha_0=1$ for the lowest Landau level and
$\alpha_l=2$ for $l=1,2\dots$. Substituting the filling factor
$\nu_h$ back to eq. (\ref{conductivity2}), the Hall conductivity
can be written as
\begin{equation}
\sigma_{xy}=\frac{\rho}{h},
\label{hall-conductivity}
\end{equation}
where $\rho$ is the charge density in the boundary theory, and
both the charge $q$ and the magnetic field $h$ carry a sign (the
prefactor $(-h_1v_F)$ comes from the normalization choice in the
fermion propagator eqs.
(\ref{green-function},\ref{spectral-function}) as given in
\cite{Faulkner:2009}, which can be regarded as a factor
contributing to the effective charge and is not important for
further considerations). The Hall conductivity eq.
(\ref{hall-conductivity}) has been obtained using the AdS/CFT
duality for the Lorentz invariant $2+1$-dimensional boundary field
theories in \cite{HallCondBH}. We recover this formula because in
our case the translational invariance is maintained in the $x$ and
$y$ directions of the boundary theory.

Low frequencies give the main contribution in the integrand of eq.
(\ref{conductivity2}). Since the self-energy satisfies
$\Sigma_1(\omega)\sim \Sigma_2(\omega)\sim \omega^{2\nu}$ and we
consider the regime $\nu> \frac{1}{2}$, we have
$\Sigma_1\sim\Sigma_2\rightarrow 0$ at $\omega\sim 0$ (self-energy
goes to zero faster than the $\omega$ term). Therefore, only the
simple poles in the upper half-plane $\omega_0=-\varepsilon_F\pm
E_l+\Sigma_1 + i\Sigma_2$ contribute to the conductivity where
$\Sigma_1\sim \Sigma_2\sim (-\varepsilon_F\pm E_l)^{2\nu}$ are
small. The same logic of calculation has been used in
\cite{Shovkovy:2006}. We obtain for the longitudinal and Hall
conductivities
\begin{eqnarray}
\sigma_{xx} &=& \frac{2q^2(h_1v_F)^2\Sigma_2}{\pi
T}\times\left(\frac{1}{1+\cosh\frac{\varepsilon_F}{T}}
+\sum_{l=1}^{\infty}4l\frac{1+\cosh\frac{\varepsilon_F}{T}\cosh\frac{E_l}{T}}{(\cosh\frac{\varepsilon_F}{T}+\cosh\frac{E_l}{T})^2}\right)
\\
\sigma_{xy} &=& \frac{q^2(h_1v_F)^2{\rm sgn}(qh)}{\pi} \times 2\left(\tanh\frac{\varepsilon_F}{2T}
+\sum_{l=1}^{\infty}
(\tanh\frac{\varepsilon_F+E_l}{2T}+\tanh\frac{\varepsilon_F-E_l}{2T})\right),
\label{conductivity3}
\end{eqnarray}
where the Fermi energy is $\varepsilon_F=v_Fk_F$ and the energy of
the Landau level is $E_l=v_F\sqrt{2|qh|l}$. Similar expressions
were obtained in \cite{Shovkovy:2006}. However, in our case the
filling of the Landau levels is controlled by the magnetic field
$h$ through the field-dependent Fermi energy $v_F(h)k_F(h)$
instead of the chemical potential $\mu$.

\begin{figure}[ht!]
\begin{center}
\includegraphics[width=8.5cm]{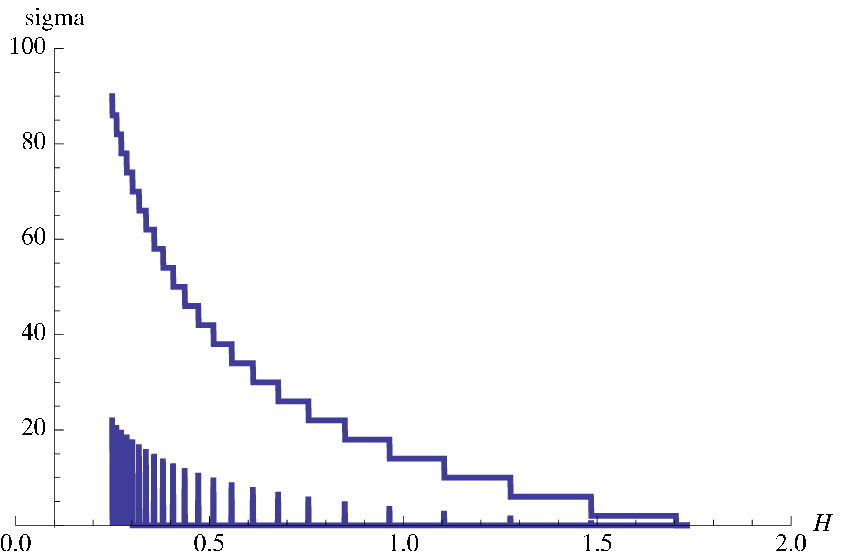}
\includegraphics[width=8.5cm]{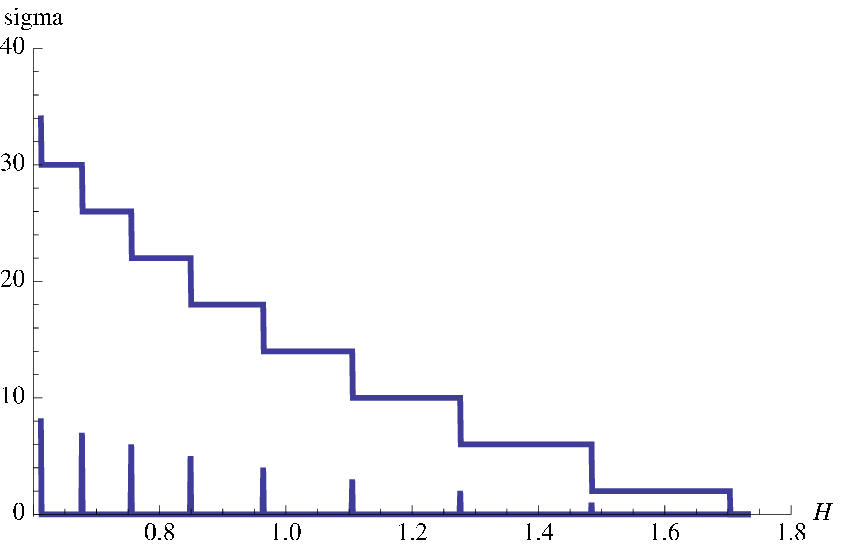}
\caption{Hall conductivity $\sigma_{xy}$ and longitudinal
conductivity $\sigma_{xx}$ vs. the magnetic field $h\rightarrow H$
at $T=0$ (we set $g_F=1$, $q=\frac{15}{\sqrt{3}}$). Contribution
from the first Fermi surface is taken. By decreasing the magnetic
field, the Fermi surface crosses the Landau levels producing the
Hall conductivity plateaus characteristic for IQHE. Longitudinal
conductivity has picks at the beginning of each plateau. The right
panel is a zoom-in for large $h$ of the left one.}
\label{plot-hall-conductivity-1fs}
\end{center}
\end{figure}

At $T=0$, $\cosh\frac{\omega}{T}\rightarrow \frac{1}{2}{\rm
e}^{\frac{\omega}{T}}$ and
$\tanh\frac{\omega}{2T}=1-2n_\mathrm{FD}(\omega)\rightarrow {\rm
sgn}\omega$. Therefore the longitudinal and Hall conductivities
are
\begin{eqnarray}
\sigma_{xx} &=& \frac{2q^2(h_1v_F)^2\Sigma_2}{\pi T}\sum_{l=1}^{\infty}l\delta_{\varepsilon_F,E_l}
= \frac{2q^2(h_1v_F)^2\Sigma_2}{\pi T}\times n\delta_{\varepsilon_F,E_n},
\\
\sigma_{xy} &=& \frac{q^2(h_1v_F)^2{\rm sgn}(qh)}{\pi}\;2\left(1+2\sum_{l=1}^{\infty}
\theta(\varepsilon_F-E_l)\right)\nonumber\\
&=& \frac{q^2(h_1v_F)^2{\rm sgn}(qh)}{\pi} \times 2(1+2n)
\theta(\varepsilon_F-E_n)\theta(E_{n+1}-\varepsilon_F),
\label{conductivity4}
\end{eqnarray}
where the Landau level index runs $n=0,1,\dots$. It can be
estimated as $n=\left[\frac{k_F^2}{2|qh|}\right]$ when $v_F\neq 0$
($[\;]$ denotes the integer part), with the average spacing
between the Landau levels given by the Landau energy
$v_F\sqrt{2|qh|}$. Note that $\varepsilon_F\equiv
\varepsilon_F(h)$. We can see that eq. (\ref{conductivity4})
expresses the integer quantum Hall effect (IQHE). At zero
temperature, as we dial the magnetic field, the Hall conductivity
jumps from one quantized level to another, forming plateaus given
by the filling factor
\begin{eqnarray}
\nu_h=\pm 2(1+2n)=\pm 4(n+\frac{1}{2}),
\label{integer}
\end{eqnarray}
with $n=0,1,\dots$. (Compare to the conventional Hall quantization
$\nu_h=\pm 4n$, that appears in thick graphene). Plateaus of the
Hall conductivity at $T=0$ follow from the stepwise behavior of
the charge density $\rho$ in eq.(\ref{hall-conductivity}):
\begin{eqnarray}
\rho\sim 4(n+\frac{1}{2})\theta(\varepsilon_F-E_n)\theta(E_{n+1}-\varepsilon_F),
\end{eqnarray}
where $n$ Landau levels are filled and contribute to $\rho$.
The longitudinal
conductivity vanishes except precisely at the transition point
between the plateaus. In Fig. \ref{plot-hall-conductivity-1fs}, we
plot the longitudinal and Hall conductivities at $T=0$, using only
the terms after $\times$ sign in eq. (\ref{conductivity3}). In the
Hall conductivity, plateau transition occurs when the Fermi level
(in Fig. \ref{plot-hall-conductivity-1fs}) of the first Fermi
surface $\varepsilon_F=v_F(h)k_F(h)$ (Figs.
\ref{plot-fermi-momentum-one},\ref{plot-fermi-velocity-one})
crosses the Landau level energy as we vary the magnetic field. By
decreasing the magnetic field, the plateaus become shorter and
increasingly more Landau levels contribute to the Hall
conductivity. This happens because of two factors: the Fermi level
moves up and the spacing between the Landau levels becomes
smaller. This picture does not depend on the Fermi velocity as
long as it is nonzero.

In the boundary field theory, we express the charge density of the
carriers (difference between the densities of "electrons" and
"holes") through the Fermi energy $\varepsilon_F$:
\begin{eqnarray}
n = {\rm tr}\left(\gamma^0 \tilde{G}(\tau,{\bf 0}) \right),\;\;\tau\rightarrow 0,
\label{density}
\end{eqnarray}
where $\tilde{G}(\tau,\vec{x})$ is the translation-invariant part
of the Green's function $G(\tau,\vec{x})$ from eq.
(\ref{spectral}) \cite{Shovkovy:2006}. Using the spectral function
representation eq. (\ref{spectral-function}), the charge density
reads
\begin{eqnarray}
n=T\sum_{n=-\infty}^{\infty}\int\frac{d^2k}{(2\pi)^2}\int_{-\infty}^{\infty}\frac{d\omega}{2\pi}
\frac{{\rm tr}\left(\gamma^0 A(\omega,\vec{k})\right)}{\omega-i\omega_n}.
\label{charge-density}
\end{eqnarray}
We express the Matsubara sum in terms of the contour integral over
real frequencies:
\begin{eqnarray}
T \sum_{n=-\infty}^{\infty}F(i\omega_n)\rightarrow -\frac{i}{4\pi}\int_C dz
\tanh\frac{z}{2T}F(z),
\end{eqnarray}
where $C$ runs anti-clockwise and encircles the poles of $\tanh$
along the upper and lower half imaginary axis. We have for the
charge density
\begin{eqnarray}
n=\frac{1}{2}\int\frac{d^2k}{(2\pi)^2}\int_{-\infty}^{\infty} \frac{d\omega}{2\pi} \tanh\frac{\omega}{2T}
{\rm tr}\left(\gamma^0 A(\omega,\vec{k})\right).
\end{eqnarray}
Substituting the spectral function eq. (\ref{spectral-function})
and integrating over momenta, we obtain
\begin{eqnarray}
n=-\frac{2|qh|h_1v_F}{\pi}\int_{-\infty}^{\infty}\frac{d\omega}{2\pi}\tanh\frac{\omega}{2T}\;\Sigma_2(\omega)
\sum_{l=0}^{\infty}\alpha_l\left(\frac{1}{(\tilde{\omega}-E_l)^2+\Sigma_2^2(\omega)}+(E_l\rightarrow -E_l)\right),
\label{density-of-carriers}
\end{eqnarray}
where the degeneracy factor is $\alpha_0=1$ for the lowest Landau
level and $\alpha_l=2$ for the higher Landau levels $l\geq 1$,
$\tilde{\omega}=\omega+\varepsilon_F+\Sigma_1(\omega)$.
Integrating over frequencies and taking into account that
$\Sigma_2$ is effectively very small near the Fermi surface, we
obtain
\begin{eqnarray}
n=\frac{|qh|h_1v_F}{\pi}\times 2\left(\tanh\frac{\varepsilon_F}{2T}
+\sum_{l=1}^{\infty}(\tanh\frac{\varepsilon_F+E_l}{2T}+\tanh\frac{\varepsilon_F-E_l}{2T})
\right).
\label{result-n}
\end{eqnarray}
Comparing with eq. (\ref{conductivity3}), we obtain the relation
$|\nu_h|=\frac{\pi}{|qh|h_1v_F}n$. When the Fermi energy vanishes
($\varepsilon_F=0$), the spectral function eq.
(\ref{spectral-function}) is even in $\omega$. From eq.
(\ref{density-of-carriers}), the carrier density of stable
quasiparticles vanishes when $\varepsilon_F=0$. At the end of this
section, we discuss a situation with no stable charge carriers and
physical consequences of it.

Eqs. (\ref{conductivity3}-\ref{result-n}) are obtained assuming
that the states are localized around the Landau levels. In QHE
models, impurities are added to prevent the states from 'spilling'
between the Landau levels and to provide the necessary occupation
number of the levels. In our holographic calculations, however, the complex
self-energy arises not from the impurities but from various
scattering processes into the black hole.
Here, the limit ${\rm Im}\Sigma\rightarrow 0$
has been considered, which
corresponds to a simplified field theory model
\cite{Shovkovy:2006} (the cited reference also considers the case
with impurities). This approximation suffices to obtain the
integer QHE \cite{Shovkovy:2006} and for our initial studies of
the fractional QHE. We leave the implementation of a physical
model with impurities for future work.

\subsection{Fractional quantum Hall effect}

In a holographic setting, using the AdS geometry is equivalent to
a calculation in a box. Therefore, for large enough fermion charge
$q$, there are multiple Fermi surfaces as shown in Figs.
\ref{plot-fermi-momentum},\ref{plot-fermi-velocity}. Labelling the
Fermi surfaces with $\nu>\frac{1}{2}$ by $m=1,2,\dots$, we
represent the spectral function $A(\omega,\vec{k})$ as a sum over
the spectral functions of individual Fermi surfaces given by eq.
(\ref{spectral-function}) \cite{Hartman:2010}. Ignoring the mixing
term, the DC conductivity becomes a direct sum over the individual
conductivities. By decreasing the magnetic field, new Fermi
surfaces gradually appear as can be seen in Figs.
\ref{plot-fermi-momentum},\ref{plot-landau-level}. Therefore the
conductivity tensor is
\begin{eqnarray}
\sigma_{ij}=\sum_{m}\sigma_{ij}^{(m)}\theta(h_{max}^{(m)}-h),
\label{sigma-multiple}
\end{eqnarray}
where $\sigma_{ij}^{(m)}$ involves the Fermi momentum $k_F^{(m)}$
and velocity $v_F^{(m)}$, respectively; at the maximum magnetic
field $h_{max}^{(m)}$ a new $k_F^{(m)}$ opens up; $h_{max}^{(m)}$
is found numerically.

Including one, two, three and four Fermi surfaces, we obtain the
following quantization rule for the filling factor in the Hall
conductivity:
\begin{eqnarray}
&& {\rm 1\;FS:}\;\; \nu_h=2(1+2n),\;\; {\rm plateaus}\rightarrow 2,6,10,\dots,\nonumber\\
&& {\rm 2\;FS's:}\;\; \nu_h=4(1+n+k),\;\; {\rm plateaus}\rightarrow 4,8,12,\dots,\nonumber\\
&& {\rm 3\;FS's:}\;\; \nu_h=2(3+2(n+k+p)),\;\; {\rm plateaus}\rightarrow 6,10,14,\dots,\nonumber\\
&& {\rm 4\;FS's:}\;\; \nu_h=4(2+n+k+p+r),\;\; {\rm plateaus}\rightarrow 8,12,16,\dots,
\label{fractional}
\end{eqnarray}
with $n,k,p,r=0,1,\dots$. An odd number of Fermi surfaces produces
the plateaus present in the IQHE, while an even number of Fermi
surfaces produces the additional plateaus appearing in the
fractional QHE (FQHE). For a large enough fermion charge $q$, many
Fermi surfaces contribute, and the primary effect of the change in
$H$ is the opening of a new Fermi surface, rather than the
occupation of the next plateau. Thus at large $q$ we expect a
filling fraction pattern at large $h$ to become
\begin{eqnarray}
\nu_h=\pm 2j, \label{fractional2}
\end{eqnarray}
where $j=1,2,\dots$ is the effective Landau level index counting the number of contributing Landau levels.
This is indeed observed in the FQHE at strong magnetic fields. The
quantization rule (\ref{fractional}) persists as long as new Fermi
surfaces open up with decreasing $h$. However, the first two
plateaus present in the FQHE $\nu_h=0, \pm 1$ are absent in eq.
(\ref{fractional2}). In order to get the Hall plateau $\nu_h=\pm
1$, the mixing term between two Fermi surfaces should probably be
taken into account (incoherent superposition), whereas the
conductivity (\ref{sigma-multiple}) includes the diagonal terms
only. We discuss the issue with $\nu_h=0$ further.

\begin{figure}[ht!]
\begin{center}
\includegraphics[width=8.5cm]{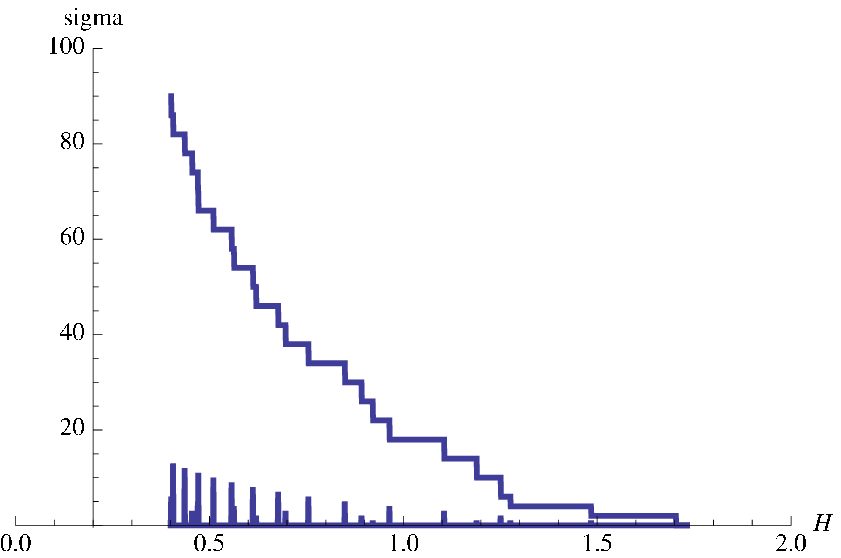}
\includegraphics[width=8.5cm]{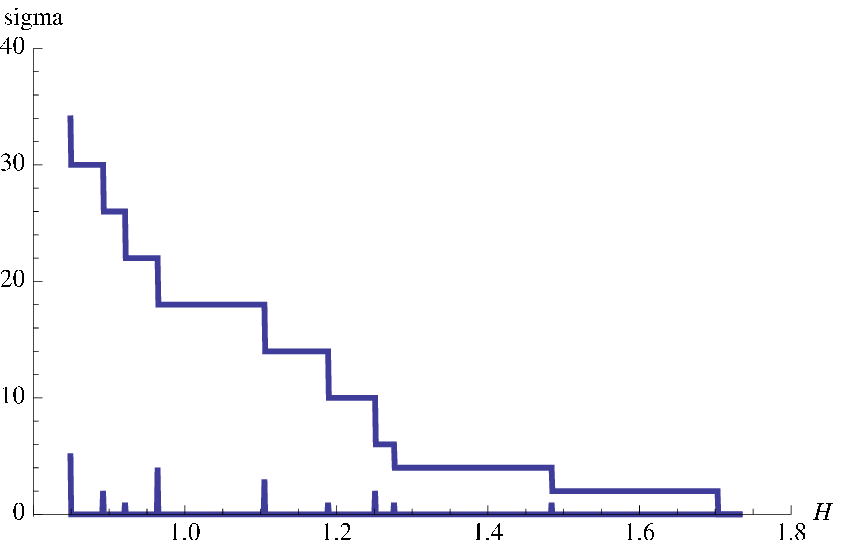}
\caption{Hall conductivity $\sigma_{xy}$ and longitudinal
conductivity $\sigma_{xx}$ vs. the magnetic field $h\rightarrow H$
at $T=0$ (we set $g_F=1$, $q=\frac{15}{\sqrt{3}}$). Contribution
from the first three Fermi surfaces are taken. At strong magnetic
fields, the Hall conductivity plateau $\nu_h=4$ appears from two
Fermi surfaces together with plateaus $\nu_h=2$ and $\nu_h=6$ when
one and three Fermi surfaces contribute, respectively. This
quantization rule is characteristic for the FQHE. At intermediate
and weak magnetic fields, the Hall conductivity plateaus are
produced as one of the three Fermi surfaces crosses the Landau
levels, resulting in the quantization rule of the IQHE. Irregular
pattern in the length of the plateaus is observed in the
experiment on thin films of graphite at strong magnetic fields
\cite{experiment}. The right panel is a zoom-in for large $h$ of
the left one.} \label{plot-hall-conductivity-3fs}
\end{center}
\end{figure}

In Fig. (\ref{plot-hall-conductivity-3fs}), we plot the Hall and
longitudinal conductivities at $T=0$ with three Fermi surfaces
contributing (eq. \ref{sigma-multiple}), where the individual
conductivities $\sigma^{(m)}$ are given by eq.
(\ref{conductivity3}). We fit the Fermi momenta by
\begin{eqnarray}
k_F^{(m)} = k_{F\;max}^{(m)}\sqrt{1-\frac{h^2}{3}}+\delta^{(m)},
\label{kF-multiple}
\end{eqnarray}
with $k_{F\;max}^{(1)}= 12.96,\; \delta^{(1)}=0.,\;
k_{F\;max}^{(2)}=10.29,\; \delta^{(2)}=1.5,\;
k_{F\;max}^{(3)}=9.75,\; \delta^{(3)}=3$, and use eq.
(\ref{kF-multiple}) together with the numerical solutions for
$v_F^{(m)}$ in Fig. \ref{plot-hall-conductivity-3fs}. In Fig.
\ref{plot-hall-conductivity-3fs}, at strong magnetic fields, the
Hall conductivity plateau $\nu_h=4$ originates from two Fermi
surfaces together with the plateaus $\nu_h=2$ and $\nu_h=6$ when
one and three Fermi surfaces contribute, respectively. As we
decrease the magnetic field further, three Fermi surfaces produce
plateaus characteristic for IQHE eq. (\ref{integer}). The
longitudinal conductivity shows a Dirac delta-like peak at the
beginning of each plateau. Since a finite contribution to the
conductivity arises as one of the three Fermi surfaces crosses the
next Landau level, the pattern is less regular (i.~e., the
plateaus have changing length) than in the case when only one
Fermi surface contributes. In Fig.
\ref{plot-hall-conductivity-3fs}, we compare the Hall
conductivities with one and three Fermi surfaces participating.
The irregular behavior of the Hall conductivity is explained
naturally from the picture with multiple Fermi surfaces.
Qualitatively similar regularity of the plateaus' length is seen
in experiments on thin films of graphite at strong magnetic fields
\cite{experiment}. The actual physics behind, however, might be
quite different as in this system multiple sheets of the Fermi
surface arise due to the (hexagonal) lattice on the UV scale, an
effect which is beyond the scope of our current model.

The somewhat regular pattern behind the irregular behavior can be
understood as a consequence of the appearance of a new energy
scale: the average distance between the Fermi levels. For the case
of Fig. \ref{plot-hall-conductivity-3fs}, we estimate it to be
$<\varepsilon_F^{(m)}-\varepsilon_F^{(m+1)}>=4.9$ with $m=1,2$.
The authors of \cite{Shovkovy:2006} explain the FQHE through the
opening of a gap in the quasiparticle spectrum, which acts as an
order parameter related to the particle-hole pairing and is
enhanced by the magnetic field (magnetic catalysis). Here, the
energy gap arises due to the participation of multiple Fermi
surfaces.

A pattern for the Hall conductivity that is strikingly similar to
Fig.(\ref{plot-hall-conductivity-3fs}) arises in the AA and
AB-stacked bilayer graphene, which has different transport
properties from the monolayer graphene \cite{bilayer-graphene},
compare with Figs.(2,5) there. It is remarkable that the bilayer
graphene also exhibits the insulating behavior in a certain
parameter regime. This agrees with our findings of
metal-insulating transition in our system.

\subsection{Metal-"strange metal" phase transition}

The previous discussion of conductivities and QHE is valid
provided that the Fermi velocity is nonzero. However, we have
shown that $v_F$ vanishes at relatively strong magnetic fields
(for the first Fermi surface it happens at $h_c$ as in Fig.
\ref{plot-conformal-dimension} and \ref{plot-fermi-velocity-one}).
In the AdS/CFT setting, the Fermi velocity vanishes when the IR
anomalous dimension is $\nu=\frac{1}{2}$ signalling the onset of a
nontrivial power law dispersion in Green's function
$G^{-1}(\omega)\sim\omega-v_fk_\perp+\omega^{2\nu}$ (the pole in
the self-energy $\Sigma\rightarrow G_R^{IR}\sim \omega^{2\nu}$ and
the pole in the prefactor of the linear term $\sim\omega$
\cite{Faulkner:2009}). Vanishing of $v_F$ was observed in
\cite{Hartman:2010} at large enough fermion charge. Note that if
$v_F$ is zero for some interval of the magnetic field, it leads to
the Hall plateau with the filling factor $\nu_h=0$ present in
FQHE.

\begin{figure}[ht!]
\begin{center}
\includegraphics[width=10cm]{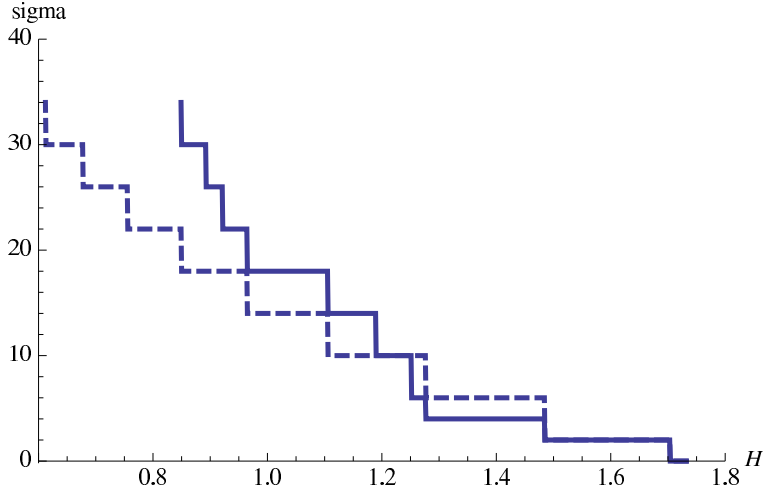}
\caption{Comparison of the Hall conductivities $\sigma_{xy}$ vs.
the magnetic field $h\rightarrow H$ from one Fermi surface (dashed
line) and from three Fermi surfaces (solid line). We set $g_F=1$
and $q=\frac{15}{\sqrt{3}}$. At strong magnetic fields, a new
plateau $\nu_h=4$} appears in the multiple Fermi surface picture,
yielding a pattern characteristic of FQHE.
\label{plot-hall-conductivity-3fs2}
\end{center}
\end{figure}

The vanishing of the Fermi velocity of the stable quasiparticle
leads to zero carrier density at leading order:
\begin{eqnarray}
v_F=0\rightarrow n=0.
\end{eqnarray}
This means that all contribution to conductivity comes from the
other terms, containing the contribution from the non-Fermi liquid
excitations and the conformal regime. This qualitatively changes
the transport properties of the system, as can be seen in Fig.
\ref{plot-hall-conductivity-3fs2}.

The finite offset magnetic field has been observed in experiments
on highly oriented pyrolitic graphite in magnetic fields
\cite{experiment-B}. In particular, analyzing the basal-plane
resistivity an approximate scaling relation between the critical
temperature of the metal-semiconducting transition and the
magnetic field has been found $T_c\sim\sqrt{h-h_c}$. It suggests
that at $T=0$, there is a threshold magnetic field $h_c$ above
which the resistivity qualitatively changes. Interestingly, the
existence of such a threshold magnetic field follows from the
AdS/CFT calculations ($h_c$ when $\nu=\frac{1}{2}$).

A phase transition is usually governed by an order parameter which
exhibits a critical behavior. In our case, there is no such order
parameter. However, it is interesting to note that the Fermi
momentum, according to eq. (\ref{kF-multiple}), behaves as
$k_F\sim \sqrt{h_{max}-h}$, which is in line with the postulated
critical behavior in the system, while the Fermi surface itself
behaves as order parameter.

To obtain a complete picture of the metal-"strange metal" phase
transition, one needs to perform calculations in the non-Fermi
liquid regime taking into account the unstable quasiparticle pole.
It is also necessary to study the temperature dependence of the DC
conductivities $\sigma_{xy}(T)$ and $\sigma_{xx}(T)$. We leave it
for the future study.

\section{Absence of the sign problem in holography.}\label{section:6}

In this section we show that the fermion determinant in the
gravity dual theory does not have a sign problem and hence can be
simulated by a lattice Monte-Carlo algorithm. Up to recently, most
of the work on AdS/CFT and applied holography focused on the
classical gravity (leading $1/N$ in field theory) limit. However,
many thermodynamic and electric properties depend on matter fields
(e.~g., the electrical conductivity depends on whether or not the
theory has a Fermi surface). In classical gravity, the
Einstein-Maxwell sector decouples, and matter fields run in loops
representing quantum oscillations. In order to include matter
fields in the bulk, one needs to calculate loop corrections, which
corresponds to going beyond the leading order in $1/N$. A study of
one-loop bulk physics was done in \cite{AdS-loops} and recently in
\cite{Denef:2009}. It shows that analytical calculations of
quantum corrections in the bulk are quite involved. The study of
quantum oscillations in the gravity dual will likely improve our
understanding of finite density systems in general.

As is well known, a finite density field theory in most cases
cannot be simulated on the lattice because of the infamous sign
problem \cite{signs}. In the field theory action, chemical potential is
introduced via the term $\bar{\psi}\mu\gamma^0\psi$ which is
Hermitian and therefore gives a complex determinant. At the same
time, in the bulk action, finite density is introduced through the
electrically charge black hole, and does not involve even matter
fields. This is the reason why the applied holography gives
universal predictions. In the leading order, the minimal
gravitational dual at finite density and temperature is the
electrically charged AdS-Reissner-Nordstr\"{o}m black hole, where
only the metric and Maxwell fields are present. Therefore, the
Einstein-Maxwell sector gives results which do not depend, for
example, on the charge and mass of matter fields in the
gravitational bulk spacetime, i.~e. are universal for a class of
field theories with different charge and scaling dimensions of the
operators. The fact that the chemical potential enters via the
electric field in the covariant derivative leads to the real and
positive definite fermion determinant which is suitable for
lattice simulations. We show it formally below.

In a semiclassical approach to gravity, the action includes the
Einstein-Maxwell sector $S_g$ with fields collectively denoted as
$g, A$ in eq. (\ref{action-g}) and the matter sector with the
fermion fields $S_{\psi}$,  eq. (\ref{action-psi}). The latter is
given as (Euclidean signature):
\begin{equation}
S_{\psi} = \int d^4x_E \sqrt{-g_E}\bar{\psi}\left(D+m\right)\psi,
\label{action-psieuc}
\end{equation}
where $D\equiv \Gamma_E^{M}{\mathcal D}_{M}$ and the covariant
derivative is ${\mathcal D}_M=\partial_M
+\frac{1}{4}\omega_{abM}\Gamma^{ab}-iqA_M$. We can always scale
away the spin connection by redefining the spinor field as in
eq.(\ref{rescale}). Finite density is described by the
electrically charged black hole with charge $Q$, that generates
the imaginary time component of the vector potential $A_{t_E}$
(eq. \ref{dimensionless3}). Radial profile of the vector potential
$A_{t_E}=\mu(1-\frac{1}{r})$ (in dimensionless units) ensures a
finite chemical potential at the field theory boundary
$A_{t_E}\rightarrow \mu$ at $r\rightarrow \infty$, where
$\mu=g_FQ$ (in dimensionless units). Integrating out the fermion
fields, the gravitational partition function can be written
schematically as
\begin{eqnarray}
Z=\sum_{g_{*},A_{*}}{\rm det}(D(g_{*},A_{*})+m){\rm e}^{-S_{g}[g_{*},A_{*}]},
\label{partition-function}
\end{eqnarray}
where $S_{g}$ is the Euclidean gravitational action at the saddle
points $g_{*},A_{*}$. The determinant describes fluctuations about
the saddle point solution $g_{*},A_{*}$ and corresponds to $1/N$
correction to the large $N$ limit of a dual gauge theory. Because
the Euclidean gamma matrices are Hermitian by convention (the
signature of the metric fixes the hermiticity), we have
$\Gamma_E^{0\dagger}=\Gamma_E^0$ and
$\Gamma_E^{i\dagger}=\Gamma_E^{i}$ with $i=1,2,3$, so the
covariant derivative is anti-Hermitian. Now it remains to show
that the determinant of this anti-Hermitian differential operator
is real and positive definite \cite{Zhang}.

Using the anti-commutation relations $\{\Gamma_E^5,\Gamma_E^{0,i}\}=0$, we have
\begin{eqnarray}
\Gamma_E^5D\Gamma_E^5=-D=D^{\dagger},
\label{antihermitian}
\end{eqnarray}
where $D\equiv D(g,A)$. Therefore the determinant
\begin{eqnarray}
{\rm det}D={\rm det}(\Gamma_E^5D\Gamma_E^5)={\rm det}D^{\dagger}=({\rm det}D)^{*},
\label{determinant}
\end{eqnarray}
is real. To show the positive definiteness, we remind the reader
that the eigenmodes of an anti-Hermitian derivative operator come
in pairs. If $(\lambda,\psi)$ is an eigenmode of $D$,
\begin{eqnarray}
D\psi=\lambda\psi,
\label{eigenvalue}
\end{eqnarray}
then, from eq. (\ref{eigenvalue}):
\begin{eqnarray}
D(\Gamma_E^5\psi) = (-\lambda)(\Gamma_E^5\psi),
\end{eqnarray}
so $(-\lambda,\Gamma_E^5\psi)$ is also an eigenmode of $D$. Due to
anti-hermiticity, from eq. (\ref{antihermitian}):
\begin{eqnarray}
D(\Gamma_E^5\psi) = \lambda^{*}(\Gamma_E^5\psi),
\end{eqnarray}
this eigenvalue is pure imaginary (or zero),
$-\lambda=\lambda^{*}$. The determinant is a product of all the
paired eigenvalues,
\begin{eqnarray}
{\rm det}(D+m)\rightarrow \Pi_i(\lambda_i+m)(-\lambda_i+m)=\Pi_i(|\lambda_i+m|^2),
\end{eqnarray}
which is positive definite (or zero).

In field theory, the eigenmodes of the operator
$D+\mu\gamma_E^4+m$ still come in pairs $(\lambda,\psi)$ and
$(-\lambda,\gamma_E^5\psi)$. However, since $\mu\gamma_E^4$ is
Hermitian, $\lambda$ is no longer pure imaginary, and therefore
${\rm det}(D+\mu\gamma_E^4+m)$ is not necessarily positive. The
sign problem occurs when ${\rm det}$ is negative for some gauge
configurations, or, in other words, it is generically present when
considering interacting matter at finite density.

\section{Conclusions}

We have studied strongly coupled electron systems in the magnetic
field focussing on the Fermi level structure, using the AdS/CFT
correspondence. These systems are dual to Dirac fermions placed in
the background of the electrically and magnetically charged
AdS-Reissner-Nordstr\"{o}m black hole. At strong magnetic fields
the dual system "lives" near the black hole horizon, which
substantially modifies the Fermi level structure. As we dial the
magnetic field higher, the system exhibits the non-Fermi liquid
behavior and then crosses back to the conformal regime. In our
analysis we have concentrated on the the Fermi liquid regime and
obtained the dependence of the Fermi momentum $k_F$ and Fermi
velocity $v_F$ on the magnetic field. Remarkably, $k_F$ exhibits
the square root behavior, with $v_F$ staying close to the speed of
light in a wide range of magnetic fields, while it rapidly
vanishes at a critical magnetic field which is relatively high.
Such behavior indicates that the system may have a phase
transition.

The magnetic system can be rescaled to a zero-field configuration
which is thermodynamically equivalent to the original one. This
simple result can actually be seen already at the level of field
theory: the additional scale brought about by the magnetic field
does not show up in thermodynamic quantities meaning, in
particular, that the behavior in the vicinity of quantum critical
points is expected to remain largely uninfluenced by the magnetic
field, retaining its conformal invariance. In the light of current
condensed matter knowledge, this is surprising and might in fact
be a good opportunity to test the applicability of the probe limit
in the real world: if this behavior is not seen, this suggests
that one has to include the backreaction to metric to arrive at a
realistic description.

In the field theory frame, we have calculated the DC conductivity
using $k_F$ and $v_F$ values extracted from holography. The
holographic calculation of conductivity that takes into account
the fermions corresponds to the corrections of subleading order in
$1/N$ in the field theory and is very involved
\cite{Faulkner:2010}. As we are not interested in the vertex
renormalization due to gravity (it does not change the magnetic
field dependence of the conductivity), we have performed our
calculations directly in the field theory with AdS gravity-dressed
fermion propagators. Instead of controlling the occupancy of the
Landau levels by changing the chemical potential (as is usual in
non-holographic setups), we have controlled the filling of the
Landau levels by varying the Fermi energy level through the
magnetic field. At zero temperature, we have reproduced the
integer QHE of the Hall conductivity, which is observed in
graphene at moderate magnetic fields. While the findings on
equilibrium physics (Landau quantization, magnetic phase
transitions and crossovers) are within expectations and indeed
corroborate the meaningfulness of the AdS/CFT approach as compared
to the well-known facts, the detection of the QHE is somewhat
surprising as the spatial boundary effects are ignored in our
setup. We plan to address this question in further work.

Interestingly, the AdS geometry produces several Fermi surfaces.
Theories where the gravity duals have larger fermion charge $q$
posses more Fermi surfaces. We find that, in a multi-Fermi surface
picture, the Hall conductivity is quantized in a way reminiscent
of fractional QHE. By reducing the magnetic field, new Fermi
surfaces open up and the quantization of Hall conductivity
alternates between two different patterns, corresponding to odd
and even number of Fermi surfaces. It turns out that odd number of
the Fermi surfaces results in IQHE plateaus, while even number of
surfaces gives new plateaus characteristic for the FQHE. In a
multi-Fermi surface picture, the quantum Hall plateaus show a less
regular pattern that agrees with experiments on thin graphite in
strong magnetic field \cite{experiment}. In our model it happens
due to the fact that as one of several Fermi surfaces crosses the
Landau level, the Hall conductivity jumps to a new plateau. This
process is not synchronized between different Fermi surfaces. We
associate the average distance between the Fermi levels with the
energy gap usually arising in the FQHE.

Notably, the AdS-Reissner-Nordstr\"{o}m black hole background
gives a vanishing Fermi velocity at high magnetic fields. It
happens at the point when the IR conformal dimension of the
corresponding field theory is $\nu=\frac{1}{2}$, which is the
borderline between the Fermi and non-Fermi liquids. Vanishing
Fermi velocity was also observed at high enough fermion charge
\cite{Hartman:2010}. As in \cite{Hartman:2010}, it is explained by
the red shift on the gravity side, because at strong magnetic
fields the fermion wavefunction is supported near the black hole
horizon modifying substantially the Fermi velocity. In our model,
vanishing Fermi velocity leads to zero occupancy of the Landau
levels by stable quasiparticles that results in vanishing regular
Fermi liquid contribution to the Hall conductivity and the
longitudinal conductivity. The dominant contribution to both now
comes from the non-Fermi liquid and conformal contributions. We
associate such change in the behavior of conductivities with a
metal-"strange metal" phase transition. Experiments on highly
oriented pyrolitic graphite support the existence of a finite
"offset" magnetic field $h_c$ at $T=0$ where the resistivity
qualitatively changes its behavior \cite{experiment-B}. At $T\neq
0$, it has been associated with the metal-semiconducting phase
transition \cite{experiment-B}. It is worthwhile to study the
temperature dependence of the conductivity in order to understand
this phase transition better.

Finally, we suggest as a possibly interesting extension of the
current AdS/CFT methodology to compute the gravity dual of the
finite density matter in Monte-Carlo lattice simulations. This is
possible since the sign problem does not arise in the holographic
setting of a finite density system. Unlike the conventional field
theory setup, finite density in holography is introduced through
an electrically charged black hole, and does not involve matter
fields (this is also the reason why holography gives universal
results: it does not depend on the expectation values of matter
fields at the leading order). In the gravity geometry, Dirac
fermions are coupled minimally to the electric field via the
covariant derivative. We have shown that the covariant derivative
is anti-Hermitian in Euclidean signature, leading to the real and
positive definite fermion determinant. This makes it possible to
simulate finite density systems on the lattice in the AdS-gravity
geometry using the curved space-time lattice formulation
\cite{curved-lattice}. The simplest holographic setup which
describes a finite charge density system includes a local $U(1)$
gauge symmetry. Finite density systems with global $U(1)$ symmetry
can not be simulated numerically in field theory due to the
problem with the Gauss law in the lattice formulation. Another
important advantage of performing Monte-Carlo simulation is that
it includes the quantum fluctuations for the gauge and
gravitational field. So far most calculations have been done in
the probe limit, with the frozen background for the metric and
gauge fields. Analytic calculations which include backreaction are
usually involved and are done in the next to leading order, e.g.
\cite{Hartnoll:2010}. Holographic lattice calculations allow to
consider dynamical gauge and gravity fields with matter, which
mimics complicated strong interactions in finite density systems
and opens a way toward studying novel state of matter and
instability mechanisms.

\section*{Acknowledgements}
We thank Tom Faulkner, Michael Fromm, Tom Hartman, Sean Hartnoll,
Nabil Iqbal, Hong Liu, Mark Mezei, Hiranmaya Mishra, Volodya
Miransky, Yusuke Nishida, Owe Philipsen, Igor Shovkovy, Tian
Zhang, Darius Sadri, Chang Yu Hou, Andrej Mesaro\v{s} and Vladimir
Juri\v{c}i\'{c} for helpful inputs and discussions. The work was
supported in part by the Alliance program of the Helmholtz
Association, contract HA216/EMMI ``Extremes of Density and
Temperature: Cosmic Matter in the Laboratory'' and by ITP of
Goethe University, Frankfurt (E.~Gubankova), by a VIDI Innovative
Research Incentive Grant (K.~Schalm) from the Netherlands
Organization for Scientific Research (NWO), by a Spinoza Award
(J.~Zaanen) from the Netherlands Organization for Scientific
Research (NWO) and the Dutch Foundation for Fundamental Research
of Matter (FOM). K.~Schalm thanks the Galileo Galilei Institute
for Theoretical Physics for the hospitality and the INFN for
partial support during the completion of this work.

\clearpage
\appendix

\section{Dirac equation in magnetic field}\label{appendix:a}

Here we solve analytically the part of the Dirac equation which
depends on magnetic field and space-time coordinates of the
boundary theory. The free spinor action in the geometry given by
eq. (\ref{ads4-metric1}) and in the presence of magnetic field
(\ref{ads4-metric2}) is given by eq. (\ref{action-psi}).

Using the translational invariance,
\begin{equation}
\psi(t,x,y,r)=\int d\omega dk {\rm e}^{-i\omega t+iky}\;\psi(\omega,k,x,r),
\end{equation}
with $k\equiv k_y$, the Dirac equation (eq. \ref{direq}) can be
written as
\begin{eqnarray}
&&\hspace{-1cm}\left(\frac{1}{\sqrt{-g_{tt}}}\Gamma^{\hat{t}}(-i\omega+\frac{1}{2}\omega_{\hat{t}\hat{r}t}
 \Gamma^{\hat{t}\hat{r}}-iqA_t(r))
+\frac{1}{\sqrt{g_{rr}}}\Gamma^{\hat{r}}\partial_r
+\frac{1}{\sqrt{g_{ii}}}\Gamma^{\hat{x}}(\partial_x+\frac{1}{2}\omega_{\hat{x}\hat{r}x}
\Gamma^{\hat{x}\hat{r}})\right.\nonumber\\
&+&\left. \frac{1}{\sqrt{g_{ii}}}\Gamma^{\hat{y}}(ik+\frac{1}{2}\omega_{\hat{y}\hat{r}y}
\Gamma^{\hat{y}\hat{r}}-iqA_y(x))
-m \right)\psi(\omega,k,x,r)=0,
\label{dirac-connection}
\end{eqnarray}
where $g_{ii}\equiv g_{xx}=g_{yy}$, and $A_t(r)=\mu(1-r_0/r)$, $A_y(x)=hx$.
From the torsion-free condition, $\omega^a_b\wedge e^b=-de^a$, we find the spin connection \cite{Carroll:2003}
for the metric \ref{ads4-metric1},
\begin{eqnarray}
\omega_{\hat{t}\hat{r}}=-\frac{\partial_r(\sqrt{-g_{tt}})}{\sqrt{g_{rr}}}dt,\;\;
\omega_{\hat{i}\hat{r}}=\frac{\partial_r(\sqrt{g_{ii}})}{\sqrt{g_{rr}}}dx^i,
\end{eqnarray}
where $i=x,y$. Note that
\begin{equation}
-\Gamma^{\hat{t}}\Gamma^{\hat{t}\hat{r}}=\Gamma^{\hat{x}}\Gamma^{\hat{x}\hat{r}}=
\Gamma^{\hat{y}}\Gamma^{\hat{y}\hat{r}}=\Gamma^{\hat{r}},
\end{equation}
and
\begin{eqnarray}
\frac{1}{4}e^{M}_{\hat{a}}\Gamma^{\hat{a}}\omega_{\hat{b}\hat{c}M}\Gamma^{\hat{b}\hat{c}} &=&
\frac{1}{4}\frac{1}{\sqrt{-g_{tt}}}\frac{\partial_r(\sqrt{-g_{tt}})}{\sqrt{g_{rr}}}\Gamma^{\hat{r}}+
\frac{2}{4}\frac{1}{\sqrt{g_{ii}}}\frac{\partial_r\sqrt{g_{ii}}}{\sqrt{g_{rr}}}\Gamma^{\hat{r}}\nonumber\\
&=& \frac{1}{\sqrt{g_{rr}}}\Gamma^{\hat{r}}\partial_r \ln \left(-\frac{g}{g_{rr}}\right)^{1/4},
\end{eqnarray}
where $g$ is the determinant of the metric. Therefore, we can
rescale the spinor field:
\begin{equation}
\psi=\left(-\frac{g}{g_{rr}}\right)^{-1/4}\Phi,
\label{rescale}
\end{equation}
and remove the spin connection completely. The new covariant
derivative does not contain the spin connection so ${\mathcal
D}_M^{\prime}=\partial_M -iqA_M$.

In new field variables, the Dirac equation is given by
\begin{eqnarray}
&& \hspace{-2cm} \left(\frac{\sqrt{g_{ii}}}{\sqrt{g_{rr}}}\Gamma^{\hat{r}}\partial_r -
\frac{\sqrt{g_{ii}}}{\sqrt{-g_{tt}}}\Gamma^{\hat{t}}\;i(\omega+\mu_q(1-\frac{r_0}{r}))
-\sqrt{g_{ii}}m +\Gamma^{\hat{x}}\;\partial_x\right.\nonumber\\
&+&\left. \Gamma^{\hat{y}}\;i(k-qhx)\right)\Phi(\omega,k,x,r)=0,
\label{dirac-equation-start}
\end{eqnarray}
with $\mu_q\equiv \mu q$. We separate the $x$- and $r$-dependent
parts:
\begin{eqnarray}
P(r) &=& \frac{\sqrt{g_{ii}}}{\sqrt{g_{rr}}}\Gamma^{\hat{r}}\partial_r
-\frac{\sqrt{g_{ii}}}{\sqrt{-g_{tt}}}\Gamma^{\hat{t}}\;
i(\omega+\mu_q(1-\frac{r_0}{r}))-\sqrt{g_{ii}}m,\nonumber\\
Q(x) &=& \Gamma^{\hat{x}}\partial_x+\Gamma^{\hat{y}}\;i(k-qhx),
\end{eqnarray}
and the Dirac equation is
\begin{equation}
(P(r)+Q(x))\Phi=0.
\end{equation}
Even though $[P(r),Q(x)]\neq 0$, one can find a transformation
matrix $U$ such that $[UP(r),UQ(x)]=0$, and then look for common
eigenvectors of $UP(r)$ and $UQ(x)$ as they are commuting
Hermitian operators, i.~e., the Dirac equation reads
\begin{equation}
 UP(r)\Phi_l=-UQ(x)\Phi_l=\lambda_l\Phi_l,
\label{dirac2}
\end{equation}
where $l$ labels the Landau levels. We use $l$ for the Landau
index, so as not to confuse it with the Matsubara frequency index
$n$. Transformation matrix $U$ should satisfy the conditions
\begin{equation}
\{U,\Gamma^{\hat{r}}\}=0,\;\,
\{U,\Gamma^{\hat{t}}\}=0,\;\;
[U,\Gamma^{\hat{x}}]=0,\;\;
[U,\Gamma^{\hat{y}}]=0,
\end{equation}
which do not fix $U$ completely. It is convenient to use the
following basis \cite{Faulkner:2009}:
\begin{eqnarray}
&& \Gamma^{\hat{r}}= \left(\begin{array}{cc}
-\sigma^3 & 0 \\
0 & -\sigma^3
\end{array}
\right),\;\;
\Gamma^{\hat{t}}= \left(\begin{array}{cc}
i\sigma^1 & 0 \\
0 & i\sigma^1
\end{array}
\right),\;\;
\Gamma^{\hat{x}}= \left(\begin{array}{cc}
-\sigma^2 & 0 \\
0 & \sigma^2
\end{array}
\right),\;\;
\nonumber\\
&& \Gamma^{\hat{y}}= \left(\begin{array}{cc}
0 & \sigma^2 \\
\sigma^2 & 0
\end{array}
\right),\;\;
\Gamma^{\hat{5}}= \left(\begin{array}{cc}
0 & i\sigma^2 \\
-i\sigma^2 & 0
\end{array}
\right).
\label{matrices}
\end{eqnarray}
Note, that the following relation holds
\begin{equation}
 \Gamma^{\hat{5}}=\Gamma^{\hat{0}}\Gamma^{\hat{1}}\Gamma^{\hat{2}}\Gamma^{\hat{3}},
\end{equation}
as expected, with $0\rightarrow t$, $1\rightarrow x$,
$2\rightarrow y$, $3\rightarrow r$. In the representation of eq.
(\ref{matrices}), we can choose
\begin{equation}
 U=\left(\begin{array}{cc}
-i\sigma^2 & 0 \\
0 & -i\sigma^2
\end{array}
\right).
\label{transform}
\end{equation}
We split the $4$-component spinors
into two $2$-component spinors (we do not write zero entries)
$F=(F_1,F_2)^{T}$ where the index $\alpha=1,2$
is the Dirac index of the boundary theory, using projectors
\begin{equation}
\Pi_{\alpha}=\frac{1}{2}(1-(-1)^{\alpha}\Gamma^{\hat{r}}\Gamma^{\hat{t}}\Gamma^{\hat{1}}),\;\;
\alpha=1,2,\;\; \Pi_1+\Pi_2=1,
\label{projection}
\end{equation}
which commute with the Dirac operator of eq. (\ref{dir2}), and
$F_{\alpha}=\Pi_{\alpha}\Phi$, $\alpha=1,2$, decouple from each
other. Gamma matrices in eq. (\ref{matrices}) were chosen in such
a way that this decoupling is possible.

Writing the Dirac equation (\ref{dirac2}) for
$F=(F_1,F_2)^{T}$, we have
\begin{eqnarray}
&& \hspace{-1cm} \left(-\frac{\sqrt{g_{ii}}}{\sqrt{g_{rr}}}\sigma^1\partial_r+\sqrt{g_{ii}}i\sigma^2 m
-\frac{\sqrt{g_{ii}}}{\sqrt{-g_{tt}}}\sigma^3(\omega+\mu_q(1-r_0/r))
-\lambda_n\right) \otimes 1
\left(\begin{array}{c}
F_1 \\
F_2
\end{array}
\right)=0  \nonumber\\
&&\hspace{-1cm}  1 \otimes \left(\begin{array}{cc}
i\partial_x+\lambda_l & (k-qhx)  \\
(k-qhx) & -i\partial_x+\lambda_l
\end{array}
\right)
\left(\begin{array}{c}
F_1 \\
F_2
\end{array}
\right)=0, \label{dirac-equationapp}
\end{eqnarray}
where in $X\otimes Y$, $X$ acts inside $F_1$ or $F_2$ and $Y$ acts
between $F_1$ and $F_2$. In eq. (\ref{dirac-equationapp}), the $1$
in the first equation shows that there is no mixing of $F_1$ and
$F_2$ by the operator $UP(r)$ and the $1$ in the second equation
shows that there is no mixing inside $F_1$ or $F_2$ by the
operator $UQ(x)$. Therefore, the solution can be represented as
\begin{equation}
\left(\begin{array}{c}
F_1\\
F_2
\end{array}
\right)=
\left(\begin{array}{c}
f^{(1)}_l(r)g^{(1)}_l(x)\\
f^{(2)}_l(r)g^{(1)}_l(x)\\
f^{(1)}_l(r)g^{(2)}_l(x)\\
f^{(2)}_l(r)g^{(2)}_l(x)
\end{array}
\right).
\end{equation}
We do not write explicitly the dependence on $\omega$ and $k$. It
is convenient to make a unitary transformation:
\begin{eqnarray}
\left(\begin{array}{c}
\zeta^{(1)}\\ \zeta^{(2)}
\end{array}\right) =M
\left(\begin{array}{c}
g^{(1)}\\ g^{(2)}
\end{array}\right),\;\;
M=\left(\begin{array}{cc}
1 & -i \\ -i & 1
\end{array}\right).
\label{basischange}
\end{eqnarray}
Dirac equations for each component are written as:
\begin{eqnarray}
\hspace{-1.5cm} \left(\frac{\sqrt{g_{ii}}}{\sqrt{g_{rr}}}\partial_r +\sqrt{g_{ii}} m \right)f^{(1)}_l(r)
+ \left(-\frac{\sqrt{g_{ii}}}{\sqrt{-g_{tt}}}(\omega+\mu_q(1-r_0/r))
+\lambda_l\right)f^{(2)}_l(r) &=& 0, \nonumber\\
\hspace{-1.5cm} \left(\frac{\sqrt{g_{ii}}}{\sqrt{g_{rr}}}\partial_r -\sqrt{g_{ii}} m \right)f^{(2)}_l(r)
+ \left(\frac{\sqrt{g_{ii}}}{\sqrt{-g_{tt}}}(\omega+\mu_q(1-r_0/r))
+\lambda_l\right) f^{(1)}_l(r) &=& 0,
\label{eqr} \\
\hspace{-1.5cm} (\partial_{\tilde{x}}-\tilde{x})\zeta^{(1)}+\tilde{\lambda}_l\zeta^{(2)} &=& 0, \nonumber\\
\hspace{-1.5cm} (\partial_{\tilde{x}}+\tilde{x})\zeta^{(2)}-\tilde{\lambda}_l\zeta^{(1)} &=& 0.
\label{eqx}
\end{eqnarray}
In equations \ref{eqx} for the $x$-dependent part, we have
rescaled $\tilde{x}=\sqrt{|qh|}\,(x-\frac{k}{qh})$ with $k\equiv
k_y$ and $\lambda_l=\sqrt{|qh|}\,\tilde{\lambda}_l$. The second
order ordinary differential equations
\begin{eqnarray}
&& -\partial^2_{\tilde{x}}\zeta^{(\rho)}+\tilde{x}^2\zeta^{(\rho)}-\tilde{\lambda}_l^2\zeta^{\rho}
-(-1)^{\rho}\zeta^{(\rho)}=0,
\end{eqnarray}
with $\rho=1,2$, are solved by substitution $\zeta^{(\rho)}={\rm
e}^{-\tilde{x}^2/2}\tilde{\zeta}^{(\rho)}$. This is exactly the
Schr\"odinger equation for a harmonic oscillator, so the
eigenfunctions are Hermite polynomials and we obtain the following
solutions, indexed by an integer $l\in\mathbb{Z}$ that is related
to the eigenvalue $\lambda_l$ by $\lambda_l = \sqrt{2|qh|l}$:
\begin{eqnarray}
\zeta_{l}^{(1)}(\tilde{x}) &=& N_{l-1} e^{-\tilde{x}^2/2}H_{l-1}(\tilde{x})\nonumber \\
\zeta_{l}^{(2)}(\tilde{x}) &=& N_{l} e^{-\tilde{x}^2/2}H_{l}(\tilde{x}).
\label{zeta}
\end{eqnarray}
The normalization constant $N_l$ is proportional to $1/\sqrt{2^l
l!}$. Substituting the solutions from eq. (\ref{zeta}) into the
first order eigenvalue equation with $x$-dependence gives the
following solutions:
\begin{eqnarray}
F_{l}=\left(\begin{array}{c}
   f^{(1)}_l(r)\zeta^{(1)}_l(\tilde{x})\\
   f^{(2)}_l(r)\zeta^{(1)}_l(\tilde{x})\\
   f^{(1)}_l(r)\zeta^{(2)}_l(\tilde{x})\\
   -f^{(2)}_l(r)\zeta^{(2)}_l(\tilde{x})
  \end{array}\right),\;\; \lambda_l=\sqrt{2|qh|l},
\label{solution1}
 \end{eqnarray}
and
\begin{eqnarray}
\tilde{F}_{l}=\left(\begin{array}{c}
   \tilde{f}^{(1)}_l(r)\zeta^{(1)}_l(\tilde{x})\\
   \tilde{f}^{(2)}_l(r)\zeta^{(1)}_l(\tilde{x})\\
   -\tilde{f}^{(1)}_l(r)\zeta^{(2)}_l(\tilde{x})\\
   \tilde{f}^{(2)}_l(r)\zeta^{(2)}_l(\tilde{x})
  \end{array}\right),\;\; \lambda_l=-\sqrt{2|qh|l}.
\label{solution2}
 \end{eqnarray}
Solving the first order $x$-dependent equation, we get the same
eigenvalue, but slightly different eigenfunctions for different
signs of $qh$. In particular, e.~g., for $F$, the pairs
$f^{(1)}(\zeta^{(1)},\zeta^{(2)})^{T}$ and
$f^{(2)}(\zeta^{(1)},-\zeta^{(2)})^{T}$ correspond to $qh>0$ and
$qh<0$, respectively. Different sign of $qh$ stands for the
positive and negative Landau level index $l$.

Finally, the general solution to the Dirac equation is given by a
linear combination of eqs. (\ref{solution1}) and
(\ref{solution2}):
\begin{equation}
F_{sol}=\sum_l (a_lF_{l}+b_l\tilde{F}_l).
\label{general-solution}
\end{equation}

Using the eigenvalues determined by eqs.
(\ref{solution1}-\ref{solution2}) in the equation for the radial
part (\ref{eqr}), we get
\begin{eqnarray}
&& \hspace{-1.5cm}\left(-\frac{1}{\sqrt{g_{rr}}}\sigma^3\partial_r- m
+\frac{1}{\sqrt{-g_{tt}}}\sigma^1(\omega+\mu_q(1-r_0/r))\right.\nonumber\\
&-&\left. \frac{1}{\sqrt{g_{ii}}}i\sigma^2\sqrt{2|qh|l}\right) \otimes 1
\left(\begin{array}{c}
F_1 \\
F_2
\end{array}
\right)=0,
\end{eqnarray}
with $l=0,1,dots$; and the same for $\tilde{F}$ replacing
$\sqrt{2|qh|l}\rightarrow -\sqrt{2|qh|l}$. It coincides with eq.
(A14) in \cite{Faulkner:2009} (Dirac equation at zero magnetic
field) with the momentum replaced by the Landau level eigenvalue
\cite{Denef:2009}
\begin{equation}
k\rightarrow \pm\sqrt{2|qh|l}.
\label{replace}
\end{equation}
Equation (\ref{replace}) also gives a prescription on how to treat
the limit of zero magnetic field $h\rightarrow 0$. The limit is to
be taken keeping, e.~g., $2|qh|(l+1)\equiv k_F^2$ fixed as
$h\rightarrow 0$. In a compact form, the Dirac equation in a
magnetic field (\ref{dirac-equation-start}) is given by
\begin{eqnarray}
&& \hspace{-1.5cm}\left(\frac{1}{\sqrt{g_{rr}}}\Gamma^{\hat{r}}\partial_r
-\frac{1}{\sqrt{-g_{tt}}}\Gamma^{\hat{t}}\;i(\omega+qA_t)-m\right.\nonumber\\
&-& \left.\frac{1}{\sqrt{g_{ii}}} U^{-1}\sqrt{2|qh|l}\right)F(r)=0,
\label{dir}
\end{eqnarray}
with $F=(F_1,F_2)^{T}$, $l=0,1,\ldots$, for $\tilde{F}$ replace
$\sqrt{2|qh|l}\rightarrow -\sqrt{2|qh|l}$, and $U^{-1}$ is the
matrix inverse to the matrix given by eq. (\ref{transform}):
\begin{equation}
 U^{-1}=\left(\begin{array}{cc}
i\sigma^2 & 0 \\
0 & i\sigma^2
\end{array}
\right). \label{transforminv}
\end{equation}
which we use in the main text.

\section{Spectral function}

In what follows we use the dimensionless variables
(\ref{dimensionless1}-\ref{dimensionless3}). Following the
analysis of \cite{Faulkner:2009}, the flow of the Green's function
is determined by
\begin{equation}
\label{GreenFunction}
G_R(\omega,l) =\left. \lim_{\epsilon\to 0}\epsilon^{-2m}\left(\begin{array}{cc}
\xi_{+}^{(l)} & 0 \\
0 & \xi_{-}^{(l)} \\
\end{array}\right)\right|_{r=\frac{1}{\epsilon}},
\end{equation}
where $\xi_{+}^{(l)}(r) =\frac{f^{(2)}}{f^{(1)}}$ and
$\xi_{-}^{(l)}(r)=\frac{\tilde{f}^{(2)}}{\tilde{f}^{(1)}}$ from
the solutions (\ref{solution1}-\ref{solution2}). In obtaining this
relation, we absorbed the coefficients appearing in eq.
(\ref{general-solution}) into the definitions of the radial
functions. The functions $\xi_{\pm}^{(l)}$ satisfy the following
differential equation \cite{Faulkner:2009}:
\begin{eqnarray}
\sqrt{\frac{g_{ii}}{g_{rr}}}\partial_r\xi_{\pm}^{(l)}=-2m\sqrt{g_{ii}}\xi_{\pm}^{(l)}+(u(r)\pm\lambda_l)^2(\xi_{\pm}^{(l)})^2
+(u(r)\mp \lambda_l),
\end{eqnarray}
with $u(r)$ given by
\begin{equation}
u(r)=\sqrt{\frac{g_{ii}}{-g_{tt}}}(\omega+qA_t(r)).
\end{equation}
Writing explicitly in the metric eq.(\ref{dim-metric}), we have
\begin{equation}
\label{eq:floweqGF}
 r^2\sqrt{f}\partial_r\xi_{\pm}^{(l)} =
 -2mr\xi_{\pm}^{(l)} + (u(r)\pm\lambda_l)(\xi_{\pm}^{(l)})^2+ (u(r)\mp\lambda_l),
\end{equation}
where $u(r)$ is given by
\begin{equation}
 u(r) = \frac{1}{\sqrt{f}}(\omega + \mu_q(1-\frac{1}{r}))
\end{equation}
with $f=\frac{(r-1)^2(r^2+2r+3)}{r^4}$ at $T=0$. Near the horizon
($r=1$) the flow equation reduces to
\begin{equation}
r^2\partial_r\xi_{\pm}^{(l)}=\frac{1}{f}(\xi_{\pm}^{(l)}+1)^2,
\end{equation}
which due to the double zero in $f$ has a regular solution only if
$\xi_{\pm}(r=1)=\pm i$. Writing the radial equation in terms of
$\xi$ and choosing the infalling boundary conditions fixes
$\xi_{\pm}^{(l)}(r=1)=i$.

The key quantity that we extract from the Green's function
is the fermionic spectral function
\begin{equation}
 A(\omega,l_x,k_y) = \mathrm{Tr}\left(\mathrm{Im }\,G_R(\omega,l_x,k_y)\right),
\end{equation}
which we analyze in the main text of the paper.

\end{document}